%% file: mythesis.tex
\documentclass[doublespacing]{lsuthesis}

\usepackage{lsutitle}

\usepackage{dcolumn}
\usepackage{float}
\usepackage{bm}
\usepackage[colorlinks=true, linkcolor=black, urlcolor=black, linkbordercolor=white]{hyperref}
\usepackage[utf8]{inputenc}	
\usepackage{csquotes}
\usepackage{lmodern} 
\usepackage{siunitx}
\usepackage[version=4]{mhchem}
\usepackage{microtype}
\usepackage{amsmath}
\usepackage{ragged2e}

\usepackage{titletoc}
\usepackage{xcolor}
\usepackage{caption}
\usepackage{url}
\usepackage{diagbox}
\usepackage{bm} 
\usepackage{amsfonts} 
\usepackage{booktabs}

\justifying

\newcommand{\bra}[1]{\langle#1|}
\newcommand{\ket}[1]{|#1\rangle}
\newcommand{\braket}[1]{\langle#1\rangle}
\newcommand{\pare}[1]{\left( #1 \right)}

\newcommand{\llav}[1]{\left\lbrace #1 \right\rbrace}

\DeclareMathOperator{\sinc}{sinc}

\usepackage{tocloft}

\title{Multiphoton Quantum Sensing}
\thesistype{Dissertation}
\department{The Department of Physics and Astronomy}
\soughtdegree{Doctor of Philosophy}
\author{Fatemeh Mostafavikhatam}
\degrees{M.S., University of Texas Rio Grande Valley, 2019
  }
\graduationdate{August 2024}

\begin{document}

\frontmatter

\maketitle


\begin{centeredpage}
For those who bring love, joy, and happiness to this world.
\end{centeredpage}


\chapter{Acknowledgments}
I owe gratitude to numerous individuals who have contributed to my academic journey. Foremost, I am profoundly thankful for the continuous love, support, and encouragement of my family, particularly my sister Maryam, whose influence has been profound.\\
\noindent
My sincere appreciation extends to all my past and present educators,  who have played pivotal roles in shaping my academic path. Special appreciation goes to the late Dr. Jonathan Dowling for his early support, and to Dr. Ward Plummer, for welcoming me and giving me the opportunity to join his research group.\\
\noindent
To my dear old friends, your enduring friendship has provided immense positivity and strength.
To Paige Whittington for her assistance in the Physics department.
Additionally, I must thank my doctoral committee: Drs. Omar Magaña-Loaiza, Joyoni Dey, Juana Moreno, and Guoxiang Gu for their dedicated service.\\
\noindent
I am grateful to my fellow members of the Quantum Photonics group: Mingyuan Hong, Dr Michelle Lollie, Riley Dawkins, Jannatul Ferdous, and Josh Fabre. It has been a pleasure working with you all. Finally, I express deep appreciation to my advisor, Dr. Omar Magaña-Loaiza, and Dr. Chenglong You, for imparting the significance of diligent and independent work, and also the value of collaboration.\\
\noindent
Last but not least, I would like to acknowledge the financial support from the Physics department, the US Department of Energy, Office of Basic Energy Sciences, Division of Materials Sciences and Engineering under Award DE-SC0021069, National Science Foundation through Grant No. ECCS-2225986.

\newpage

\textcolor{black}{\tableofcontents}

\newpage


\listoftables

\newpage

\listoffigures



\chapter{Abstract}
While the fundamental principles of light-matter interaction are well-understood and drive countless technologies, the world of multiphoton processes remains a fascinating puzzle, holding the potential to drastically alter our understanding of how light interacts with matter at its most basic level \cite{DELLANNO200653,RevModPhysPan,PhysRevLettSun,Deng2006}. This rich interplay of light and matter unveils novel phenomena that can be harnessed for sensing with exceptional precision, as exemplified by multiphoton quantum sensing. This thesis delves into the applications of multiphoton quantum protocols, particularly in imaging, communication, and plasmonic sensing, to surpass classical limitations and achieve enhanced sensitivity. We explore the potential of multiphoton quantum processes, particularly in the nanoscale regime and within subsystems of macroscopic systems, where novel and ultra-sensitive sensing methodologies emerge. Subsequent chapters of this thesis demonstrate the transformative potential of multiphoton quantum sensing, elucidating the design, implementation, and experimental results of specific sensing protocols tailored to diverse applications. Our analysis combines experimental observations and theoretical predictions to assess the sensitivity and performance of these protocols. Additionally, the thesis discusses potential future directions and advancements in the field, envisioning applications in biomolecule detection, environmental monitoring, and fundamental studies of light-matter interactions at the nanoscale. Concluding reflections highlight the implications of multiphoton quantum sensing across scientific disciplines and lay the groundwork for future research endeavors.

\mainmatter

\include{chapter1}

\include{chapter2}

\include{chapter3}

\include{chapter4}


\include{chapter5} 

\include{chapter6}

\include{chapter7}



\appendix
\label{app:A}
\chapter{Supplementary Note}
\section{P Function Theory Description}
\label{app:A:PFunction}
Let us start by noticing from Fig. \ref{fig:figure1-ch2}b of Chapter \ref{chap:chapter2} that the detected field after the slits is the result of three contributing sub-fields. The first two contributions correspond to the horizontally- and vertically-polarized fields that traverse the illuminated slit, with mean photon numbers $ \eta\Bar{n}_{\text{s}}$, and $(1-\eta) \Bar{n}_{\text{s}}$, respectively. Note that $ \Bar{n}_{\textbf{s}}$ is the total mean photon number of the field after it has traversed the slit and $ \eta=\cos^{2}{\theta}$, with $\theta$ describing the polarization angle of the initial illuminating photons, defined with respect to the vertical axis. The third contribution is the horizontally-polarized field produced by the plasmon that is coupled to the first slit and scattered by the second. We identify the mean photon number of this plasmon-induced field by $ \Bar{n}_{\text{pl}}$.
To obtain the combined photon distribution, we make use of the Glauber-Sudarshan theory of coherence \cite{PhysRevCoherentIncoherent,PhysRevLettSudarshan}. Thus, we start by writing the P-function associated to the field produced by the two independently-generated, indistinguishable horizontally-polarized modes. These represent either the photons or plasmons emerging through each of the slits, so without loss of generality, we label the P function of this combined state as
\begin{equation}
     P_{\text{pl}}(\alpha)= \int   P_{1}(\alpha-\alpha^{'})P_{2}(\alpha^{'})d^{2}\alpha^{'}.
     \label{eq1sup-ch2}
    \end{equation}
    where the P function of the thermal light fields is given by
    \begin{equation}
    P_{i}(n) =(\pi\Bar{n}_{i})^{-1} \text{exp}( \frac{-|\alpha|^{2}}{\Bar{n}_{i}})
    \label{eq2sup-ch2}
    \end{equation}
    The mean photon number of the two modes is represented by $ \Bar{n}_{1}=\eta\Bar{n}_{\text{s}}$ and $ \Bar{n}_{2}=\Bar{n}_{\text{pl}}$, whereas $\alpha$ stands for the complex amplitude as defined for coherent states $ |\alpha\rangle$. By 
     \begin{equation}
    P_{i}(n) =\frac{\pi}{(\Bar{n}_{1}+\Bar{n}_{2})} \text{exp}( \frac{-|\alpha|^{2}}{\Bar{n}_{1}+\Bar{n}_{2})})
    \label{eq3sup-ch2}
    \end{equation}
    substituting Eq. (\ref{eq2sup-ch2}) into Eq. (\ref{eq1sup-ch2}), we find that the P function of the combined horizontally-polarized field is given by
whose photon distribution can readily be evaluated as
\begin{equation}
  p_{\text{pl}}=  \langle n|\hat{\rho}_{\text{pl}} |n\rangle
\label{eq4sup-ch2}
    \end{equation}
    with
\begin{equation}
\hat{\rho}_{\text{pl}}= \int P_{\text{pl}}(\alpha)|\alpha\rangle\langle 
\alpha|d^{2}\alpha.
\label{eq5sup-ch2}
   \end{equation}
   By substituting Eq. (\ref{eq5sup-ch2}) into Eq. (\ref{eq4sup-ch2}), we can readily find that the photon distribution for the scattered photons and plasmons with horizontal polarization is given by
   \begin{equation}
   p_{\text{pl}}(n)= \sum^{n}_{m=0} \frac{(\Bar{n}_{\text{pl}}+\eta\Bar{n}_{\text{s}})^{n}}{(\Bar{n}_{\text{pl}}+\eta\Bar{n}_{\text{s}}+1)^{n+1} }.
   \label{eq6sup-ch2}
\end{equation}
We can then introduce the vertically polarized multiphoton contribution by writing the photon-number distribution at the detector as $ p_{\text{det}}= \sum^{n}_{m=0} p_{\text{pl}} (n-m) p_{\text{ph}}(m) $, where the distribution $p_{\text{ph}}(m)$ accounts for the vertical polarization component of the illuminating thermal field. Note that the distinguishability, i.e. statistical independence of the two thermal sources allows us to write the combined photon distribution as the convolution of the two independent probability distributions \cite{Hogg}. Thus, we can describe the final photon-number distribution after the plasmonic structure as
     \begin{equation}
      p_{det}(n)= \sum^{n}_{m=0} \frac{(\Bar{n}_{\text{pl}}+\eta\Bar{n}_{\text{s}})^{(n-m)}[(1-\eta)\Bar{n}_{\text{s}}]^{m}}{(\Bar{n}_{\text{pl}}+\eta\Bar{n}_{\text{s}}+1)^{(n-m+1)} [(1-\eta)\Bar{n}_{\text{s}}+1]^{m+1}}.
     \label{eq7sup-ch2}
    \end{equation}
    which is the result shown in Eq. (\ref{eq2-ch2}) in Chapter \ref{chap:chapter2}. Note that Eq. (\ref{eq7sup-ch2}) is valid only when the two sources, i.e. the two slits, are active and contribute to the resulting combined field at the detector.
  \section{Intensity Independence of the  Modification of Quantum Statistics}
  \label{app:A:Intensity}

  \begin{figure*}
	\centering
	\includegraphics[width=0.95\textwidth]{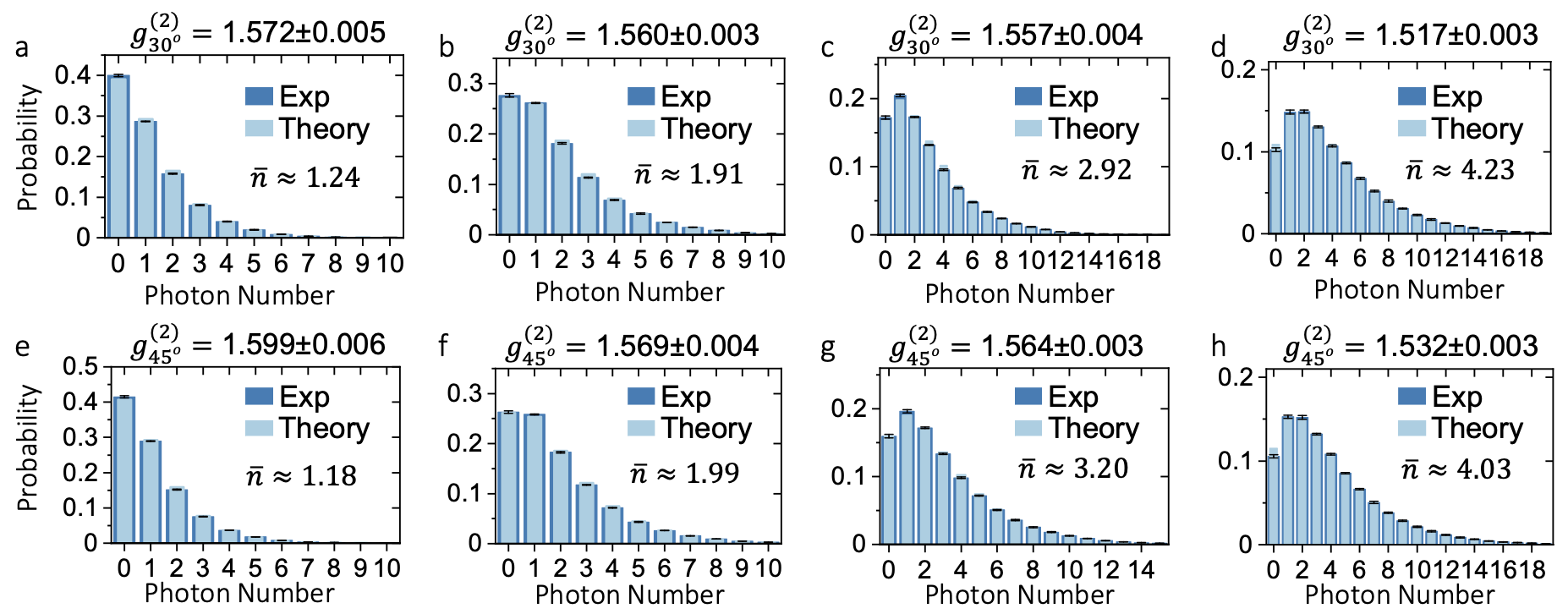}
	\mycaption{Probability Distribution from the Illuminated Sample with Polarized Light at 30$^{\circ}$.}{ Intensity independence of the modification of quantum statistics for our plasmonic system. Panels (a) to (d) show the probability distribution and the value of $g^{(2)}$ function for a situation in which the sample is illuminated with linearly polarized light at 30$^{\circ}$. While the photon number distribution changes with the brightness of the source, the value of the $g^{(2)}$ remains unchainged. This behavior shows the relevance of the $g^{(2)}$ function as a metric to quantify the quantum statistical fluctuation of a physical system. For sake of completeness, panels (e) to (h) show similar trend for the case in which the sample is illuminated with diagonally polarized light. These probability distributions demonstrate that the second-order quantum coherence function $g^{(2)}$ does not change with respect to the brightness of the experiment. The error bars represent the standard deviation of ten realizations of the experiment. Each experiment consists of approximately 100000 photon-number-resolving measurements. These figures are taken from \cite{chenglongnature}.} 
	\label{fig:fig1sup-ch2}
\end{figure*} 
To further demonstrate that the second-order quantum coherence function $g^{(2)}$ does not depend on the brightness of the source, we provide additional data for the experiment reported in Fig. \ref{fig:fig2sup-ch2} f and g. Here, we vary the brightness of $ \Bar{n}_{\text{s}}$ and $ \Bar{n}_{\text{pl}}$, while keeping the ratio $ \Bar{n}_{\text{s}}= \Bar{n}_{\text{pl}} $ unchanged. As shown in Fig. \ref{fig:fig1sup-ch2}, the $g^{(2)}$ indeed is independent of the mean photon number of the illuminating multiphoton system. Furthermore, we note that the theoretical calculation predicts $g^{(2)}_{30^{\circ}}=1.508$ and $g^{(2)}_{45^{\circ}}=1.531$, which is also independent of the mean photon number. As described in Chapter \ref{chap:chapter2}, this number is established by the ratio between $ \Bar{n}_{\text{s}}$ and $ \Bar{n}_{\text{pl}}$.        

\section{Characterization of Thermal Light Sources}
\label{app:A:Characterization}

\begin{figure*}[!ht]
	\centering
	\includegraphics[width=0.95\textwidth]{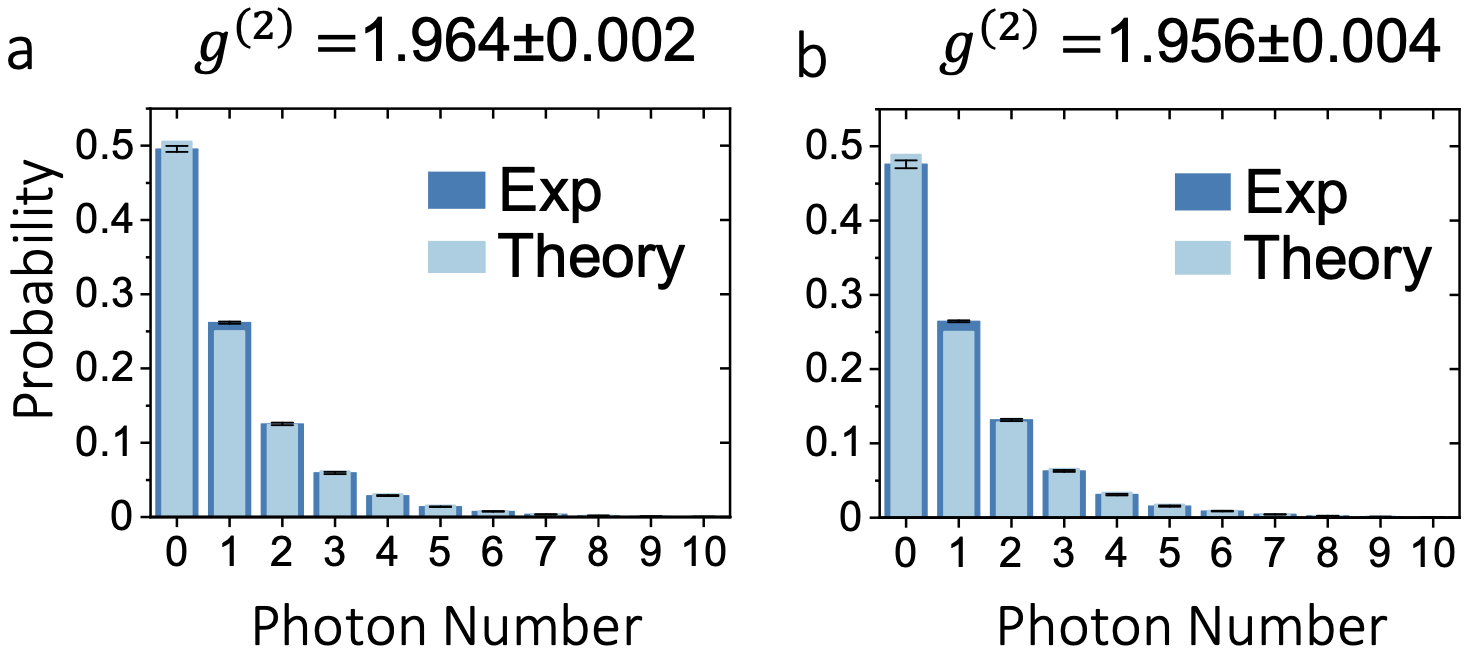}
	\mycaption{Probability Distributions of Pseudo-thermal Light Source.}{ Histogram displaying theoretical and experimental photon number probability distributions for our pseudo-thermal light source. The calculated second-order correlation function $g^{(2)}$ certify the thermal nature of our sources. The error bars represent the standard deviation of ten realizations of the experiment. Each experiment consists of approximately 100000 photon-number-resolving measurements. These figures are taken from \cite{chenglongnature}.} 
	\label{fig:fig2sup-ch2}
\end{figure*}

In our Chapter \ref{chap:chapter2} experiment, we generate pseudo-thermal light by focusing laser onto a rotating ground glass. In order to perform photon counting with our SNSPDs, we use the surjective photon counting method described in Ref. \cite{HashemiRafsanjani:17,you2020identification}. The data was divided in time bins of $1 \mu$s , which corresponds to the coherence time of our CW laser. Moreover, the $20 $ ns recovery time of our SNSPDs ensured that we perform measurements on a single-temporal-mode field. As shown in Fig. \ref{fig:fig2sup-ch2}, our pseudo-thermal light mimics the photon statistics of a thermal light source.
\section{Photon Statistics Produced by One Slit}
\label{app:A:PHOTONSTATISTICSSlit}
\begin{figure*}[!ht]
	\centering
	\includegraphics[width=0.95\textwidth]{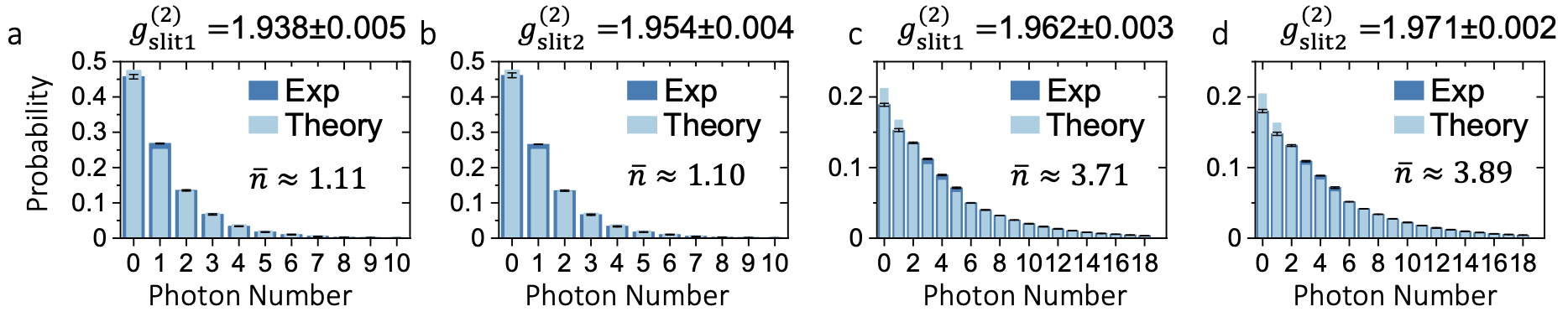}
	\mycaption{Photon-Number Probability Distributions from Either Slit.}{ Histograms displaying theoretical and experimental photon-number probability distributions of the output from either slit. In (a) and (c), the second slit, where the SPPs are transmitted, is blocked. Therefore, the probability distribution corresponds to the transmitted thermal beam. In b and d, the first slit is blocked, thus the probability distribution represents the quantum statistics of the plasmonic mode. The error bars represent the standard deviation of ten realizations of the experiment. Each experiment consists of approximitely 100000 photon-number resolving measurements. These figures are taken from \cite{chenglongnature}.} 
	\label{fig:fig3sup-ch2}
\end{figure*}
\begin{figure*}[!ht]
	\centering
	\includegraphics[width=0.95\textwidth]{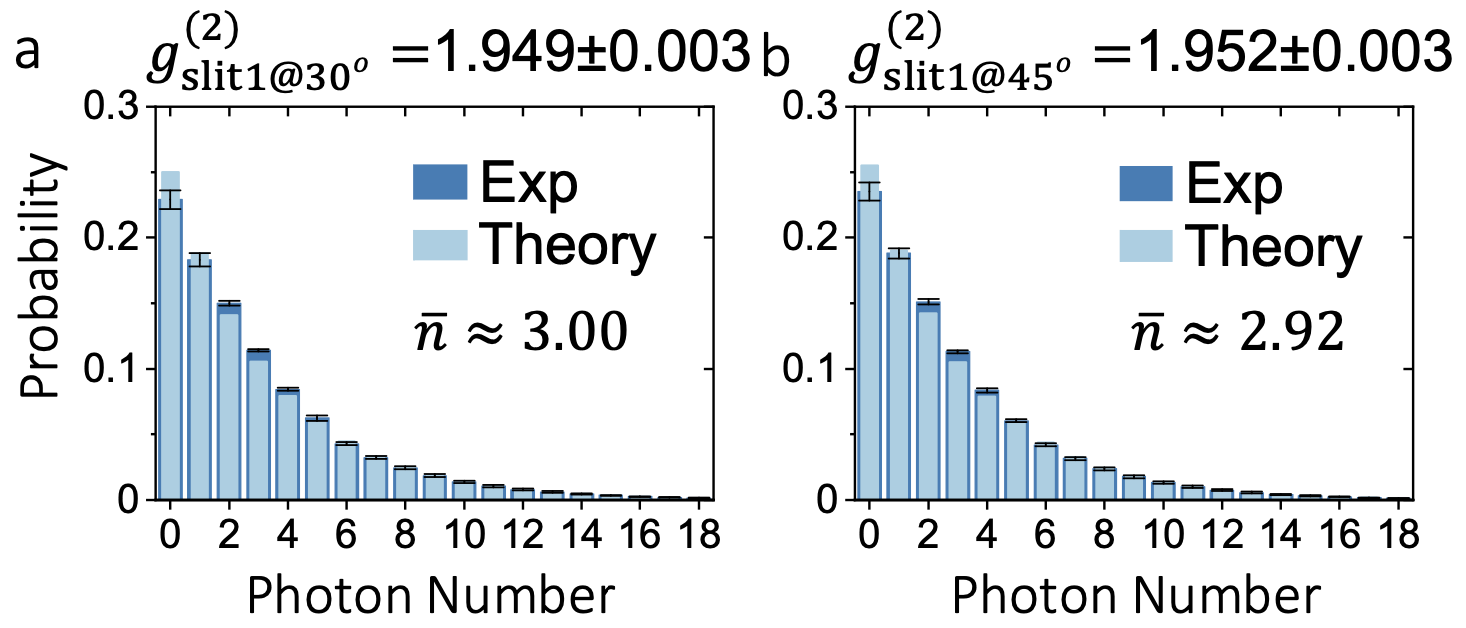}
	\mycaption{Photon-Number Probability Distributions from One Slit.}{ Histograms displaying theoretical and experimental photon-number probability distributions of the output from the first slit. The probability distribution corresponds to the transmitted thermal beam, and it is independent from the beam’s polarization. The error bars represent the standard deviation of ten realizations of the experiment. Each
experiment consists of approximately 100000 photon-number-resolving measurements. These figures are taken from \cite{chenglongnature}.} 
	\label{fig:figs4up-ch2}
\end{figure*}

\noindent
Here, we discuss the photon statistics produced by a single plasmonic slit. For this purpose, we measure  the photon statistics when blocking either slits in our plasmonic structure. In Fig. \ref{fig:fig3sup-ch2}, we plot the measured photon number distribution when blocking either slit. As expected, when measuring the transmitted light from the first slit (while keeping the second blocked), $ g^{(2)}_{\text{slit}}$ is close to 2, which is similar to $ g^{(2)}$ of our thermal source. Similarly, when measuring only the second slit(while keeping the first blocked), where the SPPs are generated, the $ g^{(2)}_{\text{slit}}$ is also close to 2. As discussed in Chapter \ref{chap:chapter2}, these results validate that the plasmonic structure indeed preserves quantum statistics through the simple single-particle dynamics. However, the additional scattering
paths supported by our sample induce complex multiparticle interactions, the resulting multiparticle dynamics can, in turn, lead to the modification of the excitation mode of plasmonic systems. Furthermore, we plot the measured
photon-number distributions for the first slit with different polarizations. As shown in Fig. \ref{fig:figs4up-ch2}, the $ g^{(2)}$ is still close to 2, and is independent of the polarization angle. Note that the theoretical prediction in slit1
 Fig. \ref{fig:figs4up-ch2} is obtained by using the analytical expression for a thermal photon distribution, rather than Eq. (\ref{eq7sup-ch2}). This shows that one cannot describe the statistics of the photons emitted by a single slit as the statistical combination of two independent polarized fields. More importantly, this result supports our central claim, which states that it is only when a plasmon i.e., a second field source is excited that photon statistics of the initially thermal field are modified.

\section{FDTD Simulation}
\label{app:A:FDTD}
The design of the plasmonic structure given in Fig. \ref{fig:figure1}b is simulated with a 2-D FDTD simulations by a \SI{130}{\micro\meter} domain in $x$ direction and  \SI{8}{\micro\meter} along the $y$ direction. The boundary condition is satisfied via the perfect matching layers to efficiently absorb the light scattred by the strucutre. Besides, the simulations time was long enough so that all energy in the simulation domain was completely decayed. 
The upper clad is made of CYTOP, a polymer with refractive index that closely matches the refractive index of 1.33. The mesh size was as small as $0.03 $ nm \SI{}{\micro\meter}
along $x$ direction and where we have highly confined field propagation. To create the propagating plasmonic modes, we use a pair of mode sources in both sides of the central slit. The generated SP modes propagate toward the central slit where they interfere. The near-fields along a linear line underneath the nanostrucutre were extracted and used for the far-field analysis. The coupled light to the mode $\hat{e}$, i.e. $T_{\text{ph}}$, was  calculated by the power flow through to the same linear line beneath the slit normalized to the input power.   To have a realistic estimation of the subtracted light, the mode $\hat{d}$  was first propagated for a distance of  10$\lambda$ (8.1 \SI{}{\micro\meter}) along the gold-glass interface and then a grating coupling efficiency of 36\% was considered to out couple the plasmonic mode to the free space  \cite{biosensordeleon}. The out-coupling was done far from the slit to avoid interactions of slit near-fields with fields of the assumed grating.

\section{Signal-to-Noise Derivation}
\label{app:A:SNR}
First, we calculate the second-order correlation function  $g^{(2)}_L (0)$ associated to the L-plasmon-subtracted light field. 
We assume a thermal light field with Bose-Einstein statistics described by 
$\hat{\rho}_{\text{th}}= \sum_{n=0}^{\infty} \text{p}_{\text{pl}}(n) {\ket n} {\bra n}$, where $\text{p}_\text{pl}(n)=\bar n^{n}/(1+\bar n)^{1+n}$. The subtraction of L-plasmon(s) from a single-mode thermal field gives
\begin{equation}
\hat{\rho}_{L}=\frac{(\hat{a})^{L} \hat{\rho} (\hat{a}^{\dagger})^{L} }{\text{Tr}((\hat{a})^{L} \hat{\rho} (\hat{a}^{\dagger})^{L})} = \sum_{n=0}^{\infty} \frac{(n+L)! }{n! L!} \frac{\bar{n}^{n} }{(1+\bar{n})^{L+n+1}} {\ket n} {\bra n} = \text{p}_\text{pl}(n) {\ket n} {\bra n}  
\label{eqgsup1}
\end{equation}
 The second-order correlation function of a single-mode field is given by
 \begin{equation}
g^{(2)}(0)=\frac{\langle \hat{a}^{\dagger} \hat{a}^{\dagger} \hat{a} \hat{a}\rangle}{\langle\hat{a}^{\dagger}\hat{a}\rangle^{2}} = \frac{\langle\hat{n}(\hat{n}-1)\rangle } {\langle\hat{n}\rangle^{2}}=\frac{\langle \hat{n}^{2}\rangle - \langle \hat{n}\rangle } {\langle\hat{n}\rangle^{2}}
\label{eqgsup2}
\end{equation}
\noindent
We can now calculate each element in Eq. (\ref{eqgsup2}). We have
\begin{equation}
\langle \hat{n}^{2}\rangle = \sum_{n=0}^{\infty} n^{2} \text{p}_\text{pl}(n) = (L+1)\Bar{n}[(L+2)\Bar{n}+1].
\label{eqgsup3}
\end{equation}
Similarly,
\begin{equation}
\langle \hat{n}\rangle = \sum_{n=0}^{\infty} n^{2} \text{p}_\text{pl}(n) =  (L+1)\Bar{n}.
\label{eqgsup4}
\end{equation}
Combining Eq. (\ref{eqgsup2}), Eq. (\ref{eqgsup3}), and Eq. (\ref{eqgsup4}), we obtain
\begin{equation}
g^{(2)}_L (0) = {\frac{L+2}{L+1}}
\label{eqgsup5}
\end{equation}
\noindent 
which is independent of the mean occupation number $ \Bar{n}$ of the input thermal field.
Now we derive Eq. (\ref{eqg2}) in Chapter 3. First, we note that in our calculation, we assume that mode $ \hat{a}$ and $ \hat{c}$ come from the same input source. Following similar approaches to those presented in \cite{HashemiRafsanjani:17}, for the lossless case, the mean occupation number of mode $ \hat{e}$ is given by $ \Bar{n}_{e} = \Bar{n}\xi \cos^{2} (\frac{\varphi}{2})$. Here, $ \Bar{n}$ is the mean occupation number in the input modes $ \hat{a}$ and $ \hat{c}$, and $\xi$ represents the normalized transmission of the plasmonic tritter \cite{Safari}. However, we need to consider that the plasmonic structure induces loss, and we have non-unity detection efficiency. As discussed in Chapter 3, conditional measurements will change the mean occupation number of the mode $ \hat{e}$. We first consider the situation in which no plasmons are subtracted (no conditional measurement is implemented). In this case, the average occupation number of mode $ \hat{e}$ is simply modulated by the loss $ \gamma$ of the plasmonic tritter, and the quantum efficiency  $\eta_{\text{ph}}$ of the detector,

\begin{equation}
 \Bar{n}_{e} = \Bar{n} \gamma \xi \eta_{\text{ph}} \cos^{2} (\frac{\phi}{2}).
\label{eqgsup6}
\end{equation}
\noindent
In this case, since no conditional measurement is made, the particle statistics are preserved. Therefore, the standard deviation is the same to that of a thermal field,
\begin{equation}
 \Delta{n}_{e} = \sqrt{ \Bar{n}_{e} + \Bar{n}_{e}^{2}}.
\label{eqgsup7}
\end{equation}
\noindent
Therefore, the signal-to-noise ratio (SNR) is given by
\begin{equation}
SNR = \frac{\Bar{n}_{e}}{\Delta{n}_{e} }= \frac{\Bar{n}_{e}}{\sqrt{ \Bar{n}_{e} + \Bar{n}_{e}^{2}}}=\frac{\sqrt{\Bar{n} \gamma \xi \eta_{\text{ph}} \cos^{2} (\frac{\phi}{2})}}{\sqrt{(1+\Bar{n} \gamma \xi \eta_{\text{ph}} \cos^{2} (\frac{\phi}{2}))}}.
\label{eqgsup8}
\end{equation}
\noindent
Now we consider the conditional subtraction of plasmons. The $ L$-plasmon subtracted state $ \hat{\rho_{e}}(L)$ of mode $ \hat{e}$ is conditioned on detection of $L$ plasmon(s) in mode $ \hat{e}$ \cite{PhysRevAalvi},
\begin{equation}
\hat{\rho_{e}}(L) = \frac{1}{\text{p}_{d}(L)} \text{Tr}_{d} [\rho \mathbb{I} \otimes \Pi_L (\eta_{\text{pl}})].
\label{eqgsup9}
\end{equation}
\noindent
Specifically, $\text{p}_{d}(L)$ is the probability of measuring $L$ plasmon(s) in mode $ \hat{e}$. Since the transformation of the plasmonic tritter preserves the particle statistics, mode $ \hat{e}$ still possess thermal statistics, 
\begin{equation}
\text{p}_{d}(L) = \frac{(\Bar{n}_{d})^n}{(1+\Bar{n}_{d})^{n+1}} 
\label{eqgsup10}
\end{equation}
\noindent
where $\Bar{n}_{d}= \Bar{n} \gamma \xi \eta_{\text{pl}} \sin^{2} (\frac{\phi}{2})$. Additionally, without loss of generality, we describe the initial state $\hat{\rho}$ before conditional
measurements as 
\begin{equation}
\hat{\rho }= \sum_{n=0}^{\infty} {\text{p}_{pl}}(n) \sum_{k,l=0}^{\infty}{\text{A}_{k}^{n}}(\xi) \text{A}_{l}^{n}(\xi)|n - k\rangle\langle n - l| \otimes |k\rangle\langle l|
\label{eqgsup11}
\end{equation}
\noindent
which describes the two-mode state after the reduced plasmonic tritter transformation. We note that this reduced plasmonic tritter transformation is similar to the beam splitter transformation, therefore $ \text{A}_{k}^{n}(\xi)= \sqrt{\binom{n}{k} \xi^{n-k}(1-\xi)^{k}} $. Finally, the positive-operator-valued measure (POVM) of a realistic photon-counting device with quantum efficiency $\eta$ is given by \cite{PhysRevAalvi}: 
\begin{equation}
\Pi_L(\eta) =\sum_{m=L}^{\infty} {B_{m,L}}(\eta)|m\rangle\langle m|
\label{eqgsup12}
\end{equation}
\noindent
in which $ B_{m,L}(\eta)= \binom{m}{L} \eta^{L} (1-\eta)^{m-L}$. Combining the above equations, we have
\begin{equation}
\hat{\rho}_{e}(L) = \frac{1}{\text{p}_{d}(L)} \sum_{m=L}^{\infty} \sum_{n=0}^{\infty}  {B_{m,L}} (\eta_{\text{pl}})\text{p}_{\text{pl}}(m+n)[A_{m}^{m+n}(\xi)]^{2}|n\rangle\langle n|.
\label{eqgsup13}
\end{equation}
\noindent
Then we can calculate the conditional mean occupation number using Eq. (\ref{eqgsup13}),
\begin{equation}
 \Bar{n}_{e} = \Bar{n} \gamma \xi \eta_{\text{ph}} \cos^{2} (\frac{\phi}{2}) \big( \frac{L+1}{1+\Bar{n}\gamma(1-\xi)\eta_{\text{pl}} \cos^{2} (\frac{\phi}{2})} \big).
\label{eqgsup14}
\end{equation}
\noindent
Similarly, one can calculate the standard deviation of the number of detection events of mode $ \hat{e}$, when conditioned on the detection of $L$ plasmons,
\begin{equation}
\Delta n_{e} = \frac{\Bar{n}_{e}}{\sqrt{\frac{(1+L)\Bar{n}\gamma\eta_{\text{ph}} \xi\cos^{2} (\frac{\phi}{2})}{1+\Bar{n}\gamma(\xi\eta_{\text{ph}}+(1-\xi)\eta_{\text{pl}})\cos^{2} (\frac{\phi}{2})}}}.
\label{eqgsup15}
\end{equation}
\noindent
Finally, the $L$-plasmon subtracted signal-to-noise ratio (SNR) is given by
\begin{equation}
\begin{split} 
\text{SNR} =  \sqrt[]{\frac{(1+L) \bar n \gamma\eta_{\text{ph}} \xi \cos^{2}\frac{\varphi}{2}}{1+ \bar n  \gamma(\xi\eta_{\text{ph}}+(1-\xi)\eta_{\text{pl}}) \cos^{2}\frac{\varphi}{2} }}.
\end{split} 
\label{eqgsup16}
\end{equation}

\section{NN Training}
\label{app:A:Training}
In what follows, we describe technical aspects of the neural networks developed in this work. The acquired sets of images in combination with machine learning algorithms enable the identification of distorted LG modes. This renders the originally encoded modes (message). The machine learning algorithms are characterized by solve tasks where conventional algorithms offer low performances or limited efficiencies. Typically, these solve tasks exploit a given collection of labeled examples or \enquote{past experiences} to predict the outcome for new data \cite{bishop2006pattern, Goodfellow2016}. We implement feed-forward neural networks with sigmoid neurons in the single hidden layer and softmax neurons in the output layer, to identify spatial modes transmitted through a multimode fiber. In this architecture, each neuron in a specific layer is connected to each neuron of the next layer through a synaptic weight. These synaptic weights are optimized by using the scaled conjugate gradient back-propagation algorithm \cite{moller1993scaled} in a direction that minimizes the cross-entropy \cite{shore1981properties, de2005tutorial}. Because sigmoid neurons are ranged in the interval [0,1], the cross-entropy is used as the loss function, as it has been shown to be ideal for classification tasks \cite{shore1981properties}.
As is standard in artificial neural networks, these algorithms undergo two stages, training and test.  We use a batch size of 10500 samples for individual modes if the communication is bit-by-bit and a batch size of 2000 samples for superposition modes if the communication is byte-by-byte. In both cases, we devoted 70\% of the dataset for training, 15\% to validation, and 15\% to testing. To avoid overfitting, the algorithm implements an early stopping technique that stops training once 1) the model performance stops improving on the validation dataset, 2) the model reaches a maximum number of 1000 epochs, and 3) the model performance achieves a performance gradient less than $10^{-6}$. 
In all cases, the networks were trained and tested with balanced data to avoid bias in identification, and the testing data was always excluded from the training stage. More specifically, we train our neural networks \cite{svozil1997introduction} from distorted modes collected by the CCD camera after the propagation through the fiber. The collection of RGB high-resolution images (1200$\times$1024 pixels) are converted into grayscale images by eliminating the hue and saturation information but retaining the luminance. To reduce the data dimension, a down-sampling process is performed on the resulting monochromatic images by averaging small clusters of 140$\times$140 pixels to form images of 9$\times$7 pixels. In this way, the feature vector is obtained by reorganizing the pixels of the resulting images as a column vector. At this point, it is important to stress that the proper choice of the feature vector can have a dramatic effect on the performance results. As shown in the main manuscript, our extreme reduction in the image resolution allows us to train neural networks in a short time with low computational resources while maintaining a high recognition rate. Once the NN has been trained, Bob can utilize its high efficiency to retrieve the message sent by Alice, even if the channel is under strain, with high confidence in both message security and integrity. Interestingly, our neural networks are capable of decrypting the encoded intensity profile images at 14 milliseconds for both bit-by-bit and byte-by-byte communication protocols. In order to assess the performance of the neural networks, we compute the ratio of the sum of false negatives and false positives to the total number of input observations, the so-called accuracy. We have run all of our algorithms in a computer with an Intel Core i7–4710MQ CPU (@2.50GHz) and 32GB of RAM with MATLAB 2019a.
\section{Unpolarized Multimode Thermal Light}
\label{app:A:Unpolarized}
In our experiment in Chapter \ref{chapter:5}, we utilized unpolarized multimode thermal light. The light is thermal such that the electric field $E^{(+)}(x)$ obeys complex-Gaussian statistics at each point $x$, and that the mean $\braket{E^{(+)}(x)}$ is zero. It is unpolarized such that it is an equal mixture between two orthogonal polarizations (here we choose horizontal ($H$) and vertical ($V$)). Finally, any two spatial projections on this source will be statistically independent. It is our goal to study the quantum properties of such a source as it propagates through our experimental setup. We will now present a sufficient quantum description for such a light source before propagation. 
\noindent
The generation of unpolarized multimode thermal light is accomplished by mixing coherent states with different amplitudes \cite{ou2007multi}. We then pixelize the source and assume that each pixel obeys independent polarization statistics. Such a source can then be written as
\begin{equation}
    \hat{\rho} = \int d\Sigma \bigotimes_{\boldsymbol{s}} \Big(|\alpha\rangle\langle\alpha|_{\Sigma,H,\boldsymbol{s}}+|\alpha\rangle\langle\alpha|_{\Sigma,V,\boldsymbol{s}}\Big).
\end{equation}
Here, each $\boldsymbol{s}$ represents the position of a pixel, $\alpha$ is a coherent amplitude, and the coherent states $|\alpha\rangle_{\Sigma,B,\boldsymbol{s}}$ are defined by the modes
\begin{equation}
    \hat{a}_{\Sigma,B,\boldsymbol{s}} = \int d\boldsymbol{\rho}\text{ Rect}\left[(\boldsymbol{s}-\boldsymbol{\rho})/d\right] \Sigma(\boldsymbol{\rho})\hat{a}_{B}(\boldsymbol{\rho}),
\end{equation}
where $d$ is the side-length of each pixel.
In the integral, $\Sigma(\boldsymbol{\rho})$ represents one instance of a random complex electric field profile. Writing $\Sigma(\boldsymbol{\rho}_i) \equiv \Sigma_i$, the action of this functional integral is characterized by the formula
\begin{equation}
    \int d\Sigma\text{ } f(\Sigma_1,...,\Sigma_n) = \int d^2\Sigma_1...d^2\Sigma_n\text{ } \frac{1}{(2\pi)^n\sqrt{|\boldsymbol{\Gamma}|}}e^{-\frac{1}{2}(\boldsymbol{r}-\boldsymbol{\mu})^T \boldsymbol{\Gamma}^{-1}(\boldsymbol{r}-\boldsymbol{\mu})}f(\Sigma_1,...,\Sigma_n),
\end{equation}
where $\boldsymbol{r} \equiv \Big(\text{Re}[\Sigma_1],\text{Im}[\Sigma_1],...,\text{Re}[\Sigma_n],\text{Im}[\Sigma_n]\Big)$, $\boldsymbol{\mu} = \braket{\boldsymbol{r}}$, $\boldsymbol{\Gamma}$ is the covariance matrix of $\boldsymbol{r}$, and $|\cdot|$ represents the determinant operation. Note that, in the case of thermal statistics, $\boldsymbol{\mu} = \boldsymbol{0}$. To complete our description of unpolarized multimode thermal light, we now determine the covariance matrix $\boldsymbol{\Gamma}$. It is easy to see that $\boldsymbol{\Gamma}$ is completely determined by $\braket{\Sigma(\boldsymbol{\rho}_1)\Sigma(\boldsymbol{\rho}_2)}$ and $\braket{\Sigma^*(\boldsymbol{\rho}_1)\Sigma(\boldsymbol{\rho}_2)}$. The term $\braket{\Sigma(\boldsymbol{\rho}_1)\Sigma(\boldsymbol{\rho}_2)}$ will always be $0$ in the case of thermal light, and $\braket{\Sigma^*(\boldsymbol{\rho}_1)\Sigma(\boldsymbol{\rho}_2)}$ will have the form
\begin{equation}
    \braket{\Sigma^*(\boldsymbol{\rho}_1)\Sigma(\boldsymbol{\rho}_2)} = \sqrt{\bar{n}(\boldsymbol{\rho}_1)\bar{n}(\boldsymbol{\rho}_2)} \left[\frac{1}{\pi \sigma}e^{-\frac{|\boldsymbol{\rho}_1-\boldsymbol{\rho}_2|^2}{\sigma}}\right],
\end{equation}
where $\sigma$ is assumed to be small so that $\frac{1}{\pi \sigma}e^{-\frac{|\boldsymbol{\rho}_1-\boldsymbol{\rho}_2|^2}{\sigma}} \approx \delta(\boldsymbol{\rho}_1-\boldsymbol{\rho}_2)$. Normalization gives us the mean photon number at position $\boldsymbol{\rho}$ as $\bar{n}(\boldsymbol{\rho}) = \pi\sigma/d^2$.
\section{Propagation to the Far Field}
The temporal evolution of a photon's spatial probability distribution obeys classical physics. This behavior is a direct consequence of the free-space Hamiltonian for the electromagnetic field being quadratic in its quadrature variables \cite{combescure2012quadratic,agarwal2012quantum}.
Assuming a paraxial light source, we can utilize the Fresnel diffraction formula to determine the propagated mode-structure of the unpolarized multimode thermal light \cite{saleh2019fundamentals}. The Fresnel kernel at propagation distance $z$ is given by \cite{goodman2008introduction}
\begin{equation}
    K(\boldsymbol{r},\boldsymbol{\rho},z) = \frac{e^{i k z}}{i\lambda z} e^{\frac{ik}{2z}(\boldsymbol{r}-\boldsymbol{\rho})},
\end{equation}
where $\lambda$ is the wavelength of the light source, $k = 2\pi/\lambda$, and $\boldsymbol{r},\boldsymbol{\rho}$ represent positions in the measurement plane and source plane respectively. We can calculate the propagated mode structure as
\begin{equation}
    E^{(+)}(\boldsymbol{r},z) = \int d\boldsymbol{\rho} K(\boldsymbol{r},\boldsymbol{\rho},z)E^{(+)}(\boldsymbol{r},0).
\end{equation}
Given a mode $\hat{a}_0 = \int d\boldsymbol{r}  \text{ } f(\boldsymbol{r}) \hat{a}_i(\boldsymbol{r})$ with arbitrary polarization $i$, the resulting mode in the far-field is given by
\begin{equation}
    \hat{a}_z = \int d\boldsymbol{r}\left[\int d\boldsymbol{\rho}\text{ }K^*(\boldsymbol{r},\boldsymbol{\rho},z)f(\boldsymbol{\rho})\right]\hat{a}_i(\boldsymbol{r}).
\end{equation}
Importantly, we note that the operator-valued distribution $\hat{a}_i(\boldsymbol{r})$ is not integrated against the Fresnel kernel.

\section{Computing the Correlation Matrix}
\label{app:A:CorrelationMatrix}
In our experiment, a linear polarization grating with a position-dependent polarization angle filters the multimode unpolarized light, and the beam is propagated in free-space \cite{You2023CittertZernike}. In the far-field, we make measurements with two point-detectors which are able to post-select on a particular configuration of polarizations. To be explicit, the operator representing a measurement of the first-order correlation at position $\boldsymbol{\rho}$ is given by $\hat{a}^\dagger_i (\boldsymbol{\rho})\hat{a}_j(\boldsymbol{\rho})$ where $i,j\in \{H,V\}$. We represent the transformation of the polarization grating with the matrix
\begin{equation}
    \boldsymbol{P}(x) = \begin{pmatrix}
        \cos^2\left(\frac{\pi x}{L}\right)& \cos \left(\frac{\pi x}{L}\right) \sin \left(\frac{\pi x}{L}\right)\\
        \cos \left(\frac{\pi x}{L}\right)\sin \left(\frac{\pi x}{L}\right)& \sin^2\left(\frac{\pi x}{L}\right)\\
        \sin \left(\frac{\pi x}{L}\right)&\cos \left(\frac{\pi x}{L}\right)
    \end{pmatrix},
\end{equation}
where $L$ is the width of the polarization grating and the represented transformation is given by
\begin{equation}
    \hat{a}_B(\boldsymbol{\rho}) = P_{HB}(\rho_x)\hat{a}_H(\boldsymbol{\rho}) + P_{VB}(\rho_x)\hat{a}_V(\boldsymbol{\rho}) + P_{\emptyset B}(\rho_x)\hat{a}_\emptyset(\boldsymbol{\rho}),
\end{equation}
for $B\in \{H,V\}$ and $\boldsymbol{\rho} = \rho_x\hat{\boldsymbol{x}} + \rho_y\hat{\boldsymbol{y}}$. The $\hat{a}_\emptyset(\boldsymbol{\rho})$ mode represents photon-loss at the polarization grating. Therefore, immediately after the linear polarizer, the mode structure is given by
\begin{equation}
    \hat{a}_{\Sigma,B,\boldsymbol{s}} = \int d\boldsymbol{\rho}\text{ Rect}\left[(\boldsymbol{s}-\boldsymbol{\rho})/d\right]\Sigma(\boldsymbol{\rho}) \Big[P_{HB}(\rho_x)\hat{a}_H(\boldsymbol{\rho}) + P_{VB}(\rho_x)\hat{a}_V(\boldsymbol{\rho}) + P_{\emptyset B}(\rho_x)\hat{a}_\emptyset(\boldsymbol{\rho})\Big].
\end{equation}
Consequently, in the far-field, it is given by
\begin{equation}
\begin{aligned}
    \hat{a}_{\Sigma,B,\boldsymbol{s},z} &= \int d\boldsymbol{r} \text{ }\left(\int d\boldsymbol{\rho} K^*(\boldsymbol{r},\boldsymbol{\rho},z)\text{ Rect}\left[(\boldsymbol{s}-\boldsymbol{\rho})/d\right]\Sigma(\boldsymbol{\rho})\right.\\
    &\quad \left.\times\Big[P_{HB}(\rho_x)\hat{a}_H(\boldsymbol{r}) + P_{VB}(\rho_x)\hat{a}_V(\boldsymbol{r}) + P_{\emptyset B}(\rho_x)\hat{a}_\emptyset(\boldsymbol{r})\Big]\right).
\end{aligned}
\end{equation}
\noindent
We now make a couple of approximations to simplify our calculations. First, we suppose that $\text{ Rect}\left[(\boldsymbol{s}-\boldsymbol{\rho})/d\right]\approx \text{ Rect}\left[\boldsymbol{\rho}/L\right]$ for all pixel positions $\boldsymbol{s}$. Then, we assume that light at each position in the far-field had originated primarily from a single pixel. With these approximations, we are able to write the propagated state in the form
\begin{equation}
    \hat{\rho}_{z} = \int d\Sigma \bigotimes_{\boldsymbol{r}} \Big(|\alpha\rangle\langle\alpha|_{\Sigma,H,\boldsymbol{r},z}+|\alpha\rangle\langle\alpha|_{\Sigma,V,\boldsymbol{r},z}\Big),
\end{equation}
where the mode structure for each $\boldsymbol{r}$ is now given by
\begin{equation}
\begin{aligned}
    \hat{a}_{\Sigma,B,\boldsymbol{r},z} \approx \int d\boldsymbol{r}' & \text{ Rect}\left[\frac{(\boldsymbol{r}-\boldsymbol{r}')}{d^{'}}\right]\Bigg(\int d\boldsymbol{\rho} K^*(\boldsymbol{r}',\boldsymbol{\rho},z)\text{ Rect}\left[\frac{\boldsymbol{\rho}}{d}\right]\Sigma(\boldsymbol{\rho})\\
    &\times\Big[P_{HB}(\rho_x)\hat{a}_H(\boldsymbol{r}) + P_{VB}(\rho_x)\hat{a}_V(\boldsymbol{r}) + P_{\emptyset B}(\rho_x)\hat{a}_\emptyset(\boldsymbol{r})\Big]\Bigg),
\end{aligned} 
\end{equation}
where $d^{'}$ is the width of each pixel in the measurement plane. From here, we can compute the second-order correlation functions for various polarization projections as 
\begin{equation}
    \begin{aligned}
        G^{(2)}_{ijkl}(\boldsymbol{r}_1,\boldsymbol{r}_2,z) &= \int d\Sigma \text{ }\sum_{A,B}\Big(\bra{\alpha}_{\Sigma, A,\boldsymbol{r}_1, z}\bra{\alpha}_{\Sigma, B,\boldsymbol{r}_2, z} \Big)\hat{a}^\dagger_i(\boldsymbol{r}_1)\hat{a}^\dagger_j(\boldsymbol{r}_2)\hat{a}_k(\boldsymbol{r}_1)\hat{a}_l(\boldsymbol{r}_2)\Big(\ket{\alpha}_{\Sigma, A,\boldsymbol{r}_1, z}\ket{\alpha}_{\Sigma, B,\boldsymbol{r}_2, z} \Big)\\
        &= \int d\Sigma \text{ } |\alpha|^4 \int d\boldsymbol{\rho}_1 d\boldsymbol{\rho}_2 d\boldsymbol{\rho}_3 d\boldsymbol{\rho}_4 K^*(\boldsymbol{r}_1,\boldsymbol{\rho}_1,z)K^*(\boldsymbol{r}_2,\boldsymbol{\rho}_2,z)\\
        &\text{ }\text{ }\text{ }\times K(\boldsymbol{r}_1,\boldsymbol{\rho}_3,z)K(\boldsymbol{r}_2,\boldsymbol{\rho}_4,z)\Sigma(\boldsymbol{\rho}_1)\Sigma(\boldsymbol{\rho}_2)\Sigma^*(\boldsymbol{\rho}_3)\Sigma^*(\boldsymbol{\rho}_4)\\
        &\text{ }\text{ }\text{ }\times \sum_{A,B}P_{iA}(\rho_{1x})P_{jB}(\rho_{2x})P_{kA}(\rho_{3x})P_{lB}(\rho_{4x})\\
        &\approx |\alpha|^4\int d\boldsymbol{\rho}_1 d\boldsymbol{\rho}_2 d\boldsymbol{\rho}_3 d\boldsymbol{\rho}_4 K^*(\boldsymbol{r}_1,\boldsymbol{\rho}_1,z)K^*(\boldsymbol{r}_2,\boldsymbol{\rho}_2,z)K(\boldsymbol{r}_1,\boldsymbol{\rho}_3,z)K(\boldsymbol{r}_2,\boldsymbol{\rho}_4,z)\\
        &\text{ }\text{ }\text{ }\times \frac{\pi^2\sigma^2}{L^4}\text{Rect}(\frac{\boldsymbol{\rho}_1}{L})\text{Rect}(\frac{\boldsymbol{\rho}_2}{L})\text{Rect}(\frac{\boldsymbol{\rho}_3}{L})\text{Rect}(\frac{\boldsymbol{\rho}_4}{L})\\
        &\text{ }\text{ }\text{ }\times\Big[\delta(\boldsymbol{\rho}_1-\boldsymbol{\rho}_3)\delta(\boldsymbol{\rho}_2-\boldsymbol{\rho}_4) + \delta(\boldsymbol{\rho}_1-\boldsymbol{\rho}_4)\delta(\boldsymbol{\rho}_2-\boldsymbol{\rho}_3)\Big]\\
        &\text{ }\text{ }\text{ }\times \frac{1}{4}\Big[P_{iH}(\rho_{1x})P_{jH}(\rho_{2x})P_{kH}(\rho_{3x})P_{lH}(\rho_{4x}) + P_{iH}(\rho_{1x})P_{jH}(\rho_{2x})P_{kV}(\rho_{3x})P_{lV}(\rho_{4x})\\
        &\text{ }\text{ }\text{ } + P_{iV}(\rho_{1x})P_{jV}(\rho_{2x})P_{kH}(\rho_{3x})P_{lH}(\rho_{4x}) + P_{iV}(\rho_{1x})P_{jV}(\rho_{2x})P_{kV}(\rho_{3x})P_{lV}(\rho_{4x})\Big]\\
        &\equiv I_0 \int d\boldsymbol{\rho}_1 d\boldsymbol{\rho}_2 d\boldsymbol{\rho}_3 d\boldsymbol{\rho}_4 \text{ } F(\boldsymbol{r}_1,\boldsymbol{r}_2,\boldsymbol{\rho}_1,\boldsymbol{\rho}_2,\boldsymbol{\rho}_3,\boldsymbol{\rho}_4,z) \\
        &\text{ }\text{ }\text{ }\times\Big[\delta(\boldsymbol{\rho}_1-\boldsymbol{\rho}_3)\delta(\boldsymbol{\rho}_2-\boldsymbol{\rho}_4) + \delta(\boldsymbol{\rho}_1-\boldsymbol{\rho}_4)\delta(\boldsymbol{\rho}_2-\boldsymbol{\rho}_3)\Big],
    \end{aligned}
\end{equation}
where we have defined $I_0 = \pi^2 \sigma^2 |\alpha|^4/L^4$ and
\noindent
\begin{equation}
    \begin{aligned}
F(\boldsymbol{r}_1,\boldsymbol{r}_2,\boldsymbol{\rho}_1,\boldsymbol{\rho}_2,\boldsymbol{\rho}_3,\boldsymbol{\rho}_4,z)  =&\text{ } K^*(\boldsymbol{r}_1,\boldsymbol{\rho}_1,z)K^*(\boldsymbol{r}_2,\boldsymbol{\rho}_2,z)K(\boldsymbol{r}_1,\boldsymbol{\rho}_3,z)K(\boldsymbol{r}_2,\boldsymbol{\rho}_4,z)\\
        &\times \text{Rect}(\frac{\boldsymbol{\rho}_1}{L})\text{Rect}(\frac{\boldsymbol{\rho}_2}{L})\text{Rect}(\frac{\boldsymbol{\rho}_3}{L})\text{Rect}(\frac{\boldsymbol{\rho}_4}{L})\\
        &\times \frac{1}{4}\Big[P_{iH}(\rho_{1x})P_{jH}(\rho_{2x})P_{kH}(\rho_{3x})P_{lH}(\rho_{4x}) \\
    &\quad + P_{iH}(\rho_{1x})P_{jH}(\rho_{2x})P_{kV}(\rho_{3x})P_{lV}(\rho_{4x})\\
        &\text{ }\text{ }\text{ }\text{ }\text{ }+ P_{iV}(\rho_{1x})P_{jV}(\rho_{2x})P_{kH}(\rho_{3x})P_{lH}(\rho_{4x}) \\
    &\quad + P_{iV}(\rho_{1x})P_{jV}(\rho_{2x})P_{kV}(\rho_{3x})P_{lV}(\rho_{4x})\Big].
    \end{aligned}
\end{equation}
These definitions allow for a drastically simplified $G^{(2)}_{ijkl}(\boldsymbol{r}_1,\boldsymbol{r}_2,z)$, and they are used in the main body of Chapter \ref{chapter:5}. From here, each $G^{(2)}_{ijkl}(\boldsymbol{r}_1,\boldsymbol{r}_2,z)$ can be calculated explicitly. Furthermore, we can use a similar approach to show that $G^{(1)}_{i,j}(\boldsymbol{r},z) = \braket{\hat{a}^\dagger_i(\boldsymbol{r})\hat{a}_j(\boldsymbol{r})} = \sqrt{I_0}L/(2 z^2 \lambda^2)$. Using these, the normalized second-order correlation functions $g^{(2)}_{ijkl}(\boldsymbol{r}_1,\boldsymbol{r}_2,z)$ can be calculated, and this list is presented in the next section. Notably, these results are in agreement with our previous theoretical approach and our experimental data \cite{You2023CittertZernike}.

\section{Second-Order Correlation Functions}
\label{app:A:Second-Order}
Here we explicitly write the relevant second-order coherence functions studied in our experiment. In this section, we are using the shorthands $\sinc(\nu) \equiv \sin(\pi \nu)/(\pi \nu)$ and $\nu = L(r_{1x}-r_{2x})/(\lambda z)$.
\begin{equation}
    \begin{aligned}
        g^{(2)}_{\text{HHHH}}(\nu) =& \frac{1}{16}(10 \sinc(\nu)^2 + 2 (6\sinc(\nu + 1) + \sinc(\nu + 2) + 6\sinc(1-\nu) + \sinc(2-\nu))\sinc(\nu)\\
        & + 6\sinc(\nu+1)^2 + \sinc(\nu+2)^2 + 6\sinc(1-\nu)^2 + \sinc(2-\nu)^2 \\
        & + 4\sinc(\nu+1)\sinc(\nu+2) + 4(\sinc(\nu_1)+\sinc(2-\nu))\sinc(1-\nu)+16),\\
        g^{(2)}_{\text{HVHV}}(\nu) =& \frac{1}{16}(2\sinc(\nu)^2 - 2(\sinc(\nu+2)+\sinc(2-\nu))\sinc(\nu) + 2(\sinc(1-\nu)\\
        &- \sinc(\nu+1))^2 + \sinc(\nu+2)^2 + \sinc(2-\nu)^2 + 16),\\
        g^{(2)}_{\text{VHHV}}(\nu) =& \frac{1}{16}(6\sinc(\nu)^2 - 2(\sinc(\nu+2)+\sinc(2-\nu))\sinc(\nu) + 2(\sinc(1-\nu)\\
        &-\sinc(\nu+1))^2 - \sinc(\nu+2)^2 - \sinc(2-\nu)^2),\\
        g^{(2)}_{\text{HHVV}}(\nu) =& \frac{1}{16}(2\sinc(\nu)^2 - 2(\sinc(\nu+2)+\sinc(2-\nu))\sinc(\nu)\\
        &+ 2(\sinc(1-\nu)-\sinc(\nu+1))^2 + \sinc(\nu+2)^2 + \sinc(2-\nu)^2).
    \end{aligned}
\end{equation}
\section{Propagation of the Photon Number Distribution}
In this section, we present a method for determining the photon number distribution in different detection planes. In doing so, we can study the dynamics of multiphoton wavepackets. It will be challenging to compute the photon number distribution directly from $\hat{\rho}_z$, but we can avoid this difficulty by recognizing that
\begin{equation}
    \eta_{AB}(\boldsymbol{r}) = \int d\boldsymbol{\rho} K^*(\boldsymbol{r},\boldsymbol{\rho},z)\text{ Rect}\left[\boldsymbol{\rho}/L\right]\Sigma(\boldsymbol{\rho})P_{AB}(\rho_x)
\end{equation}
follows Gaussian statistics as a result of $\Sigma(\boldsymbol{\rho})$ obeying Gaussian statistics. Each $\eta_{AB}(\boldsymbol{r})$ represents one instance of a coherent state, and so by determining the probability distribution of the $\eta_{AB}(\boldsymbol{r})$ we can determine the effective quantum state as measured by our detectors. For post-selected polarization $ijkl$ at positions $\boldsymbol{r}_1,\boldsymbol{r_2}$, we will need the probability distribution for $\eta_{iA}(\boldsymbol{r_1}),\eta_{jB}(\boldsymbol{r_2}),\eta_{kC}^*(\boldsymbol{r_1}),\eta_{lD}^*(\boldsymbol{r_2}) \equiv \alpha_{iA},\alpha_{jB},\alpha_{kC},\alpha_{lD}$ where $A,B,C,D\in \{H,V\}$. Denoting $\boldsymbol{t} = (\text{Re}[\alpha_{iA}],\text{Im}[\alpha_{iA}],...,\text{Re}[\alpha_{lD}],\text{Im}[\alpha_{lD}])$, the desired probability distribution is given by
\begin{equation}
    P_{iAjBkClD}(\boldsymbol{t}) = \frac{1}{(2\pi)^4\sqrt{|\boldsymbol{\Gamma}|}}e^{-\frac{1}{2}(\boldsymbol{t}-\boldsymbol{\mu})^T \boldsymbol{\Gamma}^{-1}(\boldsymbol{t}-\boldsymbol{\mu})},
\end{equation}
where $\boldsymbol{\mu} = \braket{\boldsymbol{t}}$ and $\Gamma_{nm} = \braket{t_n t_m} - \braket{t_n}\braket{t_m}$. With this, the resulting state describing these statistics is now
\begin{equation}
    \hat{\rho}_{iAjBkClD}(z) = \int d^2 \alpha_{iA}d^2 \alpha_{jB}d^2 \alpha_{kC}d^2 \alpha_{lD} P_{iAjBkClD}(\alpha_{iA}, \alpha_{jB}, \alpha_{kC},\alpha_{lD}) |\alpha_{kC},\alpha_{lD}\rangle\langle \alpha_{iA}, \alpha_{jB}|.
\end{equation}
It then follows that the total state is given by
\begin{equation}
    \hat{\rho}_{ijkl}(z) = \frac{1}{4}\left[\hat{\rho}_{iHjHkHlH}(z) + \hat{\rho}_{iHjHkVlV}(z) + \hat{\rho}_{iVjVkHlH}(z) + \hat{\rho}_{iVjVkVlV}(z)\right],
\end{equation}
and thus that the photon-number distribution $p(n_1,n_2)$ can be calculated via
\begin{equation}
    p(n_1,n_2,z) = \text{Tr}\left[\hat{\rho}_{ijkl}(z) |n_1,n_2\rangle\langle n_1,n_2|\right].
\end{equation}

\section{Realization of Polarization Grating through a Spatial Light Modulator}
\label{app:A:SLM}
\begin{figure*}[!ht]
   \centering 
   \includegraphics[width=0.85\textwidth]{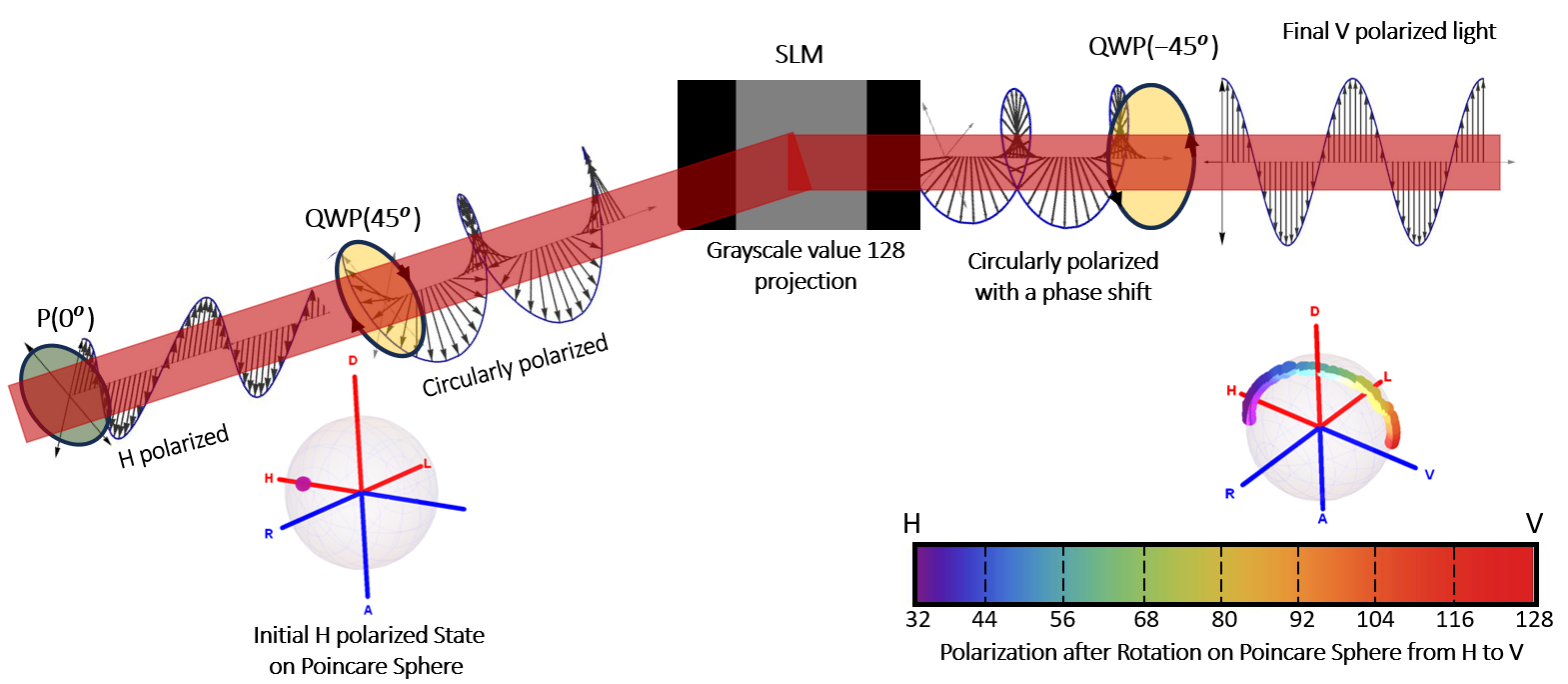}
\mycaption{Schematic Diagram for Polarization Rotation.}{ The illustration portrays a beam undergoing polarization rotation via polarization optics and a spatial light modulator (SLM). We characterize the polarization control ability of our experimental setup, and the corresponding results are displayed on the Poincare sphere on the bottom right.} 
\label{SI-1}
\end{figure*}

In this section, we describe the realization of the polarization grating using polarization optics and a spatial light modulator (SLM) \cite{PRLMirhossein2014}. As shown in Fig. \ref{SI-1}, the polarization rotation of the input beam is performed using a SLM in combination with two quarter-wave plates (QWPs). Specifically, the input beam is prepared by passing it through a polarizer aligned to the H polarization. The beam first passes through a QWP at an angle of $45^\circ$. Then, this beam is imprinted on the SLM, where a gray-value image is displayed. Finally, the reflected beam passes through another QWP at an angle of $-45^\circ$. This configuration provides the ability to rotate the polarization of the incident beam in a controlled fashion. We then characterize the relationship between the polarization rotation and the gray-value displayed on the SLM. This allows us to design a gray-scale image to implement the polarization grating. By adjusting the gray-scale values across different pixels along the $x$-axis of the SLM screen, we can control the polarization at each pixel. We can thus simulate the effect of a polarization grating on an unpolarized light source.

\section{ Probability of Observing Multiphoton Events}
\label{app:A:MultiphotonEvents}
In this section, we provide additional experimental results presented in Chapter \ref{chap:chapter6}. In Table \ref{tab:tableS1}, we report the values associated with the probability of measuring specific multiphoton events from a thermal light beam. Specifically, Table \ref{tab:tableS1} presents the probability of observing a particular number of photons under the post-selection scheme of Fig. \ref{fig:figure2-ch5}. Similarly, Table \ref{tab:tableS2} presents the joint probabilities of measuring a particular number of photons in two separate arms under the post-selection scheme. Lastly, Table \ref{tab:tableS3} presents the probability of observing a particular number of photons in the second arm under the photon-subtraction scheme of Fig. \ref{fig:figure3-ch5}.
\begin{table}[!htbp]
\centering
\caption{The Measured Probability of Post Selection.}
\begin{tabular}{|p{1.25cm}<{\centering}|p{1.25cm}<{\centering}|p{1.4cm}<{\centering}|p{1.4cm}<{\centering}|p{1.4cm}<{\centering}|p{1.4cm}<{\centering}|p{1.4cm}<{\centering}|p{1.4cm}<{\centering}|p{1.4cm}<{\centering}|}
\hline
$\bar{n}$&$|0\rangle\langle0|$& $ |1\rangle\langle1|$ &$ |2\rangle\langle2| $ &$ |3\rangle\langle3|$& $|4\rangle\langle4|$& $ |5\rangle\langle5|$& $ |6\rangle\langle6|$& $ |7\rangle\langle7|$\\
\hline
$0.8 $&$55.17\%$ &$26.11 \%$ &$10.76\%$ &$4.51\%$& $ 1.95\%$& $ 0.87\%$ & $ 0.40\%$&$ 0.18\%$\\
\hline
\end{tabular}
\label{tab:tableS1}
\end{table}

\begin{table}[!htbp]
\centering
\caption{The Measured Probability of Correlation Between the Two Arm in the Source.}
\begin{tabular}{|p{1.4cm}<{\centering}|p{1.4cm}<{\centering}<{\centering}|p{1.5cm}<{\centering}|p{1.4cm}<{\centering}|p{1.4cm}<{\centering}|p{1.4cm}<{\centering}|p{1.4cm}<{\centering}|p{1.4cm}<{\centering}|p{1.4cm}<{\centering}|p{1.4cm}<{\centering}|}
\hline
\diagbox[width=2cm]{Arm1}{Arm2} & $|0\rangle\langle0|$& $ |1\rangle\langle1|$ &$ |2\rangle\langle2| $ &$ |3\rangle\langle3|$& $|4\rangle\langle4|$& $ |5\rangle\langle5|$& $ |6\rangle\langle6|$& $ |7\rangle\langle7|$\\
\hline
$|0\rangle\langle0|$&$15.99\%$ &$9.40 \%$ &$4.26\%$ &$1.78\%$& $ 0.73\%$& $ 0.28\%$ & $ 0.11\%$&$ 0.04\%$\\
\hline
$ |1\rangle\langle1|$ &$7.53\%$ &$7.05 \%$ &$4.64\%$ &$2.65\%$& $ 1.41\%$& $ 0.68\%$ & $ 0.33\%$&$ 0.14\%$\\
\hline
$ |2\rangle\langle2| $ &$2.70\%$ &$3.68 \%$ &$3.25\%$ &$2.39\%$& $ 1.54\%$& $ 0.92\%$ & $ 0.50\%$&$ 0.28\%$\\
\hline
$ |3\rangle\langle3|$&$0.89\%$ &$1.64 \%$ &$1.86\%$ &$1.70\%$& $ 1.33\%$& $ 0.94\%$ & $ 0.61\%$&$ 0.37\%$\\
\hline
$|4\rangle\langle4|$&$0.27\%$ &$0.65 \%$ &$0.93\%$ &$1.02\%$& $ 0.93\%$& 
$ 0.77\%$ & $ 0.57\%$&$ 0.40\%$\\
\hline
$ |5\rangle\langle5|$&$0.08\%$ &$0.24 \%$ &$0.42\%$ &$0.54\%$& $ 0.58\%$& 
$ 0.54\%$ & $ 0.46\%$&$ 0.36\%$\\
\hline
$ |6\rangle\langle6|$&$0.02\%$ &$0.08 \%$ &$0.18\%$ &$0.26\%$& $ 0.32\%$& 
$ 0.35\%$ & $ 0.34\%$&$ 0.29\%$\\
\hline
$ |7\rangle\langle7|$&$0.007\%$ &$0.02 \%$ &$0.06\%$ &$0.12\%$& $ 0.16\%$& 
$ 0.19\%$ & $ 0.21\%$&$ 0.20\%$\\
\hline
\end{tabular}
\label{tab:tableS2}
\end{table}

\begin{table}[!htbp]
\centering
\caption{The Measured Probability of Photon Subtraction.}
\begin{tabular}
{|p{1.5cm}<{\centering}|p{1.5cm}<{\centering}|p{1.5cm}<{\centering}|p{1.5cm}<{\centering}|p{1.5cm}<{\centering}|}
\hline
$\bar{n}$&$N=0$&$N=1$& $ N=2$ &$ N=3 $\\
\hline
$0.08 $&$97\%$ &$2.1 \%$ &$0.3\%$& $0.05\%$\\
\hline
\end{tabular}
\label{tab:tableS3}
\end{table}

\section{ Detailed Derivation of Equations in Chapter 6}
\label{app:A:DetailedDerivation}
Here we provide a detailed derivation of the equations presented in Chapter \ref{chap:chapter6}. The initial quantum state of our signal is a weak, single-mode thermal state. Written explicitly in the Fock basis, our initial thermal state of light is represented by 
\begin{equation}
    \hat{\rho}_{0} = \sum_{n=0}^\infty \frac{\bar{n}^n}{(1+\bar{n})^{n+1}}\left|n\rangle\langle n\right|,
\end{equation}
where $\bar{n}$ is the mean number of photons of the state. In our experiment, we uniformly illuminate an object with this state, producing a signal with a new state that has a different mode structure. The mode information is contained within the annihilation operator $\hat{a}$ which obeys $\hat{a}|n\rangle = \sqrt{n}|n-1\rangle$, defined in terms of the operator-valued distribution $\hat{a}(\Vec{\boldsymbol{x}})$ by
\begin{equation}
    \hat{a} = \int d^2 x f(\Vec{\boldsymbol{x}})\hat{a}(\Vec{\boldsymbol{x}}),
\end{equation}
where $\left[\hat{a}(\Vec{\boldsymbol{x}}),\hat{a}^\dagger(\Vec{\boldsymbol{x}}')\right] = (2\pi)^2\delta(\Vec{\boldsymbol{x}}-\Vec{\boldsymbol{x}}')$ is the canonical commutation relation and $f(\Vec{\boldsymbol{x}})$ is the transverse profile of the beam. This expression assumes that the light is strongly peaked around a particular frequency, and in this case a transverse positional description can be used. In our experiment, we uniformly illuminate an object using the thermal state, which forms an image that we would like to measure. We do this by discretizing the transverse spatial profile of the mode into $X$ squares which we will call pixels. This is equivalent to the transformation taking $\hat{a}$ to $\sum_{i=1}^X \lambda_i \hat{A}_i$ where $\hat{A}_i$ is the annihilation operator for the mode at the 
$i^{\text{th}}$ pixel and $\lambda_i$ is its weight. 
Since the object was illuminated uniformly, $\lambda_i$ will be either $0$, representing a pixel with no light, or some constant value, representing a pixel with light, such that $\sum_{i=1}^X|\lambda_i|^2 = 1$. It is important to note, however, that this theory will also apply for non-uniform illuminations. This allows us to define the ideal image vector $\Vec{\boldsymbol{s}}_{0} 
\in \mathbb{R}^X$ where each component $s_{0,i}$ is equal to $|\lambda_i|^2\bar{n}$.
We now collect random combinations of these pixels onto a single-pixel camera that employs photon-number-resolving detection. We will see later how this allows for image reconstructions which use fewer measurements than traditional methods require. We will perform $M$ such measurements, and each random selection of pixels will be represented by the covector $\Vec{\boldsymbol{Q}}_t\in\mathbb{R}^{X*}$ which consists of zeros and ones. It follows that, after the signal has been filtered by this covector, the resulting mode operator of the signal will be given by $\hat{a}_t=\sum_{i=0}^X Q_{t,i}\lambda_i' \hat{A}_i$ where $\lambda_i' = \lambda_i/\sqrt{\sum_{i=0}^\infty Q_{t,i}|\lambda_i|^2}$ is the re-normalized weight of each pixel. The quantum state of the signal after this filtering process is therefore thermal, with a mean-photon-number given by $\bar{n}_t = \Vec{\boldsymbol{Q}}_t\cdot \Vec{\boldsymbol{s}}_0$, and can be written as
\begin{equation}
    \hat{\rho}_{\boldsymbol{Q},t} = \sum_{n=0}^\infty \frac{\bar{n_t}^n}{(1+\bar{n}_t)^{n+1}}\left|n\rangle\langle n\right|.
\end{equation}
Here we are using the label $\boldsymbol{Q}$, which represents the matrix of pixel filtrations and is defined by $\boldsymbol{Q} = \bigoplus_{t=1}^M \Vec{\boldsymbol{Q}}_t$. We can simultaneously write all such density matrices as
\begin{equation}
    \boldsymbol{\hat{\rho}}_{\boldsymbol{Q}} = \bigoplus_{t=1}^M \hat{\rho}_{\boldsymbol{Q},t},
\end{equation}
which can be thought of as a vector of density matrices. 
We now use the Gluaber-Sudarshan $P$ function representation of the quantum state, written in terms of coherent states $|\alpha\rangle$, and given by
\begin{equation}
    \boldsymbol{\hat{\rho}}_{\boldsymbol{Q}} = \bigoplus_{t=1}^M \int d^2\alpha \frac{1}{\pi \bar{n}_t}e^{-\frac{|\alpha|^2}{\bar{n}_t}}\left|\alpha\rangle\langle\alpha\right|.
\end{equation}
Before measuring the state with our photon-number-resolving detector, we will send it through a fiber-coupler in order to produce a second mode that can be used for the photon-subtraction technique which we will discuss later. After this transformation, represented by taking annihilation operator $\hat{a}_t$ to annihilation operators $\hat{a}_t\cos(\theta) + i \hat{b}_t\sin(\theta)$ where $\theta$ is the beam-splitter angle, the state is given by
\begin{equation}
    \boldsymbol{\hat{\rho}}_{\boldsymbol{Q}} = \bigoplus_{t=1}^M \int d^2\alpha \frac{1}{\pi \bar{n}_t}e^{-\frac{|\alpha|^2}{\bar{n}_t}}\left|\alpha\cos(\theta),i\alpha\sin(\theta)\rangle\langle\alpha\cos(\theta),i\alpha\sin(\theta)\right|_{a,b}.
\end{equation}
We will use the labels $a,b$ to represent the two output modes of the fiber-coupler.
From here, we make use of two photon-number-resolving detectors, one in each arm, to perform measurements. The primary difficulty of this measurement scheme is that the signal's strength is comparable to the noise of our two detectors, and the measurement techniques which we will employ are meant to alleviate the effects of that noise. The impacts of noise and detector efficiencies can be modeled with the photocounting technique, by which for a given state $\hat{\rho} = \sum_{n=0}^\infty p(n,m)\left|n\rangle \langle m\right|$, its diagonal matrix elements $p_{\text{noise}}(n,n)$ with dark counts $\nu$ and detector efficiency $\eta$ accounted for can be computed by
\begin{equation}
    p_{\text{loss}}(n,n) = \left\langle:\frac{(\eta\hat{n}+\nu)^n}{n!}e^{-(\eta\hat{n} + \nu)}:\right\rangle,
\end{equation}
where $:\cdot:$ is the normal-ordering prescription. In our case, these diagonal elements can be computed for dark counts $\nu_{a/b}$ and detector efficiencies $\eta_{a/b}$ as
\begin{equation}
    \begin{aligned}
               \Vec{\boldsymbol{p}}_{\boldsymbol{Q}}(n, m)&=\bigoplus_{t=1}^M\left\langle: \frac{\left(\eta_a \hat{n}_a+\nu_a\right)^n}{n !} e^{-\left(\eta_a \hat{n}_a+\nu_a\right)} \otimes \frac{\left(\eta_b \hat{n}_b+\nu_b\right)^m}{m !} e^{-\left(\eta_b \hat{n}_b+\nu_b\right)}:\right\rangle\\
               &=\bigoplus_{t=1}^M \int d^2\alpha \frac{1}{\pi \bar{n}_t}e^{-\frac{|\alpha|^2}{\bar{n}_t}} frac{\left(\eta_a |\alpha|^2\cos^2(\theta)+\nu_a\right)^n}{n !} \\
               &\text{ }\text{ }\text{ }\times
               e^{-\left(\eta_a |\alpha|^2\cos^2(\theta)+\nu_a\right)}\frac{\left(\eta_b |\alpha|^2\sin^2(\theta)+\nu_b\right)^m}{m !} e^{-\left(\eta_b |\alpha|^2\sin^2(\theta)+\nu_b\right)} \\
               &= \bigoplus_{t=1}^M\frac{e^{-\nu_a-\nu_b}}{n! m!}\sum_{i=0}^n\sum_{j=0}^m \binom{n}{i}\binom{m}{j}\eta_a^i\eta_b^j\nu_a^{n-i}\nu_b^{m-j}\cos^{2i}(\theta)\sin^{2j}(\theta)\\
               &\text{ }\text{ }\text{ }\times
               \int d^2\alpha \frac{|\alpha|^{2i+2j}}{\pi \bar{n}_t}e^{-\frac{|\alpha|^2}{\bar{n}_t} - \eta_a|\alpha|^2\cos^2(\theta) - \eta_b|\alpha|^2\sin^2(\theta)} \\
               &=\bigoplus_{t=1}^M\frac{e^{-\nu_a-\nu_b}}{\bar{n}_t n! m!}\sum_{i=0}^n\sum_{j=0}^m \binom{n}{i}\binom{m}{j}(i+j)!\frac{\eta_a^i\eta_b^j\nu_a^{n-i}\nu_b^{m-j}}{\left(\frac{1}{\bar{n}_t}+\eta_a\cos^2(\theta)+\eta_b\sin^2(\theta)\right)^{1+i+j}}\cos^{2i}(\theta)\sin^{2j}(\theta).
   \end{aligned}
\end{equation}
Unfortunately, the finite sum in the last line of this expression does not have a nice analytical form. However, since it is a finite sum, these diagonal matrix elements can be easily calculated numerically. 
When the signal $\bar{n}_t$ is absent, we will detect only the noise. In this case, the joint probability of noise event is given by $p_n(k,l) = p_{n,a}(k)p_{n,b}(l)$, where $p_{n,i}(k) = e^{-\nu_i}\frac{\nu_i^k}{k!}$. We note that our ability to reliably reconstruct the signal's mode profile from our measurements hinges on each of the $M$ measurements in arm $a$ being clearly distinguishable from its background noise. In other words, the signal-to-noise ratio for each measurement should be as high as possible. Let us now discuss two methods for accomplishing this.
\noindent
The first method is that of post-selection (Fock-projection) in arm $a$. This method does not utilize arm $b$, so that arm will always be traced out here. The signal-to-noise ratio in the case where we post-select on $N$ photons in arm $a$ can be represented by a vector, and is written as
\begin{equation}
    \overrightarrow{\textbf{SNR}}_{\text{post}}(N) = \frac{\sum_{m=0}^\infty \Vec{\boldsymbol{p}}_{\boldsymbol{Q}}(N,m)}{p_{n,a}(N)}=\frac{\Vec{\boldsymbol{p}}_{\boldsymbol{Q}}(N)}{p_{n,a}(N)}.
\end{equation}
Numerical evaluations of this quantity show that each component of the signal-to-noise ratio vector is increasing in an approximately exponential fashion with respect to $N$. This shows that we can greatly reduce the impact of noise on our data by post-selecting on high photon numbers.\\
\noindent
The other method showcased in Chapter \ref{chap:chapter6} is that of photon-subtraction, by which we first make a post-selective measurement in arm $b$ on $N$ photons and then measure the photon events in arm $a$. The conditional intensity in arm $a$ can then be written as $\langle\hat{\boldsymbol{n}}_a\rangle_N = \bigoplus_{t=0}^M\left(\sum_{k=0}^\infty k p_{\boldsymbol{Q},t}(k,N)\right)/\left(\sum_{k=0}^\infty p_{\boldsymbol{Q},t}(k,N)\right)$, where the factor in the denominator is due to the renormalization of the state after the measurement in arm $b$. Similarly, the noise measurement can be written as $\langle\hat{n}_a\rangle_{N,0} = \bigoplus_{t=0}^M\left(\sum_{k=0}^\infty k p_n(k,N)\right)/\left(\sum_{k=0}^\infty p_n(k,N)\right)$. By taking this approach, the resulting signal-to-noise ratio seen in arm $a$ can be represented by
\begin{equation}
    \overrightarrow{\textbf{SNR}}_{\text{sub}}(N) = \frac{\langle\hat{\boldsymbol{n}}_a\rangle_N}{\langle\hat{n}_a\rangle_{N,0}}.
\end{equation}
In contrast to the post-selection case, each component in this vector increases in an approximately linear fashion with respect to $N$. While this may be less desirable when compared to the exponential trend of post-selection in arm $a$, it is useful when precise post-selective measurements in arm $a$ cannot be made. For instance, if $\bar{n}_t$ is very large, then we can choose $\theta$ to be very small so that photon-number-resolution can be made accurate in arm $b$. This would allow us to increase the signal-to-noise ratio in arm $a$ through photon subtraction while making the more-precise measurement of intensity in that arm.\\
\noindent
Finally, our measurements in arm $a$ will be used to form a reconstruction of the signal vector, $\Vec{\boldsymbol{s}}_0$. This is accomplished using the compressive sensing (CS) technique, by which the reconstructed image, represented by $\Vec{\boldsymbol{s}}\in \mathbb{R}^X$, is found by minimizing the following quantity with respect to the dummy-vector $\Vec{\boldsymbol{s}}'\in \mathbb{R}^X$:
\begin{equation}
     \sum_{i=0}^X \lVert\nabla s_i'\rVert_{l_1} + \frac{\mu}{2}\lVert \boldsymbol{Q}\Vec{\boldsymbol{s}}' - \langle\hat{\boldsymbol{n}}\rangle\rVert_{l_2}.
\end{equation}
Here, $\langle\hat{\boldsymbol{n}}\rangle$ could be replaced with either of the previously-described quantities,  $\Vec{\boldsymbol{p}}_{\boldsymbol{Q}}(N)$ or $\langle\hat{\boldsymbol{n}}_a\rangle_N$. Moreover, the $1$- and $2$-norm are denoted by $\lVert\cdot\rVert_{l_1}$ and $\lVert\cdot\rVert_{l_2}$, respectively. The discrete gradient operator is described by $\nabla$, and the penalty factor by $\mu$. The value of $\Vec{\boldsymbol{s}}'$ which minimizes this quantity is then the value which we ascribe to $\Vec{\boldsymbol{s}}$. Accurate reconstruction of this image vector, such that $\Vec{\boldsymbol{s}}$ agrees with $\Vec{\boldsymbol{s}}_0$, is sensitive to background noise, and so by reducing the impact of that noise as much as possible via either of the two methods described above, we can attain a more reliable image of the signal.






\backmatter

\chapter{Works Cited}
\printbibliography[heading=myheading]
\end{document}

%% file: chapter1.tex
\chapter{Introduction}
\label{chap:intro}

\section{Quantum Sensing}

 Modern technologies heavily rely on precise measurements of physical properties, driving the rapid advancement of quantum technologies across various fields \cite{Nielsen2010,RevModPhysGaussianquantum,cariolaro2015quantum,Luo2023}. Within this landscape, quantum sensing emerges as a particularly promising field within quantum information science (QIS), offering transformative potential across diverse sectors \cite{Aslam2023}. 
Harnessing the distinct properties of light, quantum sensing encompasses a range of applications, from remote target detection to data retrieval from optical memories \cite{Gefen2019,atature2018material,Giovannetti2011}. At its core, quantum sensing and metrology exploit quantum systems, properties, or protocols to push the boundaries of parameter estimation and achieve precision in measuring physical quantities beyond classical capabilities \cite{Pirandola2018,PhysRevLettExploitingEntanglement}. However, these systems are highly sensitive to environmental conditions, such as loss or decoherence, which can introduce detrimental effects into measurements. Consequently, precise management of the sensing process becomes challenging, involving thorough preparation of the probe, ensuring effective interaction with the subject, and insightful interpretation of the collected data. During this process, there may be unwanted interference, either due to insufficient control over the probes or system or due to fundamental physical principles \cite{PhysRevLettPhaseMeasurement}. Unlike classical sensing techniques, quantum sensing achieves exceptional precision by allowing for control over the probes and dynamically adjusting measurement strategies. Optical interferometry, for instance, offers a framework to reframe precision measurements in terms of phase measurements, where the Heisenberg limit sets the boundary for error in the measured phase \cite{PhysRevLettDeterministic,PhysRevLettPhaseMeasurement}. Quantum techniques, when applied to sensing, not only enhance measurement precision but also enable reducing unwanted noise below the shot limit, thereby revealing capabilities previously inaccessible through classical methods \cite{Crawford2021,Giovannetti2011}. An intriguing avenue in quantum sensing lies in the utilization of quantum states with specific properties, such as entangled states, to surpass the standard quantum limit. For instance, employing N photons in a $\text{N00N}$ state superposition can yield a measurement accuracy inversely proportional to N, representing a fundamental limit in phase measurements. The superiority of entangled sources over classical ones in enhancing precision stems from the entangled states' inherent ability to evolve more rapidly, presenting exciting prospects for further advancements in quantum sensing capabilities \cite{Giovannetti2003}. Classical methods simply cannot reach the same level of sensitivity and control, making quantum sensing a transformative approach for pushing the boundaries of precision measurement.  

\section{Multiphoton Quantum Sensing}

Photonics emerges as a promising platform for realizing quantum technologies, offering resilience against decoherence and facilitating precise control over photonic quantum states using conventional optical components \cite{Shekhar2024,christen2022integrated}. This versatility extends across various applications, from long-distance communications to simulating complex phenomena. While single photons serve as preferred carriers for quantum information across diverse fields \cite{ref1,Flamini2018,DellAnno2006,Zhang:24}, their limitations become apparent when aiming to create more intricate quantum states. Multiphoton processes offer a solution to this challenge \cite{ChenglongAIP,Zhang2021,Zhang2019}. Observed in various light-matter interactions, multiphoton processes hold significant promise for applications in quantum communication, computation, simulation, and metrology \cite{ChenglongAIP,Giovannetti2011,BarretoLemosmetrology}. Leveraging multiplexing and postselection techniques holds promise in their generation.
Advancements in integrated photonics have played a crucial role in generating multiphoton states. Compared to free-space optics, these technologies enable superior compatibility with chip-based manipulation and detection of quantum states. Waveguides and high-Q cavities provide strong mode confinement and intensified field interaction, facilitating the development of efficient on-chip sources.
Integrated photonic and plasmonic chips have emerged as effective platforms for creating multiphoton states of light, with applications spanning telecommunications, sensing, computing, and data processing \cite{Zhang2019,Hong2024,Exceptionalpoints,Tschernig2020}. 
Techniques employed for producing these states involve nonlinear optical phenomena like spontaneous parametric down-conversion (SPDC) or four-wave mixing (FWM) \cite{Zhang2019,Llewellyn2020,Guo2017,Zhang2021}. These processes utilize the interaction of light with nonlinear materials to generate correlated or entangled photons \cite{PhysRevLettScalableSpatial,Osorio2008,Edamatsu2004}. 
Quantum interferometry, a cornerstone of quantum sensing, utilizes the unique properties of quantum states to explore and exploit interference phenomena. Within quantum optics, $\text{N00N}$ states offer an innovative approach, harnessing entanglement to achieve enhanced sensitivity \cite{Dowling2008}. Unlike regular light, where photons act independently, $\text{N00N}$ states exhibit coherence between two modes, translating to a noticeable sensitivity boost for interferometry experiments. For N photons, the state has the following form
{\begin{equation}
     |\text{N00N}\rangle = \sqrt{\frac{1}{2}}( |\text{N0}\rangle + e^{i\phi}  |\text{0N}\rangle). 
     \label{eq1-1} 
    \end{equation}}
    \noindent
    These states offer a remarkable improvement in sensitivity compared to conventional interferometers using coherent light. While traditional setups scale phase sensitivity with the square root of the number of photons ($\sqrt{\text{N}}$), precise engineering of correlations between incoming photons in a $\text{N00N}$ state allows sensitivity be proportional to the number of photons $\text{N}$, reaching the Heisenberg limit. This notable enhancement in measurement precision holds potential applications in diverse fields.
\noindent
\section{Maxwell’s Equations and Quantum Optics}\label{maxwell}
Building upon the concept of engineered correlations in multiphoton states \cite{Loaiza2019}, we can explore how the electromagnetic field itself exhibits quantum properties.  Classically described by Maxwell's equations, light can be understood as a wave phenomenon \cite{wolf1954optics}.  However, when we delve into the quantum realm, the electromagnetic field can also be viewed as quantized, meaning its energy comes in discrete packets called photons \cite{Mandel1995}.  
In this section, we examine the quantization of Maxwell’s equations, focusing on scenarios without radiation sources and the propagation of electromagnetic waves in vacuum. The Maxwell’s equations take the following form under these conditions \cite{wolf1954optics}:

\begin{equation*}
   \nabla \cdot  \bold{E}  =0
    \end{equation*} 
    \begin{equation*}
     \nabla \cdot  \bold{B} =0     
    \end{equation*}
    \begin{equation*}
     \nabla \times \bold{B} = \frac{ 1}{ c^2}\frac{\partial \bold{E}}{\partial t}
    \end{equation*}
    \begin{equation}
     \nabla \times \bold{E} =-\frac{\partial \bold{B}}{\partial t}. 
     \label{eq1-2} 
    \end{equation}
    Here, $c^{2} = 1 /{\epsilon_{0} \mu_{0}}$ is the speed of light in vacuum, and $ \bold{E}$ and $\bold{B}$ represent the electric and magnetic fields, respectively. 
    These equations govern the behavior of electric and magnetic fields in a vacuum, forming the fundamentals of classical electromagnetism. However, in the realm of quantum optics, where the discrete nature of light becomes apparent, a different perspective emerges. Light can be described not only in terms of continuous electromagnetic waves but also as individual photons exhibiting quantized behavior. This duality prompts exploration into various forms of light states, ranging from coherent states to photon number states and thermal states. Each of these states offers unique characteristics and finds applications across different domains of quantum optics and quantum information science \cite{Aslam2023}. In the following, we explore the process of quantizing light and examine the resulting electric and magnetic fields, representing the single-mode field solutions of one-dimensional perfectly conducting walls of length $L$.
    \begin{equation}
   E_{x}(z,t) =\sqrt{\frac{2 w^2}{V \epsilon_{0}}} q(t) \sin(kz),
    \end{equation} 
\begin{equation}
   B_{y}(z,t) =\frac{1}{c^2}\sqrt{\frac{2 w^2}{V \epsilon_{0}}} \frac{\dot q(t)}{k} \cos(kz).
    \end{equation} 
Where $ w $ is the frequency, $k$ is the wave number satisfying the $ k= j\pi/L  \quad  (j=1,2,3,...)$, $V$ is the volume, the $ \epsilon_{0}$ is the permittivity of free space, and $q(t)$  defines a canonical position. Then, the classical field energy of the single-mode field is given by
\begin{equation}
   \text{H} =\frac{1}{2}\ \int_{}^{}( \frac{1}{\mu_{0}} {B}^2(z,t)+\epsilon_{0} {E}^2(z,t))\,dz= \frac{1}{2}(p^2+w^2 q^2) .
    \end{equation}
    This representation of the single-mode field is equivalent to a harmonic oscillator with the commutation relation as $ [\hat{q}, \hat{p}]=i \hbar$, $ \hbar= h/2 \pi$. Introducing the non Hermitian annihilation and creation operators $ \hat{a}$ and $ \hat{a}^{\dagger}$ as 
    \begin{equation}
   \hat{a}=\sqrt{\frac{1}{2  \hbar w}} (w \hat{q}+ i\hat{p}),
    \end{equation} 
    \begin{equation}
    \hat{a}^{\dagger}=\sqrt{\frac{1}{2  \hbar w}} (w \hat{q}- i\hat{p}).
    \end{equation}
    Therefore, the Hamiltonian can be written as 
    \begin{equation}
    \hat{\text{H}}=  \hbar w (\hat{a}^{\dagger}\hat{a} + \frac{1}{2}).
    \end{equation}
Having established the mathematical framework for quantized light,  we can now delve into the distinct states a quantized electromagnetic field can occupy. These states, known as photon states, characterize the probabilistic distribution of photons within the mode. Three prominent types of photon states that play a crucial role in quantum optics are coherent states, thermal states, and Fock states \cite{agarwal2012quantum}.
  \noindent
  \subsection{Fock State }
    We start by explaining a specific category of nonclassical light states known as Fock or photon number states. These states, represented as $|n\rangle$, are eigenstates of the Hamiltonian H, with $n$ denoting the number of photons within a singular mode of the electromagnetic field. Notably, Fock states possess a precisely defined number of particles, allowing us to establish the particle number operator as $ \hat{n} |n\rangle = n |n\rangle$. 
    \noindent 
    The operation of the creation and annihilation operators on the Fock state $  |n\rangle$ can be described as follows:
 \begin{equation}
     \hat{a}^{\dagger} |n\rangle = \sqrt{n+1} |n+1\rangle.
    \end{equation}
    \begin{equation}
     \hat{a} |n\rangle = \sqrt{n} |n-1\rangle. 
    \end{equation}
     \noindent 
   When all photons are in the same mode, Fock states can be classified into single-mode Fock state that can be obtained from the vacuum state $|0\rangle$ as 
   \begin{equation}
    |n\rangle = \frac{(\hat{a}^{\dagger})^{n}}{\sqrt{n!}}|0\rangle
    \label{eq1-3}
    \end{equation}
where $n$ is the total number of photons. Equation \eqref{eq1-3} reveals additional properties of these quantum states, like being orthonormal. 
 \begin{equation}
\langle n|m\rangle={\delta}_{nm}
\label{eq1-4}
    \end{equation}
    and creating the complete basis
     \begin{equation}
\sum^{\infty}_{n=1}|n\rangle\langle n|= \hat{\text{I}}
    \end{equation}

 \noindent 
\subsection{Coherent States}
 Laser beams, often described as highly directional and well-defined light sources, can be effectively modeled using coherent states in quantum mechanics. While these states exhibit classical-like behavior, they differ fundamentally at the quantum level. A coherent state, denoted by $|\alpha\rangle$, is characterized by its relationship with the annihilation operator.   
 \begin{equation}
\hat{a} |\alpha\rangle=\alpha |\alpha\rangle.
\label{eq1-5}
    \end{equation}
  Similar to Fock states, coherent states are eigenstates of the annihilation operator. These states together form a complete set of basis states, allowing for their representation in the number basis  
  \begin{equation}
    |\alpha\rangle =\exp{(-\frac{1}{2} |\alpha|^2)} \sum^{\infty}_{n=0}\frac{(\alpha^{n})}{\sqrt{n!}}|n\rangle.
    \label{eq1-6}
    \end{equation}
 This equation highlights the distinction between coherent and Fock states. Unlike specific photon number states (Fock states), coherent states represent a superposition of various Fock states with a complex number $\alpha$ determining the weight of each contribution. This superposition results in an uncertain number of photons in a coherent state. Their average photon number $\langle n  \rangle $ defines a Poissonian distribution, meaning the probability of finding a specific photon number follows a predictable bell-shaped curve centered at $\langle n  \rangle $ 
 \begin{equation}
    P(n) =e^{-|\alpha|^2} \frac{|\alpha|^{2n}}{n!}
    \label{eq1-7}
    \end{equation}
    with $\langle n  \rangle = |\alpha|^{2}$ and $ \Delta n= |\alpha|=\sqrt{\langle n  \rangle } $ .
  \noindent
  \subsection{Thermal States}
 In contrast to the well-defined properties of coherent states, everyday light sources like sunlight are characterized by thermal states. These states represent a statistical mixture of Fock states, reflecting the random nature of light emission processes in thermal environments. Mathematically, a thermal state can be expressed as a statistical average of Fock states, represented by 
 \begin{equation}
\hat{\rho}_{\text{th}}=\frac{\bar n}{1+\bar n} \sum^{\infty}_{n=0}(\frac{\bar n}{1+\bar n})^n |n\rangle\langle n|.
\label{eq1-8}
    \end{equation}
     \noindent
Here $ \bar n $ denotes the average photon number of the thermal field. 
Unlike coherent states with Poissonian photon number statistics, thermal light exhibits a distinct behavior. It possesses super-Poissonian statistics, meaning the fluctuations in photon number are larger than the Poissonian distribution. This translates to a standard deviation of $ \langle \Delta n ^2  \rangle=\bar n +\bar n^2$. \\
  \noindent
 Having explored the properties of different quantum states, we now turn our attention to how Maxwell's equations govern light propagation in two distinct regimes: the far field and the near field. The far field, describes by the Fraunhofer approximation, is crucial for understanding how light interacts with objects at a distance, a principle employed in many imaging techniques used within quantum sensing applications. Conversely, the near field explores how light interacts with matter on the nanoscale \cite{Novotny2012}. This regime plays a vital role in plasmonics and nanophotonics, enabling light-matter interaction at the nanoscale and opens doors for developing high-density and ultrasensitive quantum sensors with enhanced light-matter coupling. 
\section{Wave Propagation - Far-Field Optics}
To further investigate the properties of light traveling over long distances (far field), we need to explore the underlying principles governing this behavior. This exploration begins with Maxwell's equations, which describe the fundamental relationship between electric and magnetic fields, even in the presence of sources like currents (represented by J) and charge densities (represented by $ \rho$). By analyzing these equations in the frequency domain, we can gain valuable insights into how light propagates in the far field and pave the way for understanding the Fraunhofer approximation, a key concept for analyzing far field radiation patterns \cite{saleh2019fundamentals,goodman2008introduction,Lecture20}. These equations represented as,  
 \begin{equation}
 \nabla \times \bold{E} =- j w \mu \bold{H} 
 \label{eq1-4E}
    \end{equation}
     \begin{equation}
 \nabla \times \bold{H} =- j w \epsilon \bold{E} + \bold{J} 
 \label{eq1-4H}
    \end{equation}
     \begin{equation}
 \nabla \cdot \mu \bold{H} =0 
    \end{equation}
     \begin{equation}
 \nabla \cdot \epsilon \bold{E} =\rho 
  \label{eq1-4rho}
    \end{equation}
     \noindent
     where $ \bold{H}= \bold{B}/\mu$ is the magnetic field and $w$ is the angular frequency. The next step involves introducing the vector potential, defined as  $ \mu \bold{H} = \nabla \times \bold{A} $. By employing this substitution in Eq. (\ref{eq1-4E}), we obtain a new equation involving the vector potential and the electric field
\begin{equation}
 \nabla \times( \bold{E} + j w \bold{A})=0
    \end{equation}
     which implied $ \bold{E} = -j w \bold{A} - \nabla \phi$. 
     We can now substitute these expressions into the second Maxwell's equation. The substitution leads to a new equation involving the vector potential $\bold{A}$ and the scalar potential $ \phi$. To further simplify this equation, we can exploit the Lorenz gauge condition $  \nabla \cdot \bold{A} = -j w \mu\epsilon\phi $. Imposing this gauge and utilizing vector calculus identities, we arrive at a pair of decoupled wave equation
\begin{equation}
 \nabla^2 \bold{A} + w^2\mu\epsilon \bold{A}=-\mu \bold{J} 
 \label{eq1-4Adecoupled}
    \end{equation} 
\begin{equation}
 \nabla^2 \phi + w^2\mu\epsilon \phi=-\frac{\rho}{\epsilon}
 \label{eq1-4phidecoupled}
    \end{equation} 
     \noindent
These equations can be solved independently using Green’s  method. Notably, their solutions for point-source configurations become particularly relevant when analyzing light propagation in the far field, where the Fraunhofer approximation holds

\begin{equation}
 \bold{A}(\bold{r})=\frac{\mu}{4\pi} \iiint d\bold{r}^\prime \frac{\bold{J}(\bold{r}^\prime)}{4\pi |\bold{r}-\bold{r}^\prime|}e^{-j\beta|\bold{r}-\bold{r}^\prime|}
 \label{eq1-Asolution}
    \end{equation} 

    \begin{equation}
 \phi(\bold{r})=\frac{1}{4\pi\epsilon} \iiint d\bold{r}^\prime \frac{\rho(\bold{r}^\prime)}{4\pi |\bold{r}-\bold{r}^\prime|}e^{-j\beta|\bold{r}-\bold{r}^\prime|}
 \label{eq1-phisolution}
    \end{equation} 
where $ \beta^{2}= w^{2}\mu\epsilon$. We can approximate the integral in  Eq. (\ref{eq1-Asolution}) and  (\ref{eq1-phisolution}). When $ |\bold{r}| \gg|\bold{r}^{\prime}|$, then the $|\bold{r}-\bold{r}^{\prime}|\approx r-\bold{r}^{\prime}\cdot \hat{r} $. Thus,  Eq. (\ref{eq1-Asolution}) becomes 
\begin{equation}
 \bold{A}(\bold{r})\approx \iiint d\bold{r}^\prime \frac{\bold{J}(\bold{r}^\prime)}{ r-\bold{r}^{\prime}\cdot \hat{r}}e^{-j\beta r + j\beta \bold{r}^{\prime}\,\cdot\,\hat{r}}\approx \frac{\mu e^{-j\beta r}}{4\pi r}\iiint d\bold{r}^\prime  \bold{J}(\bold{r}^{\prime}) e^{-j\beta \bold{r}^{\prime}\,\cdot\, \hat{r}}
    \end{equation} 
This approximation leads to a simplified expression for the vector potential in the far field. Notably, this expression reveals the far-field radiation pattern of the source takes the form of a 3D fourier transform of the source current distribution, $ \bold{A}(\bold{r})\approx \mu e^{-j\beta r}/4\pi r \bold{F}(\boldsymbol{\beta})$ where $\bold{F}(\boldsymbol{\beta})$ represents the 3D fourier transform of $\bold{J}(\bold{r}^{\prime}) $ and $ \boldsymbol{\beta} = \hat{r}\beta $ with the magnitude of the vector $\beta$ fixed. In essence, the far field behavior of light translates to a spherical wave.  By expressing $\hat{r}$ and $\boldsymbol{\beta}$ in terms of spherical coordinates and  analyzing the electric E and magnetic H fields derived from the vector potential, we have 
\begin{equation}
 \bold{H}\approx -j\frac{\beta}{\mu}\hat{r}(A_{\theta}\hat{\theta}+ A_{\phi}\hat{\phi})
    \end{equation} 
   \begin{equation}
 \bold{E}\approx -jw(A_{\theta}\hat{\theta}+ A_{\phi}\hat{\phi})
    \end{equation} 
     \noindent
In the far field, light propagates as a spherical wave with minimal wavefront curvature and $ \beta r^{\prime2}/2r \ll 1$. Consequently, $ r \gg  \pi r^{\prime2}/\lambda=r_{R}$. This simplification holds when the observation distance r is significantly larger than a specific distance called the Rayleigh distance $r_{R}$. The Rayleigh distance marks the boundary between the near field (where detailed wavefront analysis is needed) and the far field (where the spherical wave approximation applies).
While the far-field analysis provides valuable insights into the far-zone radiation pattern and its connection to the source current distribution via the fourier transform, it's important to acknowledge its limitations. Techniques like the Abbe-Rayleigh criterion utilize this analysis to estimate the diffraction-limited resolution of optical systems. However, the ultimate barrier to resolution lies in the fundamental principles of quantum mechanics, particularly the Heisenberg uncertainty principle \cite{OmarAngularRotations}. Overcoming this fundamental limit is crucial for various fields like microscopy, remote sensing, and astronomy, driving research in novel methods to enhance the spatial resolution of optical systems \cite{ref1Microscopy,Di2023}. Chapter 6 focuses on one such approach, specifically addressing scenarios where background noise significantly overshadows the signal of interest.

\section{ Wave Propagation - Near-Field Optics }
In contrast to the far field, where light propagates as a spherical wave, the near field exhibits a much richer and more complex behavior. This section explores the near-field regime, its connection to plasmonic effects, and its various nanoscale applications.
While near field optics (NFO) and surface plasmon polaritons (SPPs) describe distinct physical phenomena, they share a connection – their reliance on the near field regime for manipulating light at the nanoscale. Near-field optics, as the name suggests, explores the behavior of high-frequency electromagnetic fields near structures much smaller than the wavelength of light. Here, the wavefronts have significant curvature, and the fields exhibit strong spatial variations compared to the far field's spherical wave \cite{Kawata,saleh2019fundamentals}. \\
On the other hand surface plasmons arise from the collective oscillation of electrons at the interface between a metal and a dielectric material. These oscillations, when excited by light in the near field, can couple with the electromagnetic field, leading to a type of propagating wave called a surface plasmon polariton (SPP). SPPs are confined to the interface and decay exponentially away from it, a defining characteristic of near field phenomena known as evanescent waves \cite{Maier}. \\
 \noindent
In the near field, where the characteristic dimensions of the scatterers or sources are much smaller than the wavelength of light, Maxwell's equations take a slightly different form compared to the far field. While the underlying principles remain the same, some terms become dominant, and others can be neglected due to the rapid spatial variations of the fields. A common approach for the near field involves the scalar wave approximation. This approximation assumes that the electric field can be represented by a single scalar potential ($\phi$) such that $\bold{E} = -\nabla \phi$. This simplification is valid when the magnetic field contribution is negligible compared to the electric field, which often holds for certain near field phenomena like scattering from subwavelength structures \cite{Maier}.
By applying the scalar wave approximation to the wave equation 
\begin{equation}
 \nabla^2 \bold{E} + k_{0}^{2}\epsilon \bold{E}=0
 \label{eq1-5Helmholtz}
    \end{equation} 
Here, $ k_{0}$ is the wavenumber in vacuum $k_{0} = 2\pi/\lambda_{0}$, $\lambda_{0}$ is the free-space wavelength, and $\epsilon$ is the permittivity of the medium. To analyze surface plasmon polaritons (SPPs) at a metal-dielectric interface, we can simplify the problem by assuming a one-dimensional variation of permittivity $  \epsilon = \epsilon(z)$ along a single spatial axis z with light propagation along the $x$-direction \cite{Maier}. This simplification implies the electric and magnetic field components do not vary in the $y$-direction (homogeneity). We can express the electric field as a propagating waveform: $ \bold{E}(x,y,z)= \bold{E}(z) e^{j \beta x}$. Inserting this propagating waveform into the Helmholtz equation Eq. (\ref{eq1-5Helmholtz}) and separating terms based on their $z$-dependence, we obtain an ordinary differential equation (ODE) for the electric field component $E_{z}$ in the $z$-direction: 
\begin{equation}
 \frac{\partial^{2} \bold{E}(z)}{\partial z^{2}} + (k_{0}^{2}\epsilon + \beta^{2}) \bold{E(z)}=0
 \label{eq1-5Helmholtz-1a}
    \end{equation} 
Since we're focusing on TM (P-polarization) modes, where only the $E_{x}$, $E_{z}$, and $H_{y}$ components are non-zero, we can utilize one of Maxwell's curl equations to derive the equation for the magnetic field component $H_{y}$. We obtain the ODE for the magnetic field component as
\begin{equation}
 \frac{\partial^{2} H_{y}}{\partial z^{2}} + (k_{0}^{2}\epsilon + \beta^{2}) H_{y}=0
 \label{eq1-5Helmholtz-1b}
    \end{equation}
    These coupled equations, with the appropriate boundary conditions at the interface $z = 0$, will allow us to determine the characteristic properties of TM-polarized SPP modes, such as their propagation constants $\beta$ and field distributions within the metal and dielectric regions. We can also arrive at the dispersion relation. 
    This relation typically expresses the dependence of the propagation constant on the wave number in vacuum and the permittivities of the metal $ \epsilon_{1}$ and the dielectric $ \epsilon_{2}$ as 
 \begin{equation}
\beta= k_{0} \sqrt{\frac{\epsilon_{c} \epsilon_{d}}{\epsilon_{c}+\epsilon_{d}}}.
\label{eq1-14}
    \end{equation}
After exploring the near field regime and its interaction with surface plasmon polaritons (SPPs), it becomes evident that the manipulation of light at the nanoscale holds significant promise for various applications. Recent progress in the field, particularly through studies on the behavior of quantum plasmonic systems, contributes significantly to our comprehension of these phenomena. Prime examples of such advancements are \cite{You2020,Mostafavi2022,Hong2024}.\\


\section {Outline}
 \noindent 
This thesis is organized to explore advancements in multiphoton quantum sensing and their applications. \\
Chapter 2 of this thesis establishes a comprehensive theoretical framework for understanding the nonclassical dynamics of light-matter interactions. Our experimental observation demonstrate the possibility to manipulate spatial coherence and quantum fluctuations. This chapter's discussion is based on \cite{You2020}. \\
\noindent 
Chapter 3 delves into the theoretical explanation of how conditional detection can effectively control quantum fluctuations within plasmonic fields. The chapter explores the intricacies of conditional detection in manipulating quantum fluctuations.  The analysis presented in this chapter showcases the advantages of employing conditional detection in photonic sensors that rely on plasmonic signals. This chapter's discussion is based on \cite{Mostafavi2022}. \\
\noindent 
Chapter 4 explores the ability to manipulate the spatial wavefunction of photons, providing a versatile toolbox for various quantum protocols in communication, sensing, and information processing domains. The chapter centers on harnessing individual orbital angular momentum states of light, as well as their superpositions for encoding messages, thereby facilitating secure communication. This chapter's discussion is based on \cite{Lollie2022}.\\
\noindent 
\noindent 
In Chapter 5, we investigate the modification of quantum coherence within multiphoton systems as they propagate in free space. We demonstrate the possibility of generating multiphoton systems exhibiting sub-shot-noise coherence properties through scattering processes. Finally, we discuss the potential impact of manipulating coherence in multiphoton systems on various quantum technologies. This chapter's discussion is based on \cite{quantumcoherence2024}.\\
\noindent
Chapter 6 provides an overview of the quantum imaging experiment conducted, emphasizing the utilization of natural light sources. The chapter begins with the challenges posed by background noise in quantum imaging. Then detailed description of the experimental setup is presented, outlining the innovative approach of utilizing natural light sources for quantum imaging. We show the effectiveness of the proposed scheme in overcoming background noise and achieving high-quality imaging.  This chapter's discussion is based on our " Multiphoton Quantum Imaging using Natural Light"  manuscript which is currently under review. This chapter's discussion is based on \cite{QImage}. Additionally, we are pursuing patent protection for our work on this chapter.\\
\noindent
In Chapter 7 we conclude the thesis with a summary of the impact of multiphoton quantum sensing, its implications, and overall experimental contributions to the field.

%% file: chapter2.tex
\chapter{Engineering Quantum Statistics of Multiphoton Systems in Nanostructures}
\label{chap:chapter2}
\section{Motivation}

For almost two decades, researchers have observed the preservation of the quantum statistical properties of bosons in a large variety of plasmonic systems \cite{Tame2013,You2020}. In addition, the possibility of preserving nonclassical correlations in light-matter interactions mediated by scattering among photons and plasmons stimulated the idea of the conservation of quantum statistics in plasmonic systems. It has also been assumed that similar dynamics underlie the conservation of the quantum fluctuations that define the nature of light sources \cite{PhysRevCoherentIncoherent,Mandel1979}. So far, plasmonic experiments have been performed in nanoscale systems in which complex multiparticle interactions are restrained. Here, we demonstrate that the quantum statistics of multiparticle systems are not always preserved in plasmonic platforms and report the observation of their modification. Moreover, we show that optical near fields provide additional scattering paths that can induce complex multiparticle interactions. Remarkably, the resulting multiparticle dynamics can, in turn, lead to the modification of the excitation mode of plasmonic systems. These observations are validated through the quantum theory of optical coherence for single- and multi-mode plasmonic systems. Our findings unveil the possibility of using multiparticle scattering to perform exquisite control of quantum plasmonic systems.

\section{Background}
The observation of the plasmon-assisted transmission of entangled photons gave birth to the field of quantum plasmonics almost 20 years ago \cite{Altewischer}. Years later, the coupling of single photons to collective charge oscillations at the interfaces between metals and dielectrics led to the generation of single surface plasmons \cite{Akimov}. These findings unveiled the possibility of exciting surface plasmons with quantum mechanical properties \cite{Tame2013,You2020}. In addition, these experiments demonstrated the possibility of preserving the quantum properties of individual photons as they scatter into surface plasmons and vice versa \cite{Martino,Fasel,Huck,Daniel,ExtraordinaryLawrie,10,Safari}. This research stimulated the investigation of other exotic quantum plasmonic states \cite{Martino,Fasel,ExtraordinaryLawrie,Vest,Kolesov2009}. Ever since, the preservation of the quantum statistical properties of plasmonic systems has constituted a well-accepted tenant of quantum plasmonics \cite{Tame2013,You2020}.\\
In the realm of quantum optics, the underlying statistical fluctuations of photons establish the nature of light sources \cite{PhysRevCoherentIncoherent,Mandel1979}. These quantum fluctuations are associated to distinct excitation modes of the electromagnetic field that define quantum states of photons and plasmons \cite{Tame2013,DellAnno2006,Loaiza2019}. In this regard, recent plasmonic experiments have demonstrated the preservation of quantum fluctuations while performing control of quantum interference and transduction of correlations in metallic nanostructures \cite{Fasel,Daniel,10,Safari,BenjaminVest,Cai,Dheur,Dongfang,Lee2016,Dowran2018,Holtfrerich2016}. Indeed, the idea behind the conservation of quantum statistics of plasmonic systems results from the simple single-particle dynamics supported by the plasmonic nanostructures and waveguides used in previous experiments \cite{Fasel,Daniel,10,Safari,BenjaminVest,Dheur,Dongfang,Lee2016,Dowran2018,Holtfrerich2016}. Despite the dissipative nature of plasmonic fields, the additional interference paths provided by optical near fields have enabled the harnessing of quantum correlations and the manipulation of spatial coherence \cite{BenjaminVest,Dongfang,Alexander,Schouten,maganaloaizaexotic2016}. So far, this exquisite degree of control has been assumed independent of the excitation mode of the interacting particles in a plasmonic system \cite{Tame2013}. Moreover, the quantum fluctuations of plasmonic systems have been considered independent of other properties such as polarization, temporal, and spatial coherence \cite{Martino,Dongfang,Alexander,Schouten,maganaloaizaexotic2016}. Hitherto, physicists and engineers have been relying on these assumptions to develop plasmonic systems for quantum control, sensing, and information processing \cite{Tame2013,You2020,DELLANNO200653,BenjaminVest,Dowran2018,hofer2019hermite,OBrien2009NaturePhotonics}.\\
Previous research has stimulated the idea that the quantum statistical properties of bosons are always preserved in plasmonic systems \cite{Tame2013,You2020,Martino,Fasel,Huck,ExtraordinaryLawrie}. Here, we demonstrate that this is not a universal behavior and consequently the quantum statistical fluctuations of a physical system can be modified in plasmonic structures. We reveal that scattering among photons and plasmons induces multiparticle interference effects that can lead to the modification of the excitation mode of plasmonic systems. Remarkably, the multiparticle dynamics that take place in plasmonic structures can be controlled through the strength of the optical near fields in their vicinity. We also show that plasmonic platforms enable the coupling of additional properties of photons to the excitation mode of multiparticle systems. More specifically, we demonstrate that changes in the spatial coherence of a plasmonic system can induce modifications in the quantum statistics of a bosonic field. Given the enormous interest in multimode plasmonic platforms for information processing \cite{Daniel,ExtraordinaryLawrie,Safari,BenjaminVest}, we generalize our single-mode observations to a more complex system comprising two independent multiphoton systems. We validate our experimental results through the quantum theory of optical coherence \cite{PhysRevCoherentIncoherent}. The possibility of controlling the underlying quantum fluctuations of multiparticle systems has important implications for practical quantum plasmonic devices \cite{Tame2013,You2020}.\\

\section{Theory}
\begin{figure}[!htb]
	\centering
	\includegraphics[width=0.95\textwidth]{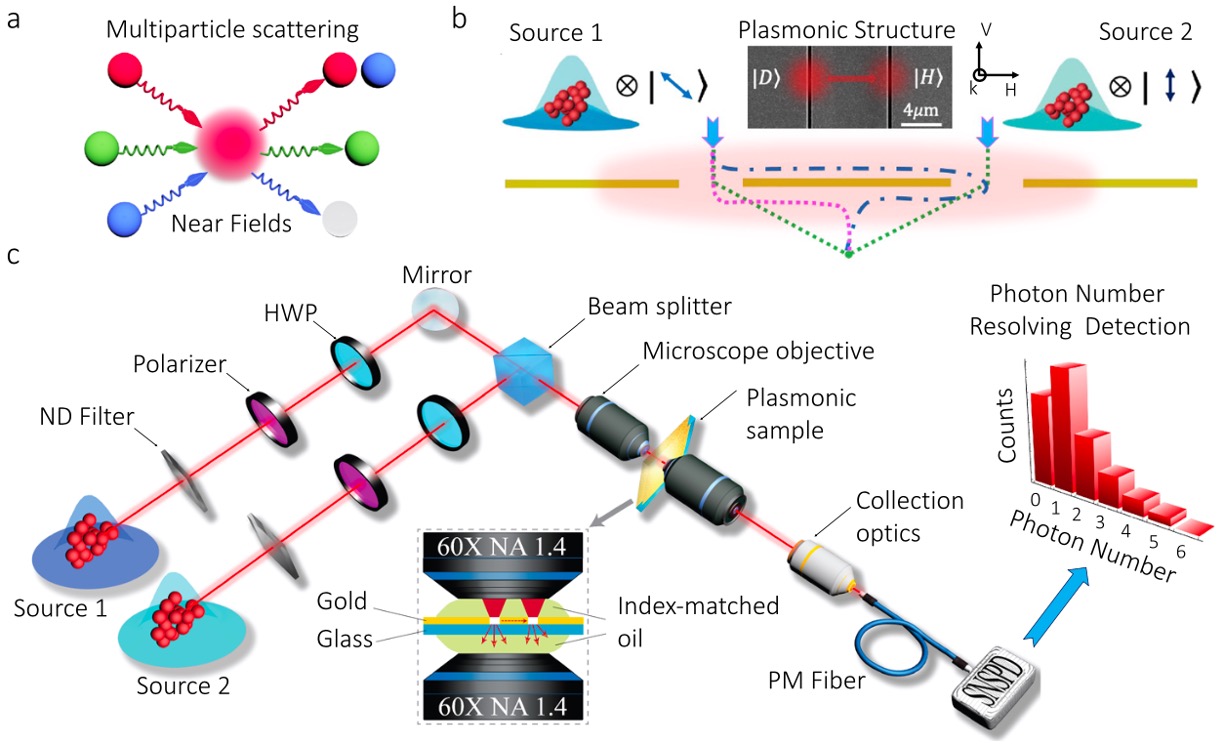}
	\mycaption{Multiparticle Scattering in Plasmonic Systems.}{ The diagram in (a) illustrates the concept of multiparticle scattering mediated by optical near fields. The additional interference paths induced by confined near fields lead to the modification of the quantum statistics of plasmonic systems. This idea is implemented through the plasmonic structure shown in (b). The dotted lines represent additional scattering paths induced by confined optical fields in the plasmonic structure \cite{maganaloaizaexotic2016}. Our metallic structure consists of a $110$ nm thick gold film with slit patterns. The width and length of each slit are $200$ nm and \SI{40}{{\micro\meter}}, respectively. The slits are separated by \SI{9.05}{{\micro\meter}}. The fabricated sample is illuminated by either one or two thermal sources of light with specific polarizations. The strength of the plasmonic near fields is controlled through the polarization of the illuminating photons. The plasmonic near fields are only excited with photons polarized along the horizontal direction. The experimental setup for the observation of the modification of quantum statistics in plasmonic systems is shown in (c). We prepare either one or two independent thermal multiphoton states with specific polarizations. The polarization state of each of the multiphoton systems is individually controlled by a polarizer (Pol) and half-wave plate (HWP). The two multiphoton states are injected into a beam splitter (BS) and then focused onto the gold sample through an oil-immersion objective. The refractive index of the immersion oil matches that of the glass substrate creating a symmetric index environment around the gold film. The transmitted photons are collected with another oil-immersion objective. We measure the photon statistics in the far field using a superconducting nanowire single-photon detector (SNSPD) that is used to perform photon-number-resolving detection \cite{you2020identification,HashemiRafsanjani:17}. These figures are taken from \cite{chenglongnature}.} 
	\label{fig:figure1-ch2}
\end{figure}

We now introduce a theoretical model to describe the global dynamics experienced by a multiphoton system as it scatters into surface plasmons and vice versa. Interestingly, these photon–plasmon interactions can modify the quantum fluctuations that define the nature of a physical system \cite{PhysRevCoherentIncoherent,DELLANNO200653,Loaiza2019}. As illustrated in Fig. \ref{fig:figure1-ch2}a, the multiparticle scattering processes that take place in plasmonic structures can be controlled through the strength of the confined near fields in their vicinity. The near field strength defines the probability of inducing individual phase jumps through scattering \cite{Safari}. These individual phases establish different conditions for the resulting multiparticle dynamics of the photonic–plasmonic system.

 \noindent
We investigate the possibility of modifying quantum statistics in the plasmonic structure shown in Fig. \ref{fig:figure1-ch2}b. The scattering processes in the vicinity of the multi-slit structure lead to additional interference paths that affect the quantum statistics of multiparticle systems \cite{BenjaminVest,maganaloaizaexotic2016}. The dynamics can be induced through either propagating or non-propagating near fields. Consequently, one could study this phenomenon using localized or propagating surface plasmons \cite{Maier}. In particular, for this study, we focus on the propagating plasmonic fields supported by the structure. The gold structure in Fig. \ref{fig:figure1-ch2}b consists of two slits aligned along the y-direction (see section \ref{Sample Design and Experiment} ). The structure is designed to excite plasmons when it is illuminated with thermal photons polarized along the x-direction \cite{Schouten,maganaloaizaexotic2016}. For simplicity, we will refer to light polarized in the x- and y-direction as horizontally $ |\text{H}\rangle$ and vertically $ |\text{V}\rangle$ polarized light, respectively. Our polarization-sensitive plasmonic structure directs a fraction of the horizontally polarized photons to the second slit when the first slit is illuminated with diagonally $ |\text{D}\rangle$ polarized photons. As depicted in Fig. \ref{fig:figure1-ch2}b, this effect is used to manipulate the quantum statistics of a mixture of photons from independent multiphoton systems \cite{Arecchi,PhysRevLett.TurbulenceFree}.
 \noindent
The modification of quantum statistics induced by the scattering paths in Fig. \ref{fig:figure1-ch2}b can be understood through the Glauber-Sudarshan theory of coherence \cite{PhysRevCoherentIncoherent,PhysRevLettSudarshan}. For this purpose, we first define the P function associated to the field produced by the indistinguishable scattering between the two independently-generated, horizontally polarized fields. These represent either photons or plasmons emerging through each of the slits.

\begin{equation}
     P_{\text{pl}}(\alpha)= \int   P_{1}(\alpha-\alpha^{'})P_{2}(\alpha^{'})d^{2}\alpha^{'}.
     \label{eq1-ch2}
    \end{equation}
    The P function for a thermal light field is given by $P_{i}(n) =(\pi\Bar{n}_{i})^{-1} e^{( \frac{-|\alpha|^{2}}{\Bar{n}_{i}})}$ . Here, $\alpha$ describes the complex amplitude as defined for coherent states $|\alpha\rangle $. The mean particle number of the two modes is represented by $ \Bar{n}_{1}=\eta\Bar{n}_{\text{s}}$ and  $ \Bar{n}_{2}=\eta\Bar{n}_{\text{pl}}$. Moreover, the mean photon number of the initial illuminating photons is represented by $ \Bar{n}_{\text{s}}$, whereas $ \Bar{n}_{\text{pl}}$ describes the mean photon number of scattered plasmonic fields. The parameter $\eta
    $ is defined as $ \cos^{2}{\theta}$. The polarization angle, $ \theta$, of the illuminating photons is defined with respect to the vertical axis. Note that the photonic modes in Eq. (\ref{eq1-ch2}) can be produced by independent multiphoton systems. Furthermore, we make use of the coherent state basis to represent the state of the combined field as $ \hat{\rho}_{\text{pl}}= \int   P_{\text{pl}}(\alpha)|\alpha\rangle\langle \alpha|d^{2}\alpha$ \cite{PhysRevLettSudarshan}. This expression enables us to obtain the probability distribution $  p_{\text{pl}}=  \langle n|\rho_{\text{pl}} |n\rangle$ for the scattered photons and plasmons with horizontal polarization. We can then write the combined number distribution for the multiparticle system at the detector as $ p_{\text{det}}(n) =\sum^{n}_{m=0} p_{\text{pl}}(n-m)p_{\text{ph}}(m)$. The distribution $ p_{\text{ph}}(m) $ accounts for the vertical polarization component of the illuminating multiphoton systems. Thus, we can describe the final photon-number distribution after the plasmonic structure as (see Appendix \ref{app:A:PFunction})
    \begin{equation}
      p_{\text{det}}(n)= \sum^{n}_{m=0} \frac{(\Bar{n}_{\text{pl}}+\eta\Bar{n}_{\text{s}})^{(n-m)}[(1-\eta)\Bar{n}_{s}]^{m}}{(\Bar{n}_{\text{pl}}+\eta\Bar{n}_{\text{s}}+1)^{(n-m+1)} [(1-\eta)\Bar{n}_{\text{s}}+1]^{m+1}}.
     \label{eq2-ch2}
    \end{equation}
   Note that the quantum statistical properties of the photons scattered from the sample are defined by the strength of the plasmonic near fields $ \Bar{n}_{\text{pl}} $. As illustrated in Fig. \ref{fig:figure1-ch2}a, the probability function in Eq. (\ref{eq2-ch2}) demonstrates the possibility of modifying the quantum statistics of photonic–plasmonic systems. It is worth remarking tha Eq. (\ref{eq2-ch2}) is valid only when the two sources, i.e., the two slits, are active and contribute to the combined field measured by the detector.

   \section{Experimental Results and Discussions}
\begin{figure}[!htb]
	\centering
	\includegraphics[width=0.95\textwidth]{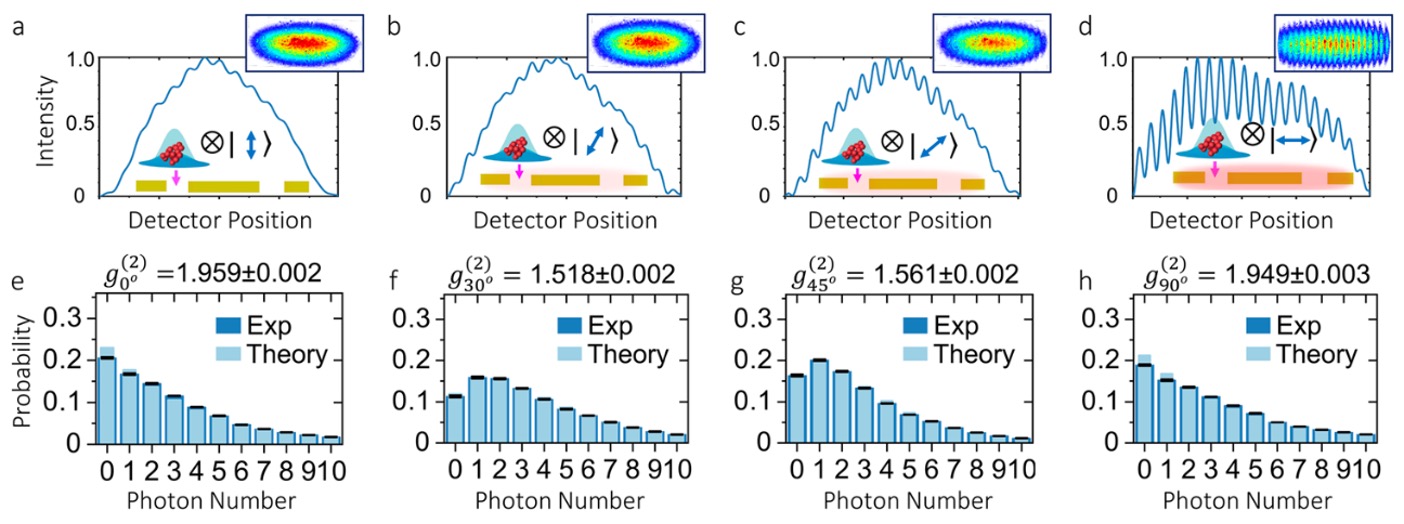}
	\mycaption{Experimental Observation of Plasmon-Induced Interference and the Modification of Quantum Statistics.}{ The formation of interference fringes in the far field of the plasmonic structure with two slits is shown in panels (a–d). Panel (a) shows the spatial distribution of a thermal multiphoton system transmitted by a single slit. In this case, the contribution from optical near fields is negligible and no photons are transmitted through the second slit. As shown in panel (b), a rotation of the photon’s polarization increases the presence of optical near fields in the plasmonic structure. In this case, photon-plasmon scattering processes induce small changes to the spatial distribution of the transmitted photons. Part (c) shows that the increasing excitation of plasmons is manifested through the increasing visibility of the interference structure. Panel (d), shows a clear modification of spatial coherence induced by optical near fields. Remarkably, the modification of spatial coherence induced by the presence of plasmons is also accompanied by the modification of the quantum statistical fluctuations of the field as indicated in panels (e–h). Each of these photon-number distributions corresponds to the spatial profiles above from (a) to (d). The photon-number distribution in (e) demonstrates that the photons transmitted by the single slit preserve their thermal statistics. Remarkably, multiparticle scattering induced by the presence of near fields modifies the photon-number distribution of the hybrid system as shown in (f) and (g). These probability distributions resemble those of coherent light sources. Interestingly, as demonstrated in (h), the photon-number distribution becomes thermal again when photons and plasmons are polarized along the same direction. The error bars represent the standard deviation of ten datasets. Each dataset consists of $ \sim  $400,000 photon-number-resolving measurements. These figures are taken from \cite{chenglongnature}.} 
	\label{fig:figure2-ch2}
\end{figure}

\begin{figure}[!ht]
	\centering
	\includegraphics[width=0.75\textwidth]{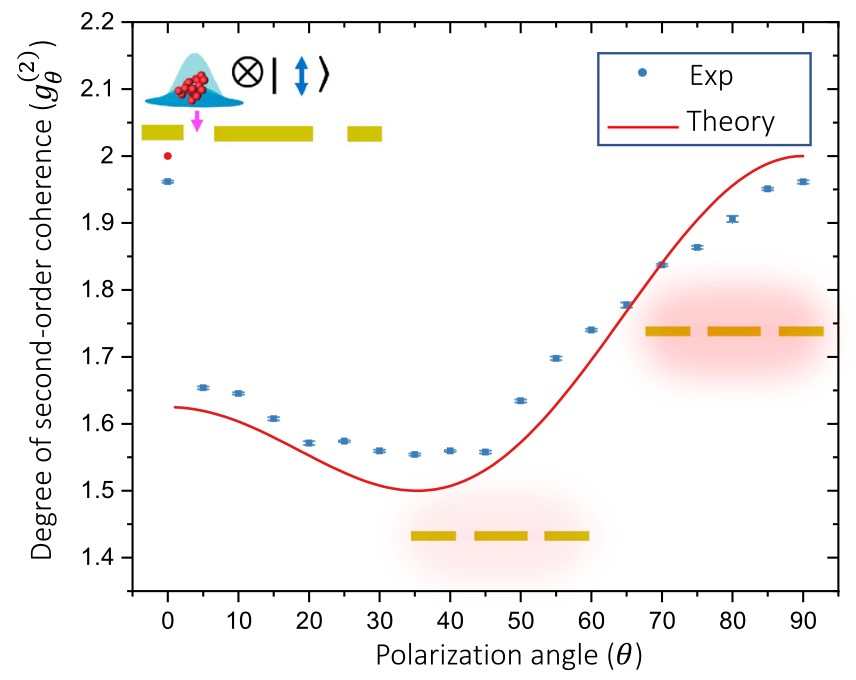}
	\mycaption{Modification of the Quantum Statistics of a Single Multiparticle System in a Plasmonic Structure.}{ The experimental data is plotted together with the theoretical prediction for the degree of the second-order correlation function. The theoretical model is based on the photon-number distribution described by Eq. (\ref{eq2-ch2}) for a situation in which $ \Bar{n}_{\text{s}} = 3 \Bar{n}_{\text{pl}}$. The error bars represent the standard deviation of ten realizations of the experiment. Each experiment consists of $\sim 100,000$ photon-number-resolving measurements. These figures are taken from \cite{chenglongnature}.} 
	\label{fig:figure3-ch2}
\end{figure}

   We first explore the modification of quantum statistics in a plasmonic double-slit structure illuminated by a thermal multiphoton system \cite{Arecchi,PhysRevLett.TurbulenceFree}. The experimental setup is depicted in Fig. \ref{fig:figure1-ch2}c. This allows us to focus thermal photons onto a single slit and measure the far-field spatial profile and the quantum statistics of the transmitted photons. As shown in Fig. \ref{fig:figure2-ch2}, we perform multiple measurements corresponding to different polarization angles of the illuminating photons. As expected, the spatial profile of the transmitted photons does not show interference fringes when the photons are transmitted by the single slit (see Fig. \ref{fig:figure2-ch2}a). However, the excitation of plasmonic fields increases the spatial coherence of the scattered photons. As indicated by Fig. \ref{fig:figure2-ch2}b–d, the increased spatial coherence leads to the formation of interference structures. This effect has been observed multiple times \cite{Daniel,Dongfang,Schouten,maganaloaizaexotic2016}. Nonetheless, it had been assumed independent of the quantum statistics of the hybrid photonic–plasmonic system \cite{Tame2013,You2020}. However, as demonstrated by the probability distributions from Fig. \ref{fig:figure2-ch2}e–h, the modification of spatial coherence is indeed accompanied by a modification of the quantum fluctuations of the plasmonic system. This effect had not been observed before as measurement devices used in previous experiments were insensitive to the multiparticle dynamics supported by this kind of plasmonic structures \cite{Fasel,ExtraordinaryLawrie,Safari,BenjaminVest,Dongfang,hofer2019hermite,Schouten,maganaloaizaexotic2016}. In our case, the modification of the quantum statistics, mediated by multiparticle scattering, is captured through a series of photon-number-resolving measurements \cite{you2020identification,HashemiRafsanjani:17}. In our experiment, the mean photon number of the photonic source  is three times the mean particle number of the plasmonic field $ \Bar{n}_{\text{pl}} $. The theoretical predictions for the photon-number distributions in Fig. \ref{fig:figure2-ch2} were obtained using Eq. (\ref{eq2-ch2}) for a situation in which $ \Bar{n}_{\text{s}} = 3 \Bar{n}_{\text{pl}}$.
 \noindent
The sub-thermal photon-number distribution shown in Fig. \ref{fig:figure2-ch2}f demonstrates that the strong confinement of plasmonic fields can induce anti-thermalization effects \cite{Kondakci2015}. Here, the scattering among photons and plasmons attenuates the chaotic fluctuations of the injected multiphoton system, characterized by a thermal photon-number distribution, as indicated by Fig. \ref{fig:figure2-ch2}e. Conversely, the transition in the photon-number distribution shown from Fig. \ref{fig:figure2-ch2}g–h is mediated by a thermalization effect. In this case, the individual phase jumps induced by photon-plasmon scattering increases the chaotic fluctuations of the multiparticle system \cite{Safari}, leading to the thermal state in Fig. \ref{fig:figure2-ch2}h. As shown in Fig. \ref{fig:figure3-ch2}, the quantum statistics of the photonic–plasmonic system show an important dependence on the strength of the optical near fields surrounding the plasmonic structure. The photon-number-distribution dependence on the polarization angle of the illuminating photons is quantified through the degree of second-order coherence $g^{(2)}$ in Fig. \ref{fig:figure3-ch2} \cite{you2020identification}. We use the measured photon statistics to evaluate $g^{(2)}$ defined as $ g^{2}(\tau)= 1+(\langle (\Delta \hat{n})^{2} \rangle - \langle  \hat{n}^{2} \rangle)/\langle \hat{n}^{2} \rangle$ \cite{Mandel1995}. This coherence function is independent of time for single-mode fields \cite{Mandel1979}. Furthermore, we point out that while Eq. (\ref{eq2-ch2})  depends on the brightness of the sources, the degree of second-order coherence $g^{(2)}$ only depends on the ratio of $ \Bar{n}_{\text{s}}$ and  $ \Bar{n}_{\text{pl}}$ (see Appendix \ref{app:A:Intensity}). In addition, we note that the data point at $\theta=0^{\circ}$ is obtained by directly calculating $g^{(2)}$ for a single-mode thermal source. In this case, a single polarized light beam cannot be treated as two independent sources, thus one cannot use Eq. (\ref{eq2-ch2}). Furthermore, we would like to point out that the identification of the transition point in Fig. \ref{fig:figure3-ch2}, specifically the transition from one-to-two source (mode) representation is an interesting but complicated task. Indeed, there have been technical studies that aim to develop tools to demonstrate full statistical mode reconstruction without a prior information \cite{Burenkov}. Nevertheless, the theoretical $g^{(2)}$ after the transition is calculated using Eq. (\ref{eq2-ch2}). The remarkable agreement between theory and experiment validates our observation of the modification of quantum statistics of plasmonic systems.

\begin{figure}[!htb]
	\includegraphics[width=0.95\textwidth]{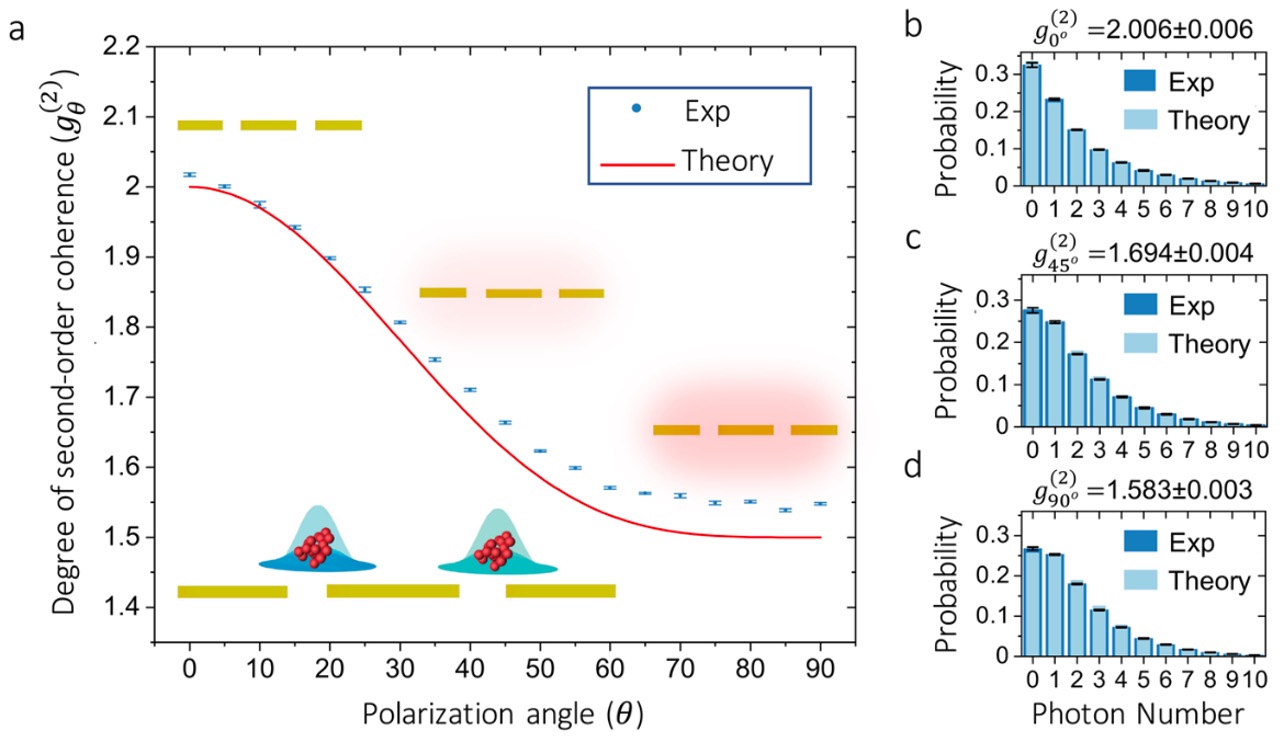}
	\mycaption{Modification of the Quantum Statistics of a Multiphoton System Comprising Two Sources.}{ The modification of the quantum statistics of a multimode plasmonic system composed of two independent multiphoton sources is shown in panel (a). Here, we plot experimental data together with our theoretical prediction for the degree of second-order coherence. The theoretical model is based on the photon-number distribution predicted by Eq. (\ref{eq2-ch2}) for two independent multiphoton systems with thermal statistics and same mean photon numbers satisfying . As demonstrated in (b), the photon-number distribution is thermal for the scattered multiphoton system in the absence of near fields ($\theta=0^{\circ}$). However, an anti-thermalization effect takes place as the strength of the plasmonic near fields is increased ($\theta=45^{\circ}$), this is indicated by the probability distribution in (c). Remarkably, as reported in (d), the degree of second-order coherence of the hybrid photonic–plasmonic system is 1.5 when plasmonic near fields are strongly confined ($\theta=90^{\circ}$). These results unveil the possibility of using plasmonic near fields to manipulate coherence and the quantum statistics of multiparticle systems. The error bars represent the standard deviation of ten realizations of the experiment. Each experiment consists of $\sim100,000$ photon-number-resolving measurements. These figures are taken from \cite{chenglongnature}.} 
	\label{fig:figure4-ch2}
\end{figure}

 \noindent
The multiparticle near-field dynamics observed for a thermal system can also induce interactions among independent multiphoton systems. We demonstrate this possibility by illuminating the plasmonic structure with two independent thermal sources. The polarization of one source is fixed to 0 whereas the polarization of the other is rotated by the angle $\theta$. This is illustrated in the central inset of Fig. \ref{fig:figure1-ch2}c. In this case, both multiphoton sources are prepared to have same mean photon numbers. In Fig. \ref{fig:figure4-ch2}, we report the modification of the quantum statistics of a multiphoton system comprising two modes that correspond to two independent systems. As shown in Fig. \ref{fig:figure4-ch2}a, the confinement of electromagnetic near fields in our plasmonic structure modifies the value of the second-order correlation function. In this case, the shape of the second-order correlation function is defined by the symmetric contributions from the two thermal multiphoton systems. As expected, the quantum statistics of the initial thermal system with two sources remain thermal (see Fig. \ref{fig:figure4-ch2}b). However, as shown in Fig. \ref{fig:figure4-ch2}b–d, this becomes sub-thermal as the strength of the plasmonic near fields increase. The additional scattering paths induced by the presence of plasmonic near fields modify the photon-number distribution as demonstrated in Fig. \ref{fig:figure4-ch2}c \cite{maganaloaizaexotic2016}. The strongest confinement of plasmonic near fields is achieved when the polarization of one of the sources is horizontal ($\theta=90^{\circ}$). As predicted by Eq. (\ref{eq2-ch2}) and reported in Fig. \ref{fig:figure4-ch2}d, the second-order coherence function for this particular case is $g^{(2)}=1.5$.
The light-matter interactions explored in this experiment demonstrate the possibility of using either coherent or incoherent bosonic scattering to modify the quantum statistical fluctuations of multiparticle systems. These mechanisms are fundamentally different to the coherent interactions induced by the indistinguishability of two bosons in Hong–Ou–Mandel interference \cite{Vest,Cai}. Indeed, the efficient preparation of indistinguishable single surface plasmons has enabled different forms of plasmonic quantum interference \cite{Kolesov2009,Vest,Cai,Dheur}. In addition, the flexibility of plasmonic platforms has led to the preparation of coherent plasmonic states with tunable probability amplitudes \cite{Kolesov2009,Dheur}. Nevertheless, the probability distribution in Eq. (\ref{eq2-ch2}) describes coherent effects, that produce interference, as well as incoherent bosonic scattering. Consequently, the plasmonic control of the quantum statistical fluctuations of a multiparticle system is achieved through both distinguishable and indistinguishable processes.
 \noindent
These results have important implications for multimode plasmonic systems \cite{Tame2013,You2020}. Recently, there has been interest in exploiting transduction of the quantum statistical fluctuations of multimode fields for applications in imaging and quantum plasmonic networks \cite{ExtraordinaryLawrie,Holtfrerich2016}. Nevertheless, the modification of quantum statistics in plasmonic systems had remained unexplored \cite{Tame2013}. Interestingly, the possibility of modifying quantum statistics and correlations through multiparticle interactions has been demonstrated in nonlinear optical systems, photonic lattices, and Bose–Einstein condensates \cite{Kondakci2015,Bromberg2010,Folling2005,Schellekens2005}. Moreover, similar quantum dynamics have been explored for electrons in solid-state devices and cold Fermi gases \cite{Folling2005,Schellekens2005,Scatteringtheory,Martin,Henny1999, Kiesel2002, Johnson}. However, our work unveils the potential of optical near fields as an additional degree of freedom to manipulate multiparticle quantum systems. This mechanism offers alternatives for the implementation of quantum control in plasmonic platforms. More specifically, our work shows that plasmonic near fields offer deterministic paths for tailoring photon statistics. In this case, the strength of plasmonic fields is deterministically controlled through polarization. Furthermore, plasmonic platforms offer practical methods for exploring controlled thermalization (and anti-thermalization) of arbitrary light fields. The modification of amplitude and phase of multiphoton systems has been explored in nonlinear optical systems \cite{Bromberg2010} and photonic lattices with statistical disorder \cite{Kondakci2015}.\\
 \noindent
The experiments performed in plasmonic platforms in which interactions are restrained to single-particle scattering led researchers to observe the preservation of the quantum statistical properties of bosons \cite{Altewischer,Akimov,Tame2013,You2020,Martino,Fasel,Huck,Daniel,ExtraordinaryLawrie,10,Safari}. These experiments triggered the idea of the conservation of quantum statistics in plasmonic systems \cite{Fasel,Huck,Daniel,10,Safari,Dheur,Dongfang,Lee2016,Dowran2018,Holtfrerich2016}. In this article, we report the observation of the modification of the quantum statistics of multiparticle systems in plasmonic platforms. We validate our experimental observations through the theory of quantum coherence and demonstrate that multiparticle scattering mediated by dissipative near fields enables the manipulation of the excitation mode of plasmonic systems. Our findings unveil mechanisms to manipulate multiphoton quantum dynamics in plasmonic platforms. These possibilities have important implications for the fields of quantum photonics, quantum many-body systems, and quantum information science \cite{Tame2013,DellAnno2006,Vest}.
\section{Sample Design and Experiment}
\label{Sample Design and Experiment}
\begin{figure*}
	\centering
	\includegraphics[width=0.9\textwidth]{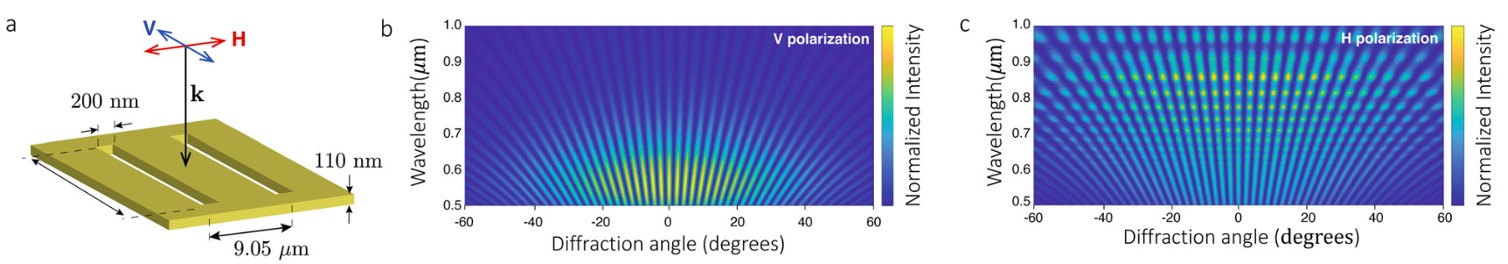}
	\mycaption{ Design of plasmonic nanostructures.}{ The design of our plasmonic sample is shown in (a). The wavelength-dependent far-field interference pattern as a function of the diffraction angle is shown in (b). In this case, the structure is illuminated with vertically-polarized photons and no plasmonic near fields are excited. The figure in (c) shows a modified interference structure due to the presence of plasmonic near fields. In this case, the illuminated photons are polarized along the horizontal direction. These figures are taken from \cite{chenglongnature}.} 
	\label{fig:fig5-ch2}
\end{figure*}

Full-wave electromagnetic simulations were conducted using a Maxwell’s equation solver based on the finite difference time domain method (Lumerical FDTD). The dispersion of the materials composing the structure was taken into account by using their frequency-dependent permittivities. The permittivity of the gold film was obtained from ref. \cite{Johnson}, the permittivity of the glass substrate (BK7) was taken from the manufacturer’s specifications, and the permittivity of the index matching fluid was obtained by extrapolation from the manufacturer’s specification.
 \noindent
As shown in Fig. \ref{fig:fig5-ch2}, our nanostructure shows multiple plasmonic resonances at different wavelengths. This enables the observation of the multiphoton effects studied in this article at multiple wavelengths.
 \noindent
As shown in Fig. \ref{fig:fig5-ch2}c, we utilized a continuous-wave (CW) laser operating at a wavelength of $780 $ nm. We generated two independent sources with thermal statistics by dividing a beam with a 50:50 beam splitter. Then, we focused the beams onto two different locations of a rotating ground-glass\cite{Arecchi}. The two beams were then coupled into single-mode fibers to extract a single transverse mode with thermal statistics (see Appendix \ref{app:A:Characterization}). In addition, we certified the thermal statistics of the photons emerging through each of the slits (see Appendix \ref{app:A:PHOTONSTATISTICSSlit}). We attenuated the two beams with neutral-density (ND) filters to tune their mean photon numbers. The polarization state of the two thermal beams was controlled by a pair of polarizers and half-wave plates. The two prepared beams were then combined using a 50:50 beam splitter. The combined beam was weakly focused to a $450 $ nm spot onto the plasmonic structure that was mounted on a motorized three-axis translation stage. This enabled us to displace the sample in small increments. Once the imaging conditions were fixed, we did not modify the position of the plasmonic sample. Furthermore, we used two infinity-corrected oil-immersion microscope objectives ($\text{NA}=1.4$, magnification of ×60 and working distance of $130 $ mm) to focus and collect light to and from the plasmonic structure. To observe the interference fringes in the far field, we built an imaging system to form the Fourier plane at $\approx40 $ cm from the plasmonic sample. Then, we characterized our plasmonic sample using horizontally polarized light. We note that the input photons can be coupled to plasmonic modes at the second slit or transmitted through the first slit. The transmission coefficient is given by the normalized transmitted intensity $I_{1}$ for the first slit. This was experimentally estimated as $I_{1}=0.608$. Similarly, the photon–plasmon conversion efficiency is given by the normalized transmitted intensity $I_{2}$ from the second slit. This coefficient was experimentally estimated as $I_{2}=0.028$. The light collected by the objective was then filtered using a 4f-imaging system to achieve specific particle number conditions for the photonic $n_{\text{s}}$ and plasmonic $n_{\text{pl}}$ modes. The experiment was formalized by coupling light, using a microscope objective ($\text{NA}=0.25$, magnification of ×10 and working distance of $5.6 $ mm), into a polarization-maintaining (PM) fiber. This fiber directs photons to a superconducting nanowire single-photon detector (SNSPD) that performs photon number resolving detection \cite{you2020identification,HashemiRafsanjani:17}.

%% file: chapter3.tex
\chapter{Multiphoton Quantum Sensing}

\section{Motivation}
The possibility of using weak optical signals to perform sensing of delicate samples constitutes one of the main goals of quantum photonic sensing. Furthermore, the nanoscale confinement of electromagnetic near fields in photonic platforms through surface plasmon polaritons has motivated the development of highly sensitive quantum plasmonic sensors \cite{NanoToday, Hwang, Maier, Lee1}. Despite the enormous potential of plasmonic platforms for sensing, this class of sensors is ultimately limited by the quantum statistical fluctuations of surface plasmons. Indeed, the fluctuations of the electromagnetic field severely limit the performance of quantum plasmonic sensing platforms in which delicate samples are characterized using weak near-field signals. Furthermore, the inherent losses associated with plasmonic fields levy additional constraints that challenge the realization of sensitivities beyond the shot-noise limit. Here, we introduce a protocol for quantum plasmonic sensing based on the conditional detection of plasmons. We demonstrate that the conditional detection of plasmonic fields, via plasmon subtraction, provides a new degree of freedom to control quantum fluctuations of plasmonic fields. This mechanism enables improvement of the signal-to-noise ratio of photonic sensors relying on plasmonic signals that are comparable to their associated field fluctuations. Consequently, the possibility of using weak plasmonic signals to sense delicate samples, while preserving the sample properties, has important implications for molecule sensing, and chemical detection.

\section{Background}
The possibility of controlling the confinement of plasmonic near-fields at the subwavelength scale has motivated the development of a variety of extremely sensitive nanosensors \cite{NanoToday, Hwang, Maier, Lee1}. Remarkably, this class of sensors offers unique resolution and sensitivity properties that cannot be achieved through conventional photonic platforms in free space \cite{ChenglongAIP,Slussarenko2017,Lee1, Polino}. In recent decades, the fabrication of metallic nanostructures has enabled the engineering of surface plasmon resonances to implement ultrasensitive optical transducers for detection of various substances ranging from gases to biochemical species \cite{NanoToday,Hwang,Lee1}. Additionally, the identification of the quantum mechanical properties of plasmonic near-fields has prompted research devoted to exploring mechanisms that boost the sensitivity of plasmonic sensors \cite {Altewischer, Akimov, Tame2013, chenglongreview, chenglongnature}. 
The scattering paths provided by plasmonic near-fields have enabled robust control of quantum dynamics \cite{BenjaminVest,Schouten,maganaloaizaexotic2016, chenglongnature}. Indeed, the additional degree of freedom provided by plasmonic fields has been used to harness the quantum correlations and quantum coherence of photonic systems \cite{Alexander,chenglongnature,Dongfang,Schouten}. Similarly, this exquisite degree of control made possible the preparation of plasmonic systems in entangled and squeezed states \cite{Vest,Huck,Fasel,Martino}. Among the large variety of quantum states that can be engineered in plasmonic platforms \cite{Tame2013,chenglongreview}, entangled systems in the form of N00N states or in diverse forms of squeezed states have been  used to develop quantum sensors \cite{Heeres2013,Chen:18,Lee1,Dowran2018,Lee2016}.
In principle, the sensitivity of these sensors is not constrained by the quantum fluctuations of the electromagnetic field that establish the shot-noise limit \cite{Lee,Polino}. However, due to inherent losses of plasmonic platforms, it is challenging to achieve sensitivities beyond the shot-noise limit under realistic conditions \cite{ChenglongAIP}. Despite existing obstacles, recent work demonstrates the potential of exploiting nonclassical properties of plasmons to develop quantum plasmonic sensors for detection of antibody complexes, single molecules, and to perform spectroscopy of biochemical substances \cite{Salazar,Kongsuwan,alashnikov,Mauranyapin2017}. 

\section{Conceptual Overview}
We explore a new scheme for quantum sensing based on plasmon-subtracted thermal states \cite{Dakna, Loaiza2019, HashemiRafsanjani:17}. Our work offers an alternative to quantum sensing protocols relying on entangled or squeezed plasmonic systems \cite{Vest,Huck,Fasel,Martino,Heeres2013,Chen:18,Lee1,Dowran2018,Lee2016}. We use a sensing architecture based on a nanoslit plasmonic interferometer \cite{biosensordeleon}. It provides a direct relationship between the light exiting the interferometer and the phase shift induced in one of its arms by the substance to be sensed (analyte). We introduce a conditional quantum measurement on the interfering plasmonic fields via the subtraction of plasmons. We show that this process enables the reduction of quantum fluctuations of the sensing field and increases the mean occupation number of the plasmonic sensing platform \cite{Loaiza2019, HashemiRafsanjani:17}. Furthermore, plasmon subtraction provides a method for manipulating the signal- to-noise ratio (SNR) associated with the measurement of phase shifts. We demonstrate that the reduced fluctuations of plasmonic fields leads to an enhancement in the estimation of a phase shift. The performance of our protocol is quantified through the uncertainty associated to phase measurements. We point out that the reduced uncertainties in the measurement of phases leads to better sensitivities of our sensing protocol. This study is conducted through a quantum mechanical model that considers the realistic losses that characterize a plasmonic nanoslit sensor. We report the probabilities of successfully implementing our protocol given the occupation number of the plasmonic sensing fields and the losses of the nanostructure. Our analysis suggests that our protocol offers practical benefits for lossy plasmonic sensors relying on  weak near-field signals \cite{Maga_a_Loaiza_2016}.  Consequently, our platform can have important implications for plasmonic sensing of delicate samples such as molecules, chemical substances or, in general, photosensitve materials \cite{Salazar,Kongsuwan,alashnikov,Mauranyapin2017}.
\section{Theory }
\begin{figure*}
	\centering
	\includegraphics[width=0.95\textwidth]{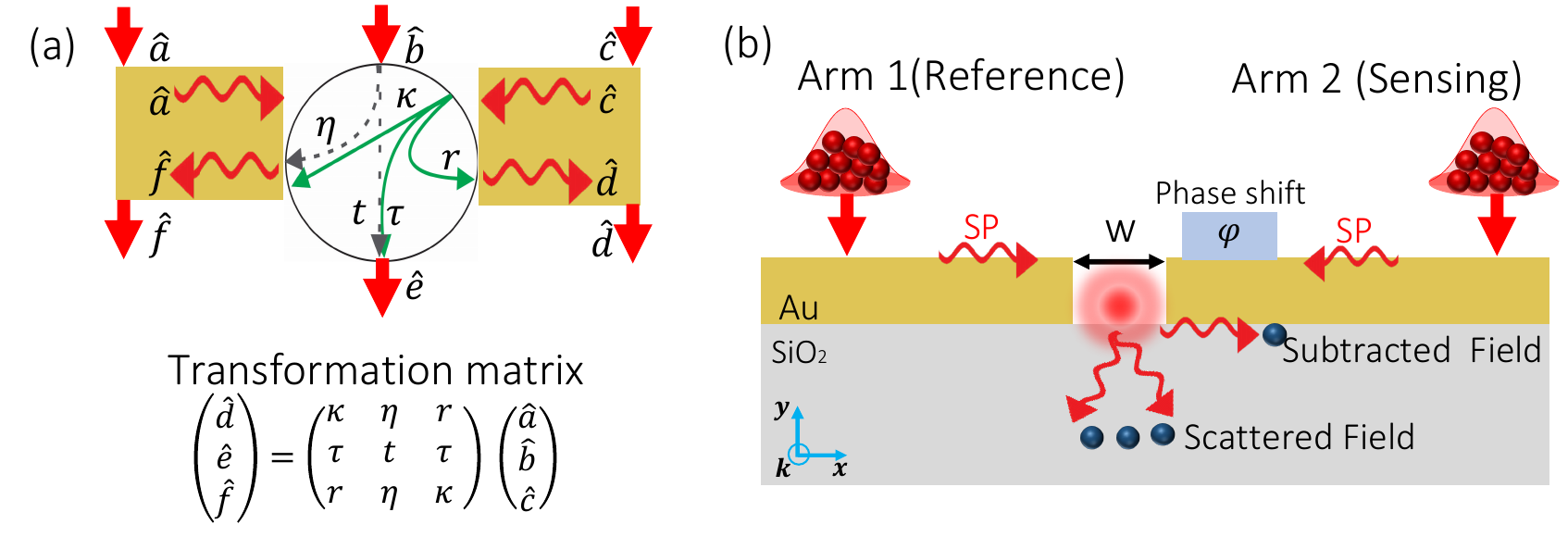}
	\mycaption{Plasmonic Nanoslits.}{ (a) Schematic diagram of the interactions in a plasmonic nanoslit. The plasmonic nanostructure has three input and three output ports. The photonic mode at the input is described by the operator $\hat{b}$, whereas the two plasmonic modes are represented by $\hat{a}$ and $\hat{c}$. These modes are coupled to the plasmonic modes $\hat{d}$ and $\hat{f}$, and to the photonic mode $\hat{e}$ at the output of the nanostructure. As described by the transformation matrix, the parameters $\kappa$, $\tau$, $r$, $\eta$, and $t$ represent the coupling coefficients among the ports of the nanostructure. For sake of clarity, the diagram only illustrates the coupling paths for the input modes $\hat{b}$ and $\hat{c}$. The diagram in (b) shows the design of our simulated plasmonic sensor, comprising a slit of width w in a $200$ nm gold thin film. Here, the plasmonic structure is illuminated by two thermal multiphoton sources that excite two plasmonic fields with super-Poissonian statistics (the input grating couplers are not shown in the figure). The two counter-propagating surface plasmon (SP) modes  interfere at the interface between the gold layer and the \ce{SiO_2} substrate. The interference conditions are defined by the phase shift $\varphi$ induced in one of the plasmonic modes by the substance that we aim to sense. These figures are taken from \cite{Mostafavi2022}.} 
	\label{fig:figure1}
\end{figure*}

We first discuss the theoretical model that we use to describe conditional quantum measurements applied to a thermal plasmonic system. Fig. \ref{fig:figure1}a describes the interactions supported by the plasmonic nanoslit under consideration \cite{Safari}. This nanostructure acts as a 
plasmonic tritter by coupling the photonic mode $\hat{b}$ and the two plasmonic modes, described by the operators $\hat{a}$ and $\hat{c}$, to three output modes \cite{Safari}. The photonic mode at the output of the nanoslit is described by $\hat{e}$, whereas the two plasmonic output modes are represented by the operators $\hat{f}$ and $\hat{d}$. As indicated in Fig. \ref{fig:figure1}b, and throughout this paper, we study the conditional detection of the output modes $\hat{d}$ and $\hat{e}$ for a situation in which only the input plasmonic modes of $\hat{a}$ and $\hat{c}$ are excited in the nanostructure. Thus, the photonic mode $\hat{b}$ is assumed to be in a vacuum state. In this case, the plasmonic tritter can be simplified to a two-port device described by the following $2\times 2$ matrix

\begin{equation}
\left(
\begin{array}{c}
\hat{d}\\
\hat{e}\\

\end{array}
\right) =  
\left(
\begin{array}{cc}
\kappa& r \\
\tau&\tau
\end{array}
\right )
\\
\left(
\begin{array}{c}
\hat{a}\\
\hat{c}
\end{array}
\right ).
\label{eq1}
\end{equation}
The photonic mode $\hat{e}$ is transmitted through the slit and its transmission probability is described by $2\lvert {\tau}\lvert ^2=T_\text{ph}$. Here, $T_\text{ph}$ represents the normalized intensity of the transmitted photons. Moreover, the plasmon-to-plasmon coupling at the output of the nanostructure is given by $\lvert{\kappa}\lvert^2+\lvert{r}\lvert^2=T_\text{pl}$. Here, the renormalized transmission (after intereference and considering loss) for the plasmonic fields is described by $T_\text{pl}$. From Fig. \ref{fig:figure1}b, we note that the interference supported by the plasmonic nanoslit shares similarities with those induced by a conventional Mach-Zehnder interferometer (MZI). More specifically, the two plasmonic modes, $\hat{a}$ and $\hat{c}$, interfere at the location of the nanoslit, which in turn scatters the field to generate the output \cite{biosensordeleon}. The interference conditions are defined by the phase shift induced by the analyte. Plasmonic sensors with nanoslits have been extensively investigated in the classical domain, showing the possibility of ultrasensitive detection using minute amounts of analyte \cite{biosensordeleon,NanoToday, Hwang, Maier, Lee1,Chen:18}.

\noindent 
We now consider a situation in which a single-mode thermal light source  is coupled to the nanostructure in Fig. \ref{fig:figure1}b  exciting two counter-propagating surface plasmon modes. This can be achieved by
using a pair of grating couplers (not shown in the figure) \cite{biosensordeleon}. The statistical properties of this thermal field can be described by the Bose-Einstein statistics as 
$\hat{\rho}_{\text{th}}= \sum_{n=0}^{\infty} \text{p}_{\text{pl}}(n) {\ket n} {\bra n}$, where $\text{p}_\text{pl}(n)=\bar n^{n}/(1+\bar n)^{1+n}$, and $\bar{n}$ represents the mean occupation number of the field. Interestingly, the super-Poissonian statistics of 
thermal light can be modified through conditional measurements \cite{Loaiza2019,Dakna,HashemiRafsanjani:17}.
As discussed below, it is also possible to modify the quantum statistics of plasmonic fields. The control of plasmonic statistics can be implemented by subtracting/adding bosons from/to thermal plasmonic systems \cite{Mizrahi,Parigi}. In this work, we subtract plasmons from the transmitted field formed by the superposition of the surface plasmon modes propagating through the reference
and sensing arms of the interferometer. This transmitted mode $\hat{e}$ is then conditioned to the output of the field $\hat{d}$. As such, the number of subtracted particles $L$ is experimentally controlled by the strength of the plasmonic near fields surrounding our sensor, and the conditional counting of particles in mode $\hat{d}$. The strength of the near field is defined by the design of the plasmonic nanostructure, whereas the conditional counting is experimentally implemented by measuring simultaneous detection events between modes $\hat{d}$ and $\hat{e}$.
It is worth mentioning that conditional measurements in photonic systems have been experimentally demonstrated in Refs. \cite{HashemiRafsanjani:17,Loaiza2019}. The successful subtraction of plasmons boosts the signal of the sensing platforms. This feature is particularly important for sensing schemes relying on dissipative plasmonic platforms. The conditional subtraction of $L$ plasmons from the mode $\hat{d}$ leads to the modification of the quantum statistics of the plasmonic system, this can be described by
\begin{equation}
\text{p}_{\text{pl}}(n) =  \frac{(n+L)!\bar n_{\text{pl}}^{n}}{ n! L!(1+\bar n_{\text{pl}})^{L+1+n}}, 
\label{eq3prime}
\end{equation} 
where $\bar{n}_\text{pl}$ represents the mean occupation number of the quasi-particles that constitute the scattered field in mode $\hat{e}$  (see Appendix \ref{app:A:SNR}). We quantify the modification of the quantum statistics through the degree of second-order correlation function $g^{(2)}(0)$ for the mode $\hat{e}$ as (see Appendix \ref{app:A:SNR})
\begin{equation}
\begin{split} 
g^{(2)}_L (0) = {\frac{L+2}{L+1}}.
\end{split} 
\label{eqg2}
\end{equation}
We note that the conditional subtraction of plasmons induces anti-thermalization effects that attenuate the fluctuations of the plasmonic thermal system used for sensing. Indeed, the $g^{(2)}_L (0)$ approaches one with the increased number of subtracted plasmons, namely large values of $L$. This effect produces bosonic distributions resembling those of coherent states \cite{Loaiza2019}. Recently, similar anti-thermalization effects have been explored in photonic lattices \cite{Kondakci2015}. 
The aforementioned plasmon subtraction can be implemented in the plasmonic nanoslit interferometer  shown in Fig. \ref{fig:figure1}b. It consists of a $200$ nm thick gold film deposited on a glass substrate \cite{biosensordeleon}. This thickness is large enough to enable decoupled plasmonic modes on the top and bottom surfaces of the film, as required.  The gold film features a $320$ nm slit, defining the
reference arm of the interferometer to its left and the sensing arm (holding the analyte) to its right. The analyte then induces a phase difference $\varphi$  relative to the reference arm, thereby creating the output ($\hat{d}, \hat{e}$ and $\hat{f}$) that depends on this parameter.
\noindent 
We note that we aim to detect two plasmonic modes $\hat{d}$ and $\hat{e}$ in the far-field plane of the sample. This detection can be achieved using similar setups to those reported in \cite{BenjaminVest, Zhu2016, Marie}. Consequently, the experimental realization of this protocol includes two output ports in the form of grating couplers placed on the bottom surface of the gold film. These gratings are used to out-couple the $\hat{d}$ and $\hat{e}$ plasmonic modes into photons propagating in the substrate in the negative y-direction. Then, an optical system is used to collect the out-coupled photons from each port and direct them into two single-photon detectors. The photon-number-resolving detection of the scattered field can be performed through transition-edge sensors \cite{Gerrits2016} or through the use of surjective photon counting techniques \cite{HashemiRafsanjani:17,you2020identification}. This scheme has been extensively used in previous quantum plasmonic experiments such as those listed in Refs. \cite{BenjaminVest, Zhu2016, Marie}.  
\section{Simulation }
To verify the feasibility of our conditional measurement approach, we perform a finite-difference time-domain (FDTD) simulation (see  Appendix \ref{app:A:FDTD}) of the plasmonic nanoslit using a wavelength of $\lambda=810$ nm for the two counter-propagating surface plasmon modes ($\hat{a}$ and $\hat{c}$). The nanoslit is designed to support two localized surface plasmon (LSP) modes, one with dipolar symmetry and other with quadrupolar symmetry. Depending on the phase difference $\varphi$, these two LSP modes can be excited with different strengths by the fields interfering at the nanoslit being the dipolar (quadrupolar) mode optimally excited with $\varphi = 0$ ($\varphi= \pi$). This is due to the fact that the near-field symmetries of the interfering field are well-matched to the dipolar and quadrupolar fields for those values of $\varphi$ \cite{biosensordeleon}. 
\begin{figure*}[!ht]
	\centering \includegraphics[width=1\textwidth]{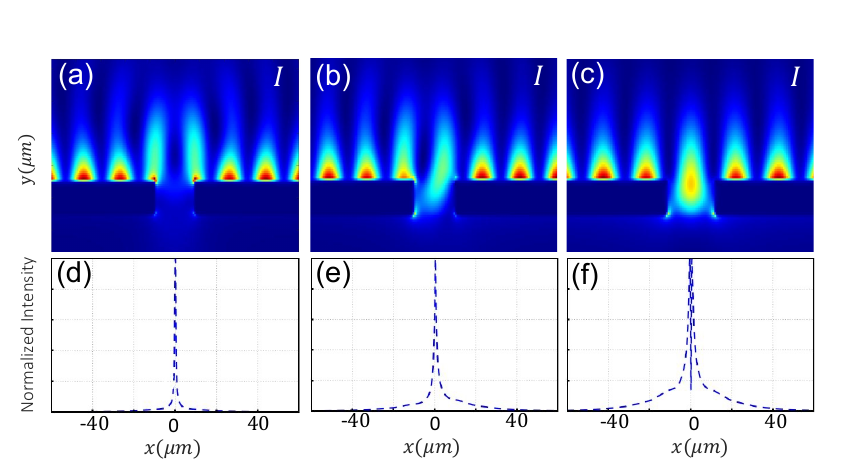}
	\mycaption{Normalized Near-Field Interference Pattern.}{ Panels (a-c) illustrate the intensity $I$ of the near-field distribution of our designed nanoslit in the x-y plane. The blue dashed line in panels (d-f) show the normalized near-field interference pattern produced by the field transmitted by a $320$ nm wide slit. This corresponds to the scattered field described by mode $\hat{e}$. The plots are obtained for $\varphi=0$, $\varphi = \pi/2$, and $\varphi = \pi$, respectively. These figures are taken from \cite{Mostafavi2022}.} 
	\label{fig:figurenearfield}
\end{figure*}
Fig. \ref{fig:figurenearfield}, panels (a) to (c), represent the near-field intensity distribution $I$ of our designed nanoslit in the x-y plane. Panels (d) to (f) show the normalized near-field interference pattern produced by the field transmitted by a $320$ nm wide slit. These near-field interference patterns are associated with the excited LSP mode corresponding to $\varphi=0$, $\varphi=\pi/2$, and $\varphi = \pi$. In this case, (d) and (f) illustrate the dipolar and quadrupolar modes of the nanostructure.  
The dashed lines in Fig. \ref{fig:figurestdv2} indicate the far-field angular distributions of the transmitted intensity associated with the dipolar LSP mode (panels a to d) and the quadrupolar LSP mode (panels i to l). Only a small angular range of the far-field distribution (range within vertical lines in Fig. \ref{fig:figurestdv2}) is used as the sensing signal. Thus, the sensing signal varies monotonically from a maximum value at $\varphi = 0$  to to a minimum value at $\varphi = \pi$ \cite{biosensordeleon}.
\begin{figure*}[!ht]
	\centering \includegraphics[width=1\textwidth]{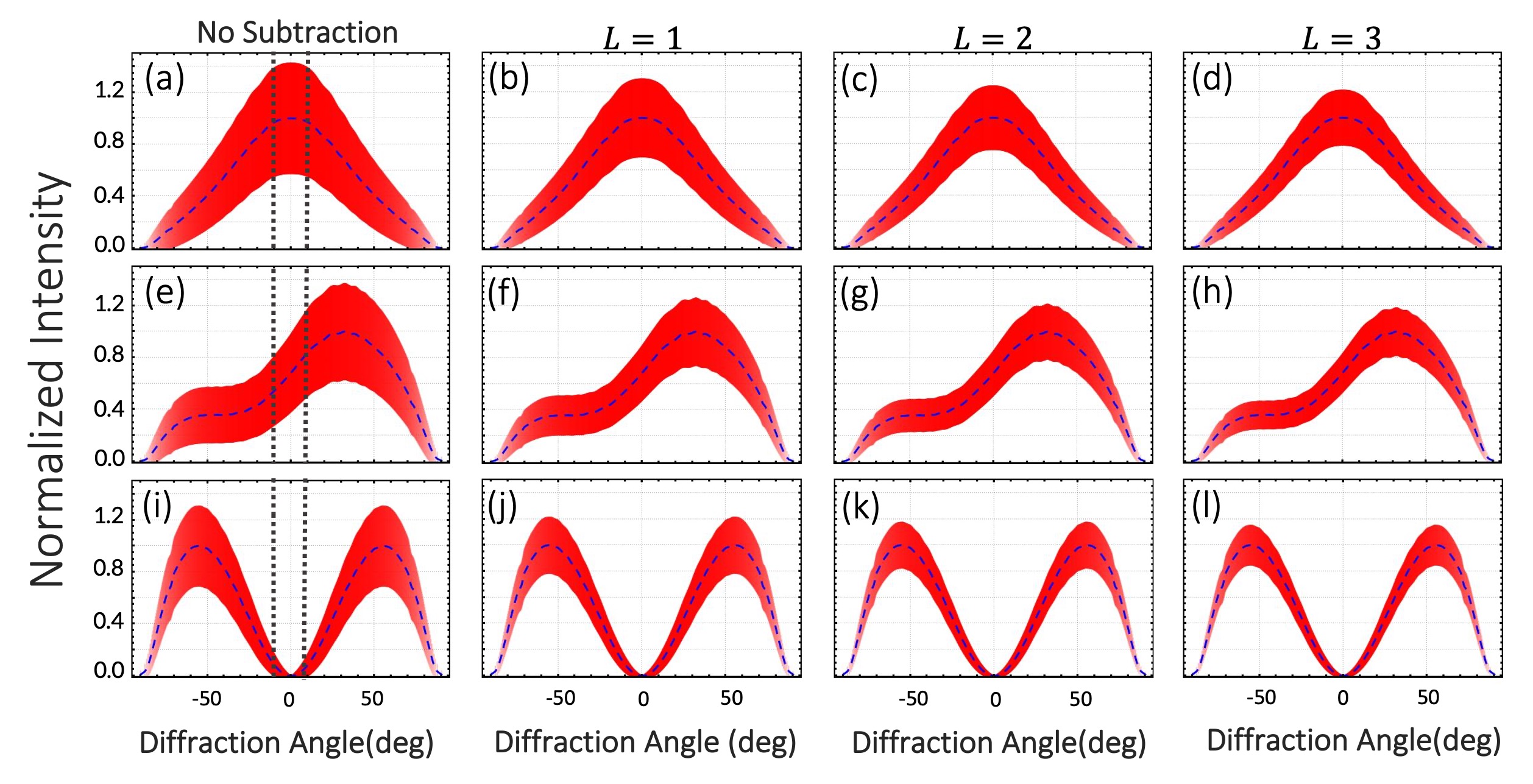}
	\mycaption{Normalized Far-Field Intensity Distribution Scattered by the Plasmonic Nanoslit.}{ Normalized far-field intensity distribution scattered by the plasmonic nanoslit. The blue dashed line indicates the interference pattern produced by the field transmitted through the $320$ nm wide slit, this corresponds to mode $\hat{e}$. The panels from (a) to (d) are obtained for $\varphi=0$, whereas those from (e) to (h) and (i) to (l) are calculated for $\varphi = \pi/2$ and $\varphi = \pi$ respectively. The dashed line in all plots represents the intensity distribution of the fields transmitted through the slit indicative of dipolar and quadrupolar near-field symmetry for $\varphi=0$ and $\varphi = \pi$. The red shaded regions correspond to the standard deviation for $\bar{n} = 3.75$. Panels (a),(e) and (i) depict the unconditional detection of the signal with its associated noise.  As displayed in panels (b) to (d), (f) to (h) and (i) to (l), the signal-to-noise ratio of the plasmonic sensor improves as the fluctuations of the field are reduced through the conditional detection of plasmons. The vertical lines on panels (a), (e) and (i) represent the angular range used for the calculation of the intensity variation with phase (i.e. sensitivity depicted in Fig. \ref{fig:SNRvag2}b). These figures are taken from \cite{Mostafavi2022}.} 
	\label{fig:figurestdv2}
\end{figure*}
The transmission parameters of our sensor are estimated from FDTD simulations. Specifically, the transmission values for the photonic and plasmonic modes are $T_{\text{ph}}\approx 0.076$ 
and $T_{\text{pl}}\approx0.0176$ 
for $\varphi= \pi$. However, our subtraction scheme is general and valid for any phase angle  $\varphi$  in the range of $0 \leq \varphi \leq   2 \pi$. Moreover, the total amount of power coupled to modes $\hat{e}$ and $\hat{d}$ normalized to the input power of the plasmonic structure is defined as $\gamma=T_{\text{ph}}+T_{\text{pl}}\approx0.0941$. 
For the results shown in Fig. \ref{fig:figurestdv2}, we assume a mean occupation number of $\bar n=3.75$ for the input beam. As shown in panels (a), (e) and (i) of Fig. \ref{fig:figurestdv2}, the output signals, calculated from Eq. \eqref{eq3prime} and represented by the red shaded region across all panels, exhibit strong quantum fluctuations. Surprisingly, after performing plasmon subtraction, the quantum fluctuations decrease, as indicated in the panels (b)-(d), (f)-(h) and (j)-(l) of Fig. \ref{fig:figurestdv2}. Evidently, this confirms that our conditional measurement protocol can indeed boost the output signal and consequently improve the sensing performance of a plasmonic device. However, due to the probabilistic nature of our protocol and the presence of losses, it is important to estimate the probability rates of successfully performing plasmon subtraction. Furthermore, we note that the structural parameters of the nanostructure will modify the scattering conditions among photons and plasmons. Consequently, the change of the structural parameters of our plasmonic nanoslit will also modify the probability of plasmon subtraction. In Table \ref{tab:table1} we list the degree of second-order correlation $ g^{(2)}_L (0)$, and the probability of successfully subtracting one, two, and three plasmons for different occupation numbers of the plasmonic fields used for sensing.
\begin{table} [!ht]
		\centering
	\mycaption{The Estimated Probability of Plasmon Subtraction.}{ The estimated probability of plasmon subtraction and the corresponding degree of second-order coherence $ g^{(2)}_L (0)$. The losses of the plasmonic nanostructure reduce the probability of subtracting multiple plasmons $L$ from the scattered field with an occupation number of $\bar{n}$. In this case, we assume $\varphi=\pi$.\\}
	\begin{tabular}{p{1.5cm}p{2cm}p{2cm}p{2cm}}
 \toprule
		$\bar{n}$& $ L=1$ &$ L=2 $ &$ L=3$\\
		\hline
		$2$&$1.0 \times10^{-2}$ &$1.0 \times10^{-4}$ &$1.1 \times10^{-6}$\\
		$1$&$5.2 \times10^{-3}$ &$2.7\times10^{-5}$ &$1.4 \times10^{-7}$\\
		$0.5$&$2.6 \times10^{-3}$&$7.0 \times10^{-6}$&$1.8 \times10^{-8}$\\
		$0.3$&$1.5\times10^{-3}$&$2.5\times10^{-6} $&$4.0 \times10^{-9} $\\
		\hline
		$ g^{(2)}_L (0)$&$1.5$&$1.33$&$1.25$\\
		\hline
	\end{tabular}
	\label{tab:table1}
\end{table}
The quantities reported in Table \ref{tab:table1} were estimated for a phase shift given by $\varphi=\pi$. This table considers realistic parameters for the losses associated to the propagation of the plasmonic sensing field, and the limited efficiency $\eta_{\text{ph}}$ and $\eta_{\text{pl}}$ of the single-photon detectors used to collect photonic and plasmonic mode respectively. In this case, we assume $\eta_{\text{ph}}=0.3$ and $\eta_{\text{pl}}=0.3$. The latter value is obtained from our simulation, whereas the former corresponds to the efficiency of commercial single-photon detectors \cite{HashemiRafsanjani:17,Marsili2013}.  In general, the value for $\varphi$ determines how strongly the dipolar and quadrupolar LSP modes are excited, and consequently their far-field angular distributions. However, the process is applicable for other phases $\varphi$. Our predictions suggest that plasmonic subtraction can be achieved at reasonable rates using a properly designed nanostructure.
 \begin{figure}[!ht]
 	\centering \includegraphics[width=0.6\columnwidth]
 	{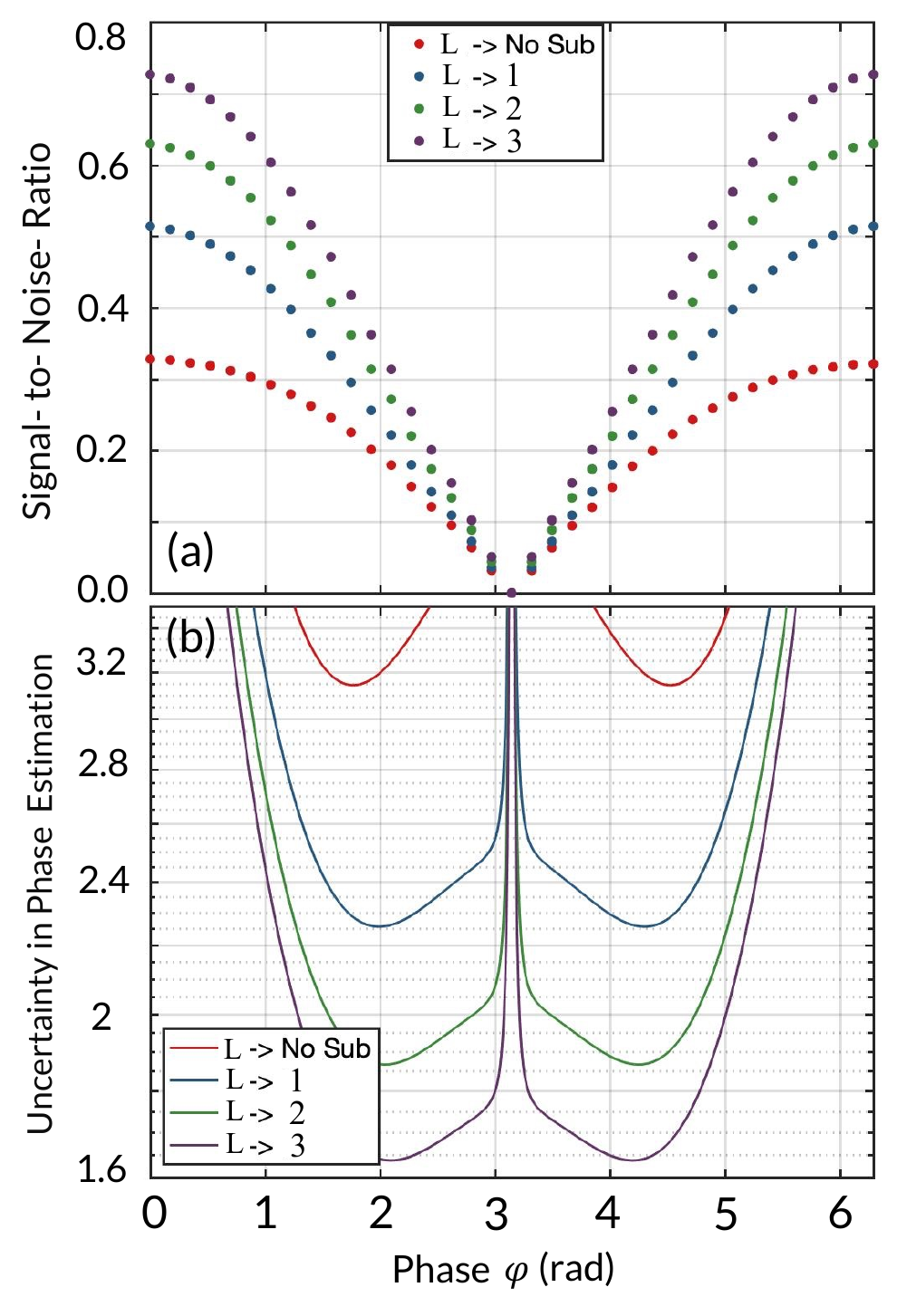}
 	\mycaption{ Signal-to-Noise Ratio.}{ The panel in (a) reports the signal-to-noise ratio (SNR) as a function of $\varphi$ for the conditional detection of the plasmonic modes transmitted by a $320$ nm nanoslit. The red dots represent the unconditional SNR. Furthermore, the blue, green, and purple dots indicate the SNR for the subtraction of one, two, and three plasmons, respectively. This plot shows the possibility of improving the SNR of our plasmonic sensor through the subtraction of plasmons. The panel in (b) indicates that an increasing SNR leads to lower uncertainties in the estimation of phase shifts induced by analytes. The lower uncertainties described by $\Delta\varphi$ imply higher sensitivities of our plasmonic sensor. These figures are taken from\cite{Mostafavi2022}. }
 	\label{fig:SNRvag2}
 \end{figure}
We now quantify the performance of our conditional scheme for plasmonic sensing through the SNR associated to the estimation of a phase shift. The SNR is estimated as the ratio of the mean occupation number to its standard deviation. This is defined as (see Appendix \ref{app:A:SNR})
\begin{equation}
\begin{split} 
\text{SNR} =  \sqrt[]{\frac{(1+L) \bar n \gamma\eta_{\text{ph}} \xi \cos^{2}\frac{\varphi}{2}}{1+ \bar n  \gamma(\xi\eta_{\text{ph}}+(1-\xi)\eta_{\text{pl}}) \cos^{2}\frac{\varphi}{2} }}.
\end{split} 
\label{eq4}
\end{equation}
\noindent 
Here, the parameter $\xi=T_{\text{ph}}/(T_{\text{ph}}+T_{\text{pl}})=0.80$ represents the normalized transmission of the photonic mode. In Fig. \ref{fig:SNRvag2}a, we report the increasing SNR of our plasmonic sensor through the process of plasmon subtraction by plotting the SNR for the subtraction of one, two, and three plasmons for different phase shifts $\varphi$. In addition, for sake of completeness, we evaluate the improvement in sensitivity using error propagation \cite{Helstrom}. More specifically, we calculate the uncertainty of a phase measurement $\Delta\varphi$. This parameter is estimated as 
\begin{equation}
\Delta \varphi=\sqrt{\left\langle \hat{n}^{2}\right\rangle-\langle \hat{n}\rangle^{2}}/\left|\frac{d\langle \hat{n}\rangle}{d \varphi}\right|
\label{eqg5}
\end{equation}
\noindent
The observable $\hat{n}=\hat{e}^{\dagger}\hat{e}$, corresponds to the conditional intensity measurement within an angular range of the far-field distribution (specified in Fig. \ref{fig:figurestdv2} with the vertical lines). In the field of quantum metrology, the reduced uncertainty of a phase measurement $\Delta\varphi$ is associated to an improvement in the sensitivity of a quantum sensor \cite{Hwanglee,phaseEstimation}. In this regard, the conditional detection of plasmons increases the sensitivity of our plasmonic sensor. This enhancement is reported in Fig. \ref{fig:SNRvag2}b. Here, we demonstrate that the attenuation of the fluctuations of a weak plasmonic field, through the subtraction of up to three plasmons, leads to lower uncertainties in the sensing of photosensitive analytes. 
We point out that our conditional measurement scheme is general and can be applied to any plasmonic sensing platform \cite{Garoli2019NanoLett}. As demonstrated through Fig. \ref{fig:SNRvag2} and Eq. (\ref{eq4}), the subtraction of plasmons boosts the signal-to-noise ratio of any electromagnetic field used for classical sensing. Thus, if a small physical parameter can be sensed by a classical scheme for plasmonic sensing, our technique will increase its signal-to-noise ratio. However, if the sample of interest acting as a phase shifter induces a change in the particle number that is smaller or equal than the statistical fluctuations ($\Delta n$) of the sensing field, our technique cannot provide any advantage. In this situation, the signal cannot be discriminated from the inherent noise of the plasmonic near fields. As such, our scheme is capable of improving current capabilities for single-molecule detection if the signal induced by the single-molecule is stronger than the fluctuations of the sensing field \cite{Garoli2019NanoLett}.
In conclusion, we have investigated a new method for quantum plasmonic sensing based on the conditional subtraction of plasmons. We have quantified the performance of this scheme, under realistic conditions of loss, by considering the design of a real plasmonic nanoslit sensor. We showed that conditional measurements offer an important path for controlling the statistical fluctuations of plasmonic fields for sensing. In our work, we considered the case for which the sensing field contains a mean plasmonic number lower than two. In this regime, we showed that the attenuation of the quantum fluctuations of plasmonic fields increases the mean occupation number of the sensing field. Interestingly, this effect leads to larger signal-to-noise ratios of our sensing protocol. Furthermore, this feature of our technique enables performing sensitive plasmonic sensing with weak signals \cite{NanoToday, Hwang, Maier, Lee1}.  We believe that our work offers an alternative approach to boost signals in quantum plasmonic platforms operating in the presence of loss at the few particle regime \cite{Tame2013, chenglongreview}.

%% file: chapter4.tex
\chapter{High-Dimensional Optical Encryption in Multimode Fibers}

\section{Motivation}
The ability to engineer the spatial wavefunction of photons has enabled a variety of quantum protocols for communication, sensing, and information processing. These protocols exploit the high dimensionality of structured light enabling the encoding of multiple bits of information in a single photon, the measurement of small physical parameters, and the achievement of unprecedented levels of security in schemes for cryptography \cite{omar:2019,willner:15,wang:2012}. Unfortunately, the potential of structured light has been restrained to free-space platforms in which the spatial profile of photons is preserved. Here, we make an important step forward to using structured light for fiber optical communication. We introduce a classical encryption protocol in which the propagation of high-dimensional spatial modes in multimode fibers is used as a natural mechanism for encryption. This provides a secure communication channel for data transmission. The information encoded in spatial modes is retrieved using artificial neural networks, which are trained from the intensity distributions of experimentally detected spatial modes. Our on-fiber communication platform allows us to use single spatial modes for information encoding as well as the high-dimensional superposition modes for bit-by-bit and byte-by-byte encoding respectively. This protocol enables one to recover messages and images with almost perfect accuracy. Our classical smart protocol for high-dimensional encryption in optical fibers provides a platform that can be adapted to address increased per-photon information capacity at the quantum level, while maintaining the fidelity of information transfer. This is key for quantum technologies relying on structured fields of light, particularly those that are challenged by free-space propagation. 
\section{Background}


Among the multiple families of structured optical beams, Laguerre-Gaussian (LG) modes have received particular attention for their ability to carry orbital angular momentum (OAM) \cite{allen1999iv,AllenOAM_PRA,yao:2011,padgett2017orbital}. 
Over the past decade, there has been an enormous interest in using photons carrying OAM for quantum communication and integrated photonics applications \cite{willner:15, WillnerLiu:2021, wang:2012, baghdady:2016, Liu:2019, Cozzolino:2019,yan2014high, Zahidy2022,zahidyQRNG2022}. These structured beams of light allow for the encoding of multiple bits of information in a single photon \cite{yao:2011, padgett2017orbital, omar:2019}. Additionally, it has been shown that high-dimensional Hilbert spaces defined in the OAM basis can increase the robustness of secure protocols for quantum communication \cite{mirho:2015,Rodenburg_2014, Forbes:2021}.  However, despite the enormous potential of structured photons, their vulnerabilities to phase distortions impose important limitations on the realistic implementation of quantum technologies \cite{malik:2012, bhusal2021spatial,willner:15, omar:2019, yao:2011, milione:2017, lavery:2013, magana2016hanbury, mirho:2015, willner:15}. 
Indeed, LG beams are not eigenmodes of commercial optical fibers and consequently their spatial profile is not preserved upon propagation. For this reason,  quantum communication with structured photons has been limited to free-space platforms \cite{gibson2004free,anguita2008turbulence,wang2012terabit,ren2016experimental, su2012demonstration, xie2015performance}. Recently, there has been an enormous interest in employing artificial neural networks to boost the functionality and robustness of quantum technologies  \cite{krenn2020computer, Beer:2020, bishop2006pattern, lecun2015deep, carleo2019machine, Walln:2020}. In the field of photonics, there has been extensive research devoted to developing artificial neural networks for the implementation of novel optical instruments \cite{you2020identification, kudyshev2020rapid, gebhart2020neural}. Indeed, convolutional neural networks (CNNs) have enabled the demonstration of new imaging schemes working at the single-photon level \cite{giordani2020machine, doster2017machine, bhusal2021spatial}. These protocols have been employed to characterize structured photons in the Laguerre-Gaussian (LG), Hermite-Gaussian (HG), and Bessel-Gaussian (BG) bases \cite{doster2017machine, sun2019identifying, giordani2020machine, hofer2019hermite, park2018multiplexing, willner:15, omar:2019, yao:2011, Huang:2021}. Here, we introduce a machine learning protocol that exploits single and superposed spatial modes of light propagating in multimode fibers for high-spatial-mode dimensional encryption. This is achieved by training artificial neural networks from experimental spatial profiles in combination with a theoretical model that describes the propagation of spatial modes in multimode fibers. The trained neural network enables us to decrypt information encoded in spatial modes of light. We demonstrate robust and efficient bit-by-bit and byte-by-byte encryption in commercial multimode fibers.

\section{Conceptual Overview}
\begin{figure}
\centering  
\includegraphics[width=0.9\linewidth]{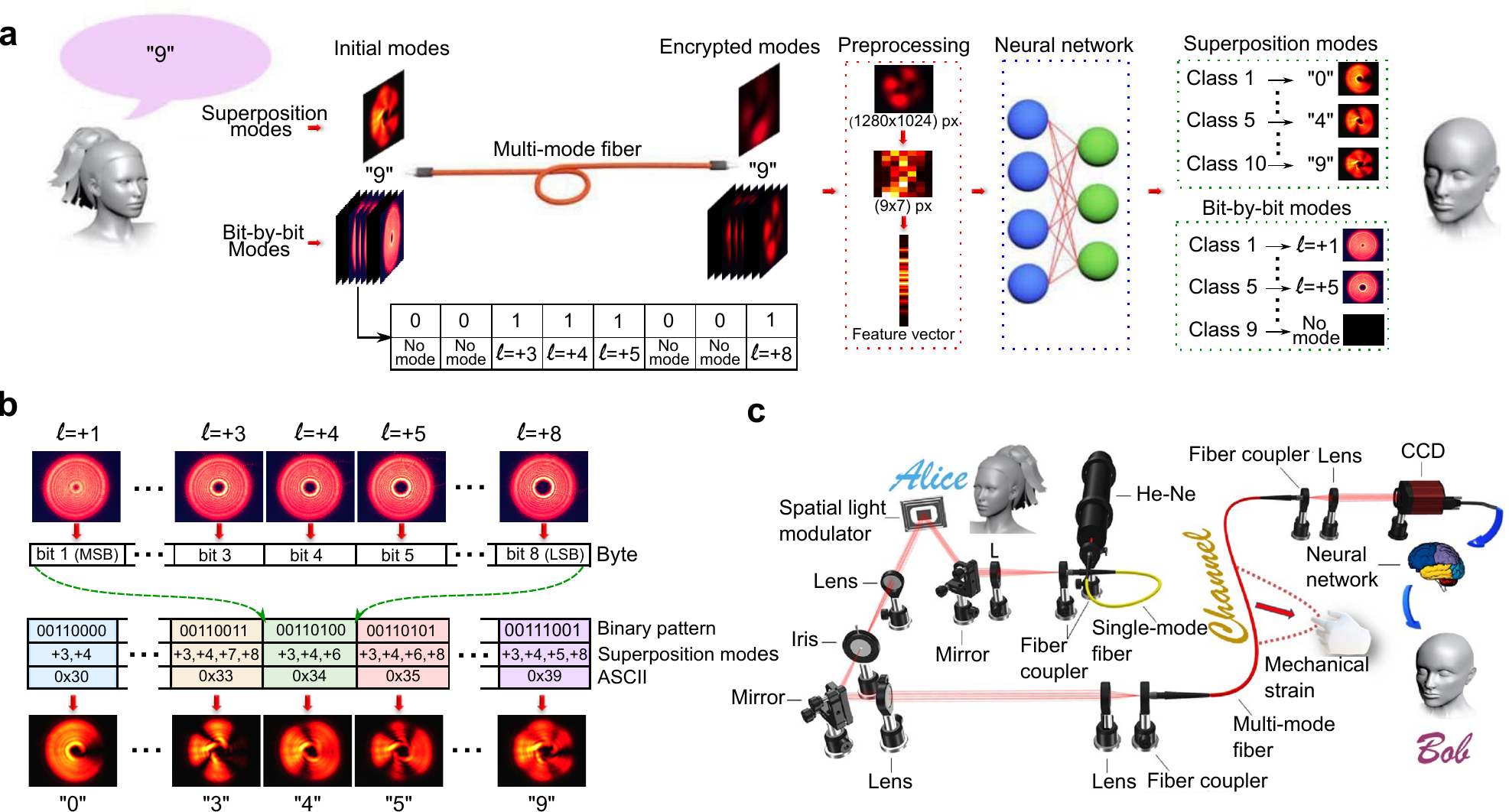}
\mycaption{ Schematic of Encryption Protocol.}{ (\text{a}) Conceptual schematic of our encryption protocol. In this case, Alice sends the message \enquote{9} to Bob in a bit-by-bit fashion or through a superposition of spatial modes (byte-by-byte). The resulting computationally efficient feature vector is used to train a neural network with high accuracy. The preprocessing details for encrypted modes can be found in the Methods section. (\text{b}) The OAM mode-to-bit-position relation is shown along with superposition states that correspond to the ASCII digits from zero to nine. This \enquote{alphabet} is used to encode information in spatial modes carrying OAM. Our experimental setup is depicted in (\text{c}). Note that the design is flexible and collimator free-space optical lenses can be used in lieu of lens/iris systems. Here, Alice encodes a message using OAM modes generated through a spatial light modulator (SLM). The spatial modes are coupled into a 1-meter long multimode fiber that is used to transmit information to Bob. In this case, we emulate multiple transmission conditions by introducing stress to the fiber via mechanical manipulation. The resulting perturbations are used to encrypt the message. We train our artificial neural network by collecting multiple spatial profiles of the distorted beams produced by the multimode fiber. Each distorted spatial profile of the optical beam corresponds to a particular condition of stress exerted on the fiber. Remarkably, our neural network is capable of recovering the initial superposition modes, converting them to the standard
alphabet characters for Bob at read out. These figures are taken from \cite{Lollie2022}.}
\label{fig:concept}
\end{figure}
The conceptual illustration of our smart encryption protocol is presented in Fig. \ref{fig:concept}a. Here, Alice prepares a message encoded in high-dimensional OAM modes that is then sent to Bob through a 1-meter long multimode fiber. From information theory, per-photon storage/transfer capacity scales as $\log_{2}\text{D}$, where D is the dimensionality of the system \cite{Shannon1948}. D is determined by the number of spatial modes in superposition when Alice encodes her message. For our protocol, the high-dimensionality encompasses up to an 8-dimensional OAM state space. The protocol entails the use of the 8-bit ASCII (American Standard Code for Information Interchange) code, allowing Alice to encode a message in two different ways using the alphabet shown in Fig. \ref{fig:concept}b. In the first approach, each character in the message is represented by a byte (eight bits). Then, the OAM modes from $\ell=+1$ to $\ell=+8$ are one-to-one correlated with the position of each bit. The mode $\ell=+1$ is the most significant bit (MSB) and $\ell=+8$ the least significant bit (LSB). Consequently, for each character in the message, Alice has an eight-bit binary string where each bit position is mapped to an LG mode and sequentially sent to Bob in a bit-by-bit fashion. In the second approach,  Alice prepares a high-dimensional one state (byte) composed of a superposition of eight bits representing a particular character. This enables Alice to send a message to Bob through a sequence of characters, or byte-by-byte, leading to a more computationally efficient process. It is worth mentioning that the use of the 8-bit ASCII code as our “alphabet” state space to encode a message from Alice to Bob is arbitrary and used for the purpose of demonstrating our protocol. This 8-bit state space provides for an eight-dimensional spatial mode OAM state to represent the byte. If the chosen encoding “alphabet” is larger, the total possible state dimensionality will be commensurate with the number of superposed OAM modes needed to encode that state space, thus increasing the dimensionality and complexity of the protocol. This scheme enables the mitigation of some errors that may be introduced during the transmission and reception of information, such as the loss of bits. 
\section{Theory}
\begin{figure}[!htb]
\centering  
\includegraphics[width=0.85\linewidth]{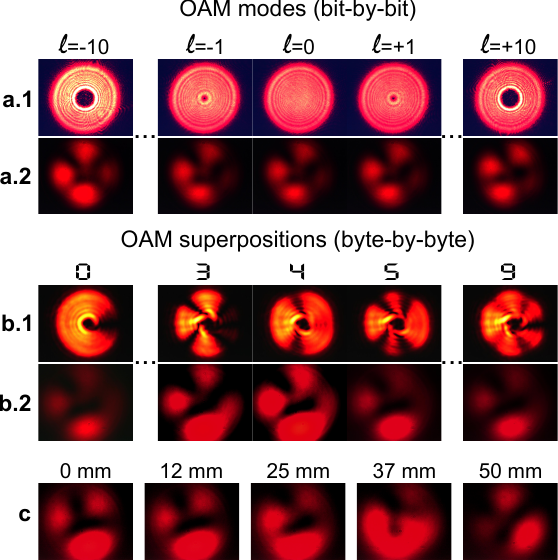}
\mycaption{Experimental Spatial Intensity Distributions and Encrypted LG Modes.}{ Spatial intensity distributions of initial and encrypted LG modes obtained experimentally for the maximum strain in the fiber (50 mm).  Intensity profiles of individual modes with azimuthal quantum numbers, $\ell=-10$, $\ell=-1$, $\ell=0$, $\ell=+1$, $\ell=+10$, before (\text{a.1}) and after (\text{a.2}) the propagation through the multimode fiber. (\text{b.1}) Superposition of LG modes representing the numeric characters 0, 3, 4, 5, and 9. Each character has been encoded using the alphabet displayed in Fig. \ref{fig:concept}. The bottom row (\text{b.2}) shows the encrypted modes corresponding to each of the superpositions. (\text{c}) Spatial profiles for the numeric character 1 obtained after propagation for different displacements of the fiber: 5, 12, 25, 37, and 50mm. Note that the fiber experiences strain due to the displacement, resulting in a dynamic intensity output. These figures are taken from \cite{Lollie2022}.}
\label{fig:modes}
\end{figure}

We decrypt messages by training artificial neural networks with experimental spatial profiles in combination with a theoretical model that describes our protocol and introduce our model to describe propagation of spatial modes in multimode fibers. For this purpose, we consider the coupling of an encoded message from free space to the transmission channel, namely the optical fiber. In this case, one has to decompose the initially injected field into the modes that are sustained by the specific features of the fiber. For the weakly guiding step-index fiber used in this experiment, the modes are described by the linearly polarized (LP) solution set. The field distribution, in polar coordinates, $\Psi\pare{r,\phi}$, is thus described by the solution of the scalar Helmholtz equation, which for a cylindrical fiber with a core radius $a$ is given by \cite{saleh2019fundamentals,fibers_book,brunin2015}
\begin{equation}
\text{LP}_{\ell p} = N_{\ell p}
\left\{
	\begin{array}{ll}
		\text{J}_{\ell}\pare{\kappa_{T\ell p}r}\exp\pare{-i\ell \phi}  & \mbox{if } r < a, \\
		\text{K}_{\ell}\pare{\gamma_{\ell p} r}\exp\pare{-i\ell \phi} & \mbox{if } r \geq a,
	\end{array}
\right.
\end{equation}
\noindent 
where $N_{\ell p}$ is a normalization constant, $J_{\ell}\pare{x}$ is the Bessel function of the first kind and order $\ell$, and $K_{\ell}\pare{x}$ is the modified Bessel function of the second kind and order $\ell$. Note that the parameters $\kappa_{T\ell p}$ and $\gamma_{\ell p}$ determine the oscillation rate of the field in the core and the cladding, respectively. These are defined by
\begin{eqnarray}
\kappa_{T\ell p}^{2} &=& n_{\text{core}}^{2}k_{0}^{2} - \beta_{\ell p}^{2}, \\
\gamma^{2}_{\ell p} &=& \beta_{\ell p}^{2} - n_{\text{cladding}}^{2}k_{0}^{2},
\end{eqnarray}
\noindent 
where $k_{0}=2\pi/\lambda_{0}$, with $\lambda_{0}$ being the vacuum wavelength of the light inside the fiber, $\beta_{\ell p}$ is the propagation constant of the $p$th guided mode for each azimuthal index $\ell$, and $n_{\text{core}}$ and $n_{\text{cladding}}$ are the refractive indices of the core and the cladding, respectively. For the description of the LP modes, the additional fiber parameter $V$ is required, which is defined as
\begin{equation}
V^{2} = \kappa_{T\ell p}^{2} + \gamma_{\ell p}^{2} = \pare{2\pi\frac{a}{\lambda_0}}^{2}\pare{n_{\text{core}}^{2} - n_{\text{cladding}}^{2}}.
\end{equation}
This fiber parameter determines the amount of modes and their propagation constants. In our experiments, we make use of a 1-meter long, 10 $\mu m$-diameter fiber, with $\text{N.A.} = \sqrt{n_{\text{core}}^{2} - n_{\text{cladding}}^{2}} = 0.1$. In these conditions, an arbitrary field propagating along the fiber may be decomposed in six LP modes with indexes $(\ell,p) \in \llav{(-2,1),(-1,1),(0,1),(1,1),(2,1),(0,2)}$. This implies that, regardless of the initial condition, the output mode of the fiber can always be written as
\begin{equation}\label{eq:coeff}
\Psi_{\text{out}}\pare{r,\phi} = \sum_{\ell,p}c_{\ell,p}\text{LP}_{\ell,p},
\end{equation}
where the coefficients $c_{\ell,p}$ are defined by the injected field and the properties of the optical fiber throughout the propagation length. A key consideration for spatial light in fibers is the toroidal structure of an OAM mode, with its optical vortex (phase signularity) along the propagation axis. The product of $\ell$ with the azimuthal index $\phi$ gives the topological charge of the mode and the diameter of the vortex scales with increased quantum number $\ell$. This limits the azimuthal diameter of modes coupled into the fiber. If the vortex diameter of single or superposed modes is larger than that of the fiber diameter, this severely attenuates the light coupled into the fiber, resulting in sub-optimal intensity profiles with which to train the neural network.
In a realistic scenario, the local random variations of the fiber properties produce significant distortions of spatial modes, thus making almost impossible to predict the spatial distribution of photons at the end of the fiber, i.e., the coefficients $c_{\ell,p}$ in Eq. (\ref{eq:coeff}). This is the main motivation behind our machine-learning protocol for  encryption in optical fibers. In our experiments, variations are produced by a mechanical strain, and in the case of superposition modes, by both strain and mixing of the modes during the propagation. Once optical spatial modes leave the fiber, the goal is to recover the sequence of transmitted modes either bit-by-bit (individual modes) or byte-by-byte (superposition modes) as the case may be and then effectively decode the optical profiles (images) to compose the message. In this respect, Bob exploits the self-learning features of artificial neural networks to decrypt the information encoded in the distorted spatial modes efficiently. To train the neural network, the data-set comprises a collection of down-sampled images, rearranged as column vectors that correspond to the aberrated optical profiles, as shown in Fig. \ref{fig:concept}a. After the training, Bob utilizes the high efficiency of the neural network to retrieve the message by identifying individual modes if the communication was bit-by-bit, or recognising superposition modes when the communication was byte-by-byte. The schematic diagram of our experimental setup is shown in Fig. \ref{fig:concept}c. Alice use a He-Ne laser with a spatial light modulator (SLM) to prepare the message to be sent using OAM states of light. The light beam is then sent to Bob through multimode fiber, and the output is measured by a camera. More details of the experiment can be found in the Methods section. Examples of different OAM intensity distributions collected experimentally are shown in Fig. \ref{fig:modes}. Spatial profiles of individual LG modes with different azimuthal quantum numbers ($\ell=-10$, $\ell=-1$, $\ell=0$, $\ell=+1$, $\ell=+10$) and LG superpositions, before the multimode fiber, are displayed in Figs. \ref{fig:modes}a.1 and b.1, respectively. Each superposition has unique intensity distribution given by combining two and up to eight OAM single modes, depending on the character to be represented (see the alphabet shown in Fig. \ref{fig:concept}b). For demonstration purposes, Fig. \ref{fig:modes}b.1 presents superpositions of LG modes for the numeric characters: 0, 3, 4, 5 and 9.

\begin{figure}[!htb]
\centering  
\includegraphics[width=0.95\linewidth]{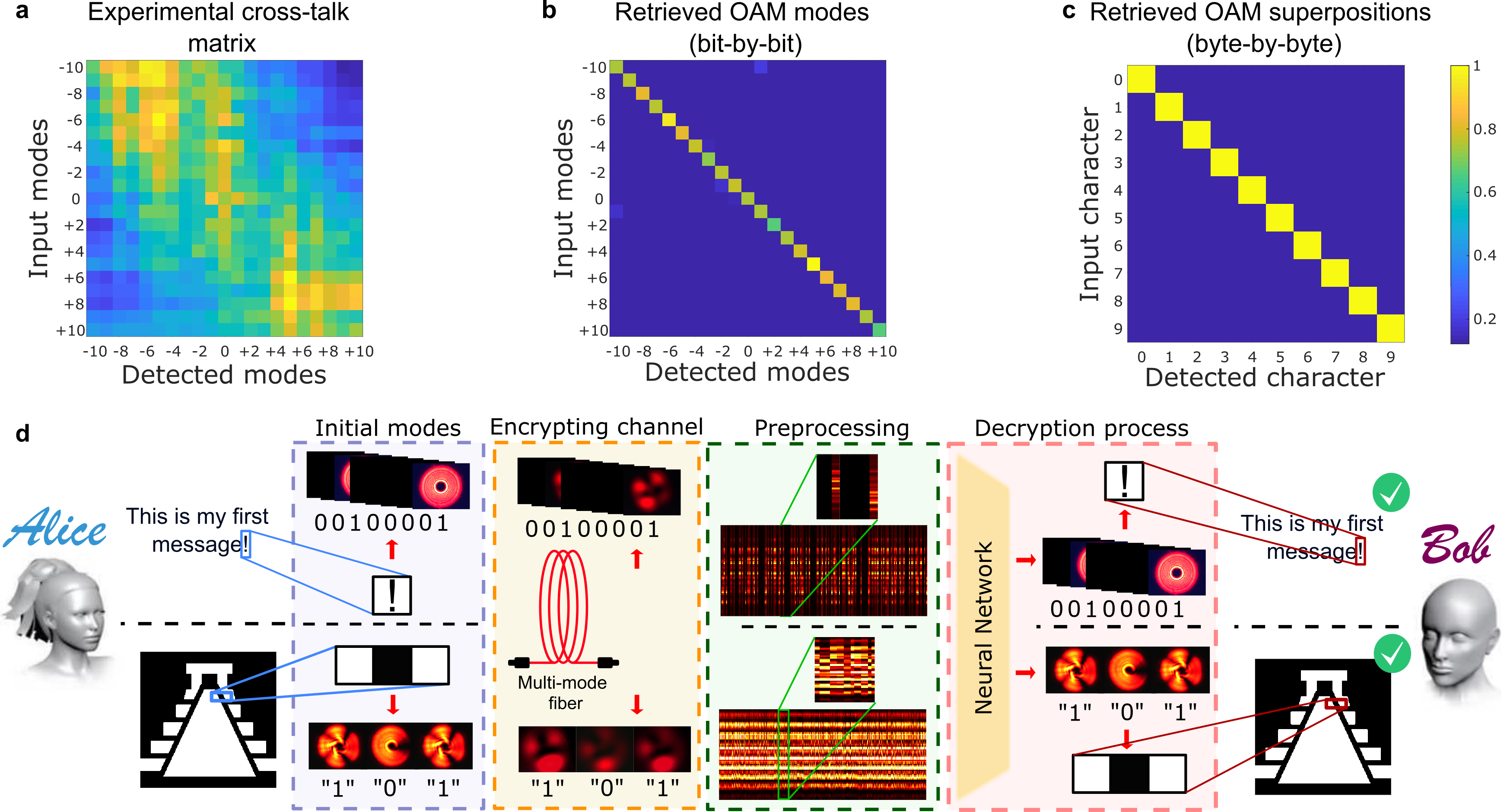}
\mycaption{Cross-Talk Matrix for LG Modes.}{ (\text{a}) Cross-talk matrix for LG modes with azimuthal quantum numbers from $\ell=-10$ to $\ell=+10$. Note that the LG modes are distorted severely, so the identification is practically impossible. (\text{b}) Diagonal cross-talk matrix obtained after applying our neural network. Our approach provides a powerful tool to recognize OAM modes after the fiber with an efficiency of 98\%. (\text{c}) The cross-talk matrix obtained for the LG superpositions that represent the numeric characters zero to nine. The diagonal elements indicate that the transmitted characters are correctly identified. This neural network exhibits a performance of 99.9\%. (\text{d}) Smart communication protocol with Alice sending the message \enquote{This is my first message!} (upper message) and the image of a Mexican pyramid (bottom message) to Bob through the multimode fiber. The transmission of the text and image is bit-by-bit and byte-by-byte, respectively. Bob deciphers both messages by using the trained NN with near-unity accuracy. These figures are taken from \cite{Lollie2022}.}
\label{fig:cross}
\end{figure}

\noindent 
As mentioned above, once a spatial mode is transmitted through the fiber, there is significant distortion from the applied tension, the local variations of the fiber properties, and even the noise of the camera sensor,  resulting in an encrypted mode. Fig. \ref{fig:modes}a.2 and b.2 show the encrypted modes corresponding to the spatial beams in Fig. \ref{fig:modes}a.1 and b.1, respectively, for a displacement of 50 mm, which represents the maximum strain that may be applied in the fiber in our experiment. Note that intensity distributions of the encrypted modes change drastically with respect to the distributions of the initial modes. Moreover, the shape of the intensity patterns can change significantly as a function of the strain experienced by the fiber. Fig. \ref{fig:modes}c displays spatial profiles for LG superposition of the character 1 with different applied tension represented by the displacements: 0, 12, 25, 37, and 50 mm. Importantly, these distortions are induced randomly, which leads to an unbounded set of encrypted modes. Nevertheless, our NN 
(see Appendix \ref{app:A:Training}) can decode these encrypted modes with high accuracy for both individual modes and mode superpositions. This effectively generalizes an unbounded set from a limited collection of labeled examples. In standard encryption techniques, the encryption key must be secret, and typically the same key is used for both encryption and decryption processes. In our smart communication protocol, the local variations of the fiber properties act as the encryption key. This key is unknown and different for each encrypted mode because distortions in the spatial profiles are induced randomly. Interestingly, our NN provides a universal decryption key to retrieve the entered modes. Note that the strength of the encryption relies on the random variations induced by the multimode fiber whereas the security of the decryption key is based on the correct optimization of the NN synaptic weights during the training stage.\\
\noindent 
To study the LG mode cross-talk when the beam propagates through the fiber, we measured the cross-correlation matrix for modes with azimuthal quantum numbers from $\ell=-10$ to $\ell=+10$. In the cross-talk matrix, the diagonal elements represent the conditional probabilities among the transmitted and detected modes that were correctly recognized. Remarkably, the diagonal elements in our cross-talk matrix are erased completely as indicated by Fig. \ref{fig:cross}a. In fact, it is practically impossible to recognize any mode. However, as shown in Figs. \ref{fig:cross}b and \ref{fig:cross}c, we exploit the functionality of machine learning algorithms to design a NN sufficiently sensitive to discern LG modes after the multimode fiber, enabling us to reconstruct a diagonal cross-talk matrix. To show the ability of our machine-learning algorithms to recognize encrypted OAM modes,  we first design, train, and test a multi-layer neural network with the capacity to identify LG beams with different positive and negative topological charges, which go from $\ell=-10$ to $\ell=+10$. It is known that two pure LG modes with identical radial numbers but with opposite topological charges are indistinguishable using intensity measurements solely because they present exactly the same distributions. However, we experimentally demonstrate that our approach enables the discrimination of oppositely charged LG modes from their intensity patterns. We exploit the fact that OAM propagation through the multimode fiber induces phase distortions. The fiber is interpreted to be a \enquote{disordered} medium due to the inherent noise and the local variations of its properties. This leads to distinct modal cross-talk for the LG modes and their conjugates, resulting in changes in the intensity profiles. Consequently, this allows the NN algorithms to distinguish opposite LG modes unequivocally.  Thus, our approach overcomes the limitations of existing strategies based on projection measurements and phase-measurement interferometry techniques. As seen in Fig. \ref{fig:cross}b, we obtain a classification accuracy of 98\%.\\
\noindent 
Now we describe the implementation of the smart communication protocol using the trained NN. For bit-by-bit communication, we select the LG modes from $\ell=+1$ to $\ell=+8$ to form 8-bit binary words that allow us to encode characters from the ASCII code. It is worth mentioning that by using these eight modes, our neural network reaches an overall accuracy of 99.9\%. Again, Alice encodes a message using the alphabet shown in Fig. \ref{fig:concept}b. This process is presented in Fig. \ref{fig:cross}d. Alice prepares the plain text \enquote{This is my first message!} that is transmitted to Bob through the multimode fiber. Note that we show the detailed encoding and decoding processes for a particular character. In the figure, we highlight the exclamation mark, however the same stages are applied for all the characters of the message. The communication channel acts as the encryption process, so Bob receives a sequence of indistinguishable intensity profiles. The goal is to recover the sequence of transmitted bits by Alice from the intensity distributions. Prior to the decryption process, Bob carries out image pre-processing that includes the transformation of an image from RGB to grayscale. This is followed by the down-sampling process and the rearrangement of the pixels from resulting matrices into column vectors. In the decryption process, Bob uses the neural network to decipher the message by identifying each received LG mode and translating it via the standard alphabet. Note that to decrypt the spatial modes that leave the multimode fiber, essentially, Bob solves a classification problem using the neural network,  where the output classes are the LG modes from $\ell=+1$ to $\ell=+8$.
To describe the implementation of our proof-of-principle smart communication protocol for byte-by-byte communication, LG superposition modes are prepared using the alphabet in Fig. \ref{fig:concept}b. We begin by using the dataset of encrypted superposition modes to train, validate and test a neural network that maps the distorted mixtures to one of the transmitted digits. Importantly, the defined output classes for byte-by-byte communication protocol are the digits formed by the superpositions of LG modes. After training, the performance of our neural network is 99.9\%. This demonstrates the ability of our neural network to discern, with near-unity accuracy, experimental superpositions of LG modes. This is highlighted via the cross-talk matrix in Fig. \ref{fig:cross}c. Furthermore, to unveil the utility and functionality of our smart communication protocol, Fig.  \ref{fig:cross}d presents a scheme where Alice sends the image of a Mexican pyramid to Bob through the multimode fiber. As in the previous case for the plain text, we emphasize the involved processes in the communication protocol for three particular pixels from the image. Here, each pixel of the image is represented by an eight-bit word whose decimal value is ``1'' for white pixels and ``0'' for black pixels. Alice can employ the superposition modes that represent the digits ``1''  and ``0'' to map the image and transmit it byte-by-byte (or equivalently pixel-by-pixel) through the communication channel. Thus, Bob receives one by one the encrypted pixels that comprise the image and preprocesses them. Then, Bob uses the neural network to identify the digits encoded in the superposition modes, after which he can retrieve the Mexican pyramid. At this point it is worth mentioning that, after the propagation, the image information cannot be inferred from the distorted beams. This decryption process requires the trained neural network to recover the plain image.
We quantify the integrity of the received information by calculating the mean squared error (MSE), defined by $\text{MSE}=\frac{1}{n}\langle \textbf{e} | \textbf{e} \rangle$ where $\textbf{e}=(\hat{\textbf{y}}-\textbf{y})$. Here, $\hat{\textbf{y}}$  and $\textbf{y}$ are vectors that contain the received and transmitted bytes, respectively. The measured MSE for both the message and image is zero. This validates the robustness and high efficiency of our protocol to decode OAM modes transmitted through the multimode fiber.
In summary, we have demonstrated a machine learning protocol that employs spatial modes of light in commercial multimode fibers for high-dimensional encryption. This classical protocol was implemented on a communication platform that utilizes LG modes for high-dimensional bit-by-bit and byte-by-byte encoding. The method relies on a theoretical model that exploits the training of artificial neural networks for identification of  spatial optical modes distorted by multimode fibers. This process allows for the recovery of encrypted messages and images with almost perfect accuracy. Our smart protocol for high-dimensional optical encryption using spatial modes in optical fibers has key implications for quantum technologies that rely on structured fields of light, especially those technologies where free-space propagation poses significant challenges.

\section{Experimental Results and Discussions}
The schematic diagram of our experimental setup is shown in Fig. \ref{fig:concept}c. We use a He-Ne laser at 633 nm that is spatially filtered by a single-mode fiber (SMF). The output beam with a Gaussian profile illuminates a spatial light modulator (SLM) displaying a computer-generated hologram. The SLM together with a 4f-optical system allows us to prepare any arbitrary spatial mode of light carrying OAM. We then use a telescope to demagnify the structured beam before coupling into a 1-meter long multimode fiber with diameter of 10 $\mu \text{m}$. The preparation of the modes used to store the message to be sent is performed by Alice. At the output of the fiber, Bob uses a camera to measure the collimated spatial profile of the communicated modes. Mechanical stress is induced in the fiber channel to generate the neural network training palette. The fiber is configured in a loop with the base secured to the optical table. The top of the loop is secured to a 3D translation stage with displacement occurring along the y-axis (orthogonal to and away from the plane of the table). Displacing the top of the loop attached to the translation stage 50 mm produces strain in the fiber. As the fiber is being pulled taut, successive images show the dynamic change of the mode, so the output at detection is now an LG mode distorted both via the multimode-fiber beam transformation as well as the applied tension. A camera is used to detect and display the output image. Two sets of data are taken: 1) The SLM is programmed to produce holograms for each OAM mode from -10 to 10, 21 modes total. For each mode, one image is captured at 0.10 mm translation intervals for a total displacement of 50 mm producing 500 images equivalent to 500 strain steps. 2) The SLM is programmed to produce holograms of OAM superposition modes for the 8-bit ASCII characters zero to nine. Each character is represented by a superposition of two, three, four, or five OAM single modes. One image is captured per 0.25 mm displacement interval over 50 mm for a total of 200 images per each superposition mode resulting in a total of 200 strain steps. The data collection was straightforward. A possible contributor to optical setup instability is power fluctuation of the laser. However, our protocol uses the Helium-Neon laser, which is ubiquitous in optics experiments. The time frame for the longest data collection, the 500 strain steps for each OAM mode at 21 total modes producing 10,500 images was 120 minutes, using MATLAB 2019a to automate the SLM programming. We found the laser power to be fairly constant with several power measurements at the beginning and end of each data collection.

%% file: chapter5.tex
\chapter{Multiphoton Quantum Coherence by Light Propagation}
\label{chapter:5}

\section{Motivation}
The modification of the quantum properties of coherence of photons through their interaction with matter lies at the heart of the quantum theory of light \cite{Mandel1995}. Indeed, the absorption and emission of photons by atoms can lead to different kinds of light with characteristic quantum statistical properties. As such, different types of light are typically associated with distinct sources \cite{GlauberPR1963, PhysRevLettSudarshan}. Here, we report on the observation of the modification of quantum coherence of multiphoton systems in free space. This surprising effect is produced by the scattering of thermal multiphoton wavepackets upon propagation. The modification of the excitation mode of a photonic system and its associated quantum fluctuations result in the formation of different light fields with distinct quantum coherence properties. Remarkably, we show that these processes of scattering can lead to multiphoton systems with sub-shot-noise quantum properties. Our observations are validated through the nonclassical formulation of the emblematic van Cittert-Zernike theorem \cite{Cittert:34, ZERNIKE:38}. We believe that the possibility of producing quantum systems with modified properties of coherence, through linear propagation, can have dramatic implications for diverse quantum technologies.

\section{Background}
The description of the evolution of spatial, temporal, and polarization properties of the light field gave birth to the classical theory of optical coherence \cite{wolf1954optics,Mandel1995, born2013principles, Dorrer:04,Gori2000, Cai2020, Cai2012}. Naturally, these properties of light can be fully characterized through the classical electromagnetic theory \cite{born2013principles}. Furthermore, there has been interest in describing the evolution of propagating quantum optical fields endowed with these classical properties \cite{Saleh05PRL, You2023CittertZernike}. This has been accomplished by virtue of the Wolf equation and the van Cittert-Zernike theorem \cite{wolf1954optics, Saleh05PRL, Cittert:34, ZERNIKE:38}. Nevertheless, there is a long-sought goal in describing the evolution of the inherent quantum mechanical properties of the light field that define its nature and kind \cite{GlauberPR1963, PhysRevLettSudarshan}. Such formalism would enable modeling the evolution of the excitation mode of propagating electromagnetic fields in the Fock number basis. Given the large number of scattering and interference processes that can take place in a quantum optical system with many photons, this ambitious description remains elusive \cite{DellAnno2006, Loaiza2019, You2020, chenglongnature, Walmsley:23}. Although, it is essential to describe the evolution of propagating multiphoton wavepackets in diverse quantum photonic devices \cite{Walmsley:23,omar:2019,npj2022, aspuru-guzik_photonic_2012}.\\
\noindent
The quantum theory of optical coherence developed by Glauber and Sudarshan provides a description of the excitation mode of the electromagnetic field \cite{GlauberPR1963, PhysRevLettSudarshan,PhysRevCoherentIncoherent}. This elegant formalism led to the identification of different kinds of light that are characterized by distinct quantum statistical fluctuations and noise levels \cite{Loaiza2019,you2020identification,GlauberPR1963,PhysRevCoherentIncoherent,YouLXAP2024}. As such, a particular quantum state of light is typically associated with a specific emission process and a light source \cite{PhysRevCoherentIncoherent}. Moreover, the quantum theory of electromagnetic radiation enables describing light-matter interactions \cite{allen1987optical}. These consist of absorption and emission processes that can lead to the modification of the excitation mode of the light field and consequently to different kinds of light \cite{PhysRevCoherentIncoherent,allen1987optical}. This possibility has triggered interest in achieving optical non-linearities at the single-photon level to engineer and control quantum states of light \cite{venkataraman_phase_2013, Zasedatelev-2021, Snijders-2016, HallajiNatPhys2017}. Thus, it is believed that the excitation mode of the light field, and its quantum properties of coherence, remain unchanged upon propagation in free space \cite{PhysRevCoherentIncoherent,allen1987optical}.
\begin{figure*}[!ht]
   \centering 
   \includegraphics[width=1.0\textwidth]{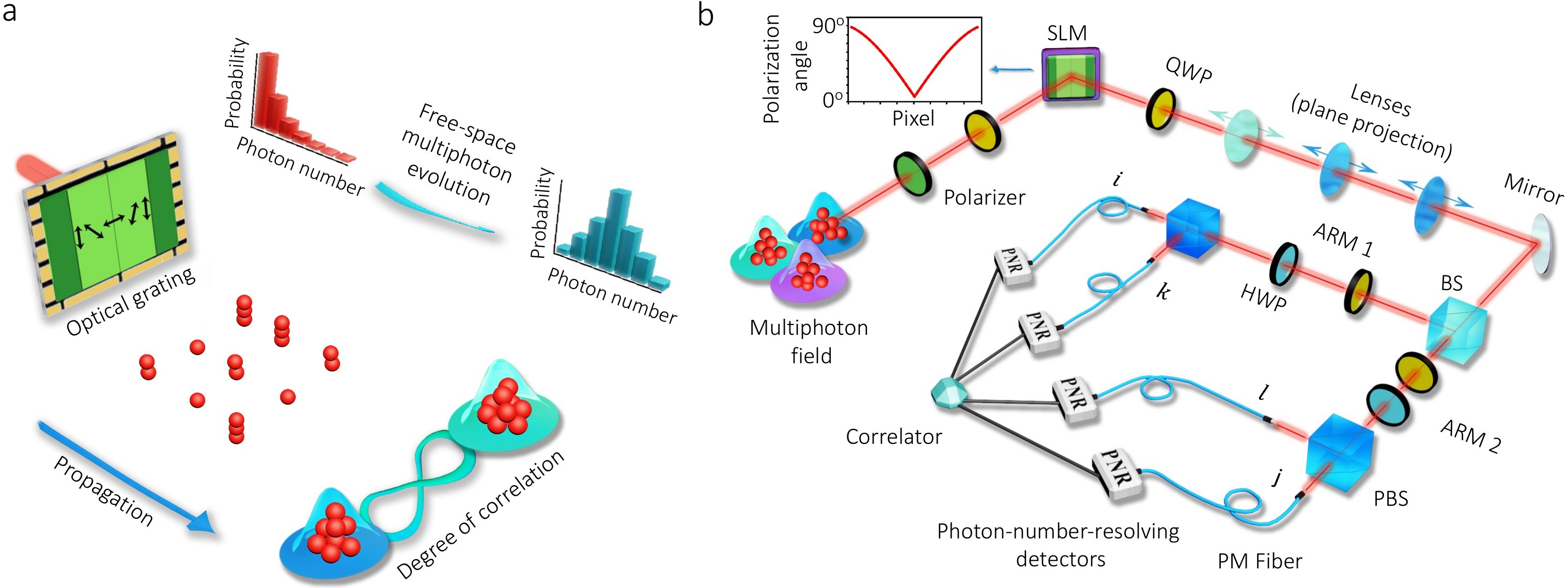}
    \mycaption{Modification of Quantum Coherence in Propagating Multiphoton Wavepackets.}{ The diagram in (\text{a}) illustrates the scattering of
    thermal multiphoton wavepackets by an optical grating. The grating modifies the polarization properties of the multiphoton wavepackets at different transverse spatial locations. The interference of the scattered multiphoton wavepackets, at different propagation planes, leads to changes in the quantum statistical properties of the thermal field. Interestingly, these interference events lead to the modification of multiphoton quantum coherence upon propagation in free space. The setup for the experimental investigation of this effect is shown in (\text{b}). Here, a multimode thermal multiphoton beam passes through a polarizer and a quarter-wave plate (QWP) to modulate its polarization. The transmitted circularly polarized photons illuminate a spatial light modulator (SLM) where we display a polarization grating. The beam reflected off the SLM is sent to another QWP to rotate its polarization at different transverse positions (details can be found in Appendix \ref{app:A:SLM}) \cite{PRLMirhossein2014}. The resulting polarization angle as a function of the transverse pixel position is depicted next to the SLM. This experimental arrangement induces partial polarization properties to the initial thermal light beam. The multiphoton field is then sent to a tunable telescope consisting of three lenses. This setup enables us to select different propagation planes of the scattered multiphoton field. We then perform polarization tomography of multiphoton wavepackets at an specific propagation plane by means of a beam splitter (BS), half-wave plates (HWPs), QWPs, and two polarizing beam splitters (PBS) \cite{Altepeter2005PhotonicST}. We use photon-number-resolving (PNR) detection to characterize the quantum coherence of propagating multiphoton systems and their quantum fluctuations \cite{you2020identification, HashemiRafsanjani:17}. These figures are taken from \cite{quantumcoherence2024}. }
   \label{Fig1-ch6}
\end{figure*} 
\noindent
We demonstrate that the statistical fluctuations of thermal light fields, and their quantum properties of coherence, can be modified upon propagation in the absence of light-matter interactions \cite{You2023CittertZernike,Hong2024}. This effect results from the scattering of multiphoton wavepackets that propagate in free space. The large number of interference effects hosted by multiphoton systems with up to twenty particles leads to a modified light field with evolving quantum statistical properties \cite{You2020,DellAnno2006}. Further, we show that the evolution of multiphoton quantum coherence can be described by the nonclassical formulation of the van Cittert-Zernike theorem \cite{You2023CittertZernike}. Interestingly, our description of propagating multiphoton quantum coherence unveils conditions under which multiphoton systems with sub-shot-noise quantum properties are formed \cite{Gerry2004}. Remarkably, these quantum multiphoton systems are produced upon propagation in the absence of optical nonlinearities \cite{venkataraman_phase_2013, Zasedatelev-2021, Snijders-2016, HallajiNatPhys2017}. As such, we believe that our findings provide an all-optical alternative for the preparation of multiphoton systems with nonclassical statistics. Given the relevance of photonic quantum control for multiple quantum technologies, similar functionalities have been explored in nonlinear optical systems, photonic lattices, plasmonic systems, and Bose-Einstein condensates \cite {OLSEN2002373, chenglongnature, Kondakci2015, venkataraman_phase_2013, Zasedatelev-2021, Snijders-2016, HallajiNatPhys2017}.
\section{Experimental Results and Discussions}

The optical system under consideration is depicted in Fig. \ref{Fig1-ch6}\text{a}. In this case, an unpolarized thermal field is scattered by an optical grating to produce multiphoton wavepackets with distinct polarization properties at different transverse spatial positions \cite{Gori2000}. The scattered photons contained in the thermal beam interfere upon propagation to change the statistical fluctuations of the field \cite{You2023CittertZernike}. Interestingly, these effects enable the modification of the quantum properties of coherence of the initial multiphoton thermal system in free space. As discussed in the Appendix \ref{app:A:Unpolarized}, we describe our initial thermal system as an incoherent superposition of coherent states \cite{PhysRevLettSudarshan, GlauberPR1963, ou2007multi} 
\begin{equation}
\label{eq1-ch6}
    \hat{\rho} = \int d\Sigma \bigotimes_{\boldsymbol{s}} \Big(|\alpha\rangle\langle\alpha|_{\Sigma,H,\boldsymbol{s}}+|\alpha\rangle\langle\alpha|_{\Sigma,V,\boldsymbol{s}}\Big),
\end{equation}
where $|\alpha\rangle_{\Sigma,B,\boldsymbol{s}}$ represents a coherent state of amplitude $\alpha$ with random mode-structure $\Sigma$, where $\hat{a}_{\Sigma,B,\boldsymbol{s}} = \int d \boldsymbol{\rho} \text{Rect}[(\boldsymbol{s}-\boldsymbol{\rho})/d]\Sigma(\boldsymbol{\rho})\hat{a}_B(\boldsymbol{\rho})$ and polarization $B\in\{H,V\}$ (see Appendix \ref{app:A:Unpolarized}). Furthermore, the tensor product over positions $\boldsymbol{s}$ represents the pixelated transverse beam profile where each pixel has sidelength $d$.
\begin{figure*}[t!]
   \centering 
   \includegraphics[width=\textwidth]{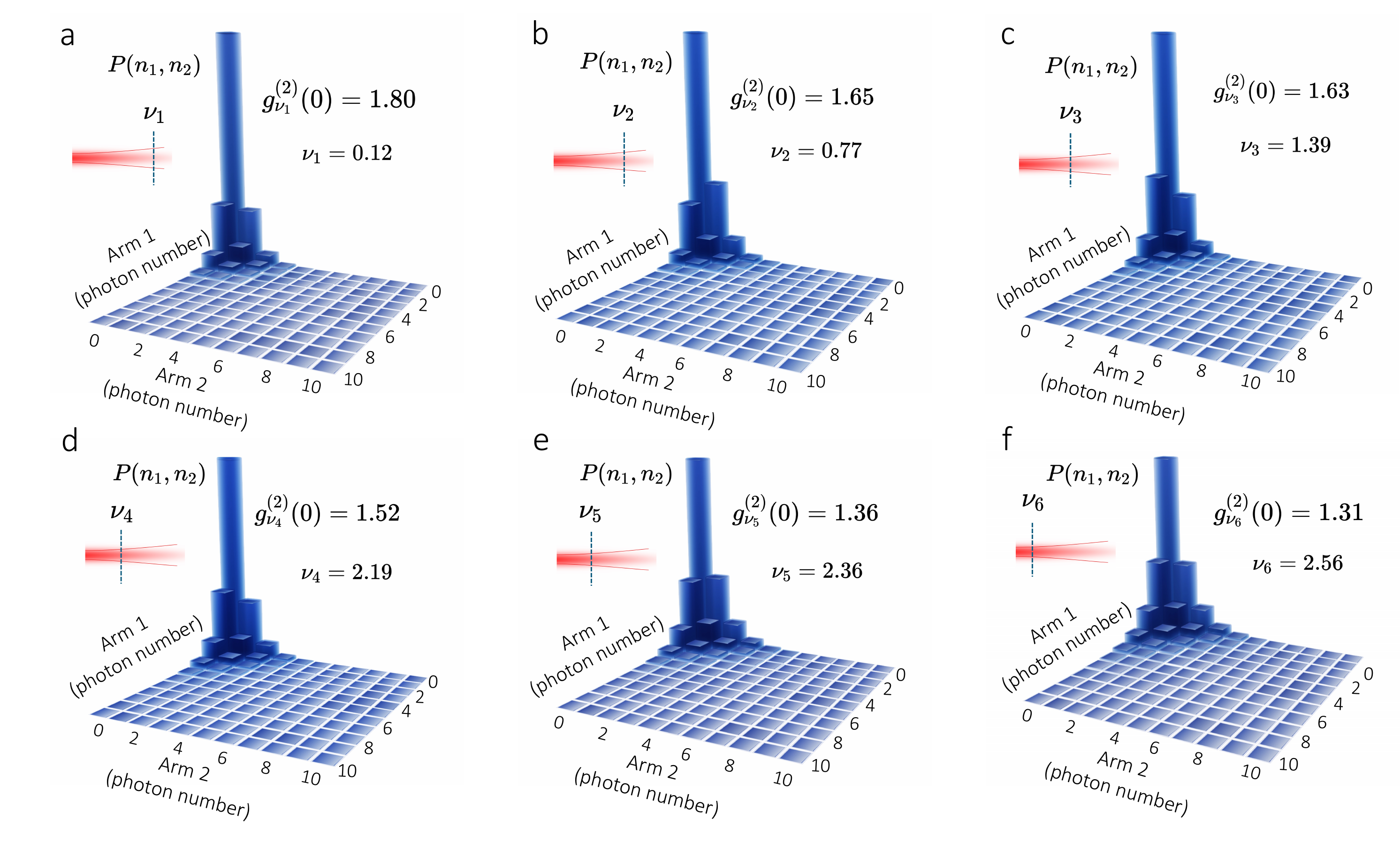}
    \mycaption{Evolving Quantum Coherence Induced by Light Propagation.}{
    The propagation of the multiphoton system reflected off the SLM induces modifications in its photon-number distribution. In this case, we focus on the horizontally-polarized component of the initial thermal beam with up to twenty particles. As shown in (\text{a}), the multiphoton system at the propagation plane characterized by $\nu_{1} = 0.12$ is nearly thermal. We define $\nu$ as $L\Delta X/(\lambda z)$, in this case $L=3 $ mm, $\Delta X=2 $ mm , $\lambda=780 $ nm and we scan the propagation distance $z$.  Interestingly, the quantum fluctuations of the multiphoton system are attenuated with $\nu$. This is quantified through the degree of second-order coherence $g^{(2)}_{\nu}(0)$, which also evolves upon propagation. The experimental results from (\text{a}) to (\text{f}) were obtained by scanning two detectors through different propagation planes. The large number of interference events upon propagation leads to the modified multiphoton system in (\text{f}), which is characterized by a $g^{(2)}_{\nu_6} (0)$ of 1.31. This multiphoton beam exhibits quantum statistical properties that approach those observed in coherent light. The evolving quantum dynamics of our multiphoton system can be modeled through Eq. (\ref{eq5-ch6}). Remarkably, the conversion processes of the multiphoton system, and its modified properties of quantum coherence, take place in free space in the absence of light-matter interactions. These figures are taken from \cite{quantumcoherence2024}. }
   \label{Fig2-ch6}
\end{figure*} 
\noindent
After the polarization grating, the resulting state is obtained via the transformation
\begin{equation}
    \begin{aligned}
        \hat{a}_B(\boldsymbol{\rho}) \rightarrow P_{HB}(\boldsymbol{\rho})\hat{a}_H(\boldsymbol{\rho}) + P_{VB}(\boldsymbol{\rho})\hat{a}_V(\boldsymbol{\rho}) + P_{\emptyset B}(\boldsymbol{\rho})\hat{a}_\emptyset(\boldsymbol{\rho}),
    \end{aligned}
\end{equation}
where $\hat{a}_\emptyset(\boldsymbol{\rho})$ is the mode for photon loss and $P_{AB}$ are the components of the transformation
\begin{equation}
    \boldsymbol{P}(x) = 
    \begin{bmatrix}
    \cos^2\left(\frac{\pi x}{L}\right) & \cos\left(\frac{\pi x}{L}\right)\sin\left(\frac{\pi x}{L}\right) \\
    \cos\left(\frac{\pi x}{L}\right)\sin\left(\frac{\pi x}{L}\right) & \sin^2\left(\frac{\pi x}{L}\right) \\
    \sin\left(\frac{\pi x}{L}\right) & \cos\left(\frac{\pi x}{L}\right)
    \end{bmatrix}
    \label{eqn:Prop} 
\end{equation}
where $A\in\{H,V,\emptyset\}$ and $L$ is the length of the polarization grating. The beam described by Eq. (\ref{eq1-ch6}) is then propagated by a distance of $z$ before being measured by two pairs of photon-number-resolving (PNR) detectors \cite{you2020identification, ChenglongAIP}. This propagation can be modeled through the Fresnel approximation on the mode structure of the initial beam \cite{goodman2008introduction}. We can then compute the second-order correlation function $G^{(2)}_{ijkl} (\boldsymbol{r}_1,\boldsymbol{r}_2,z)$ for the post-selected polarization measurements in the detection plane as \cite{Mandel1995}
\begin{equation}
    \begin{aligned}
        &G^{(2)}_{ijkl}(\boldsymbol{r}_1,\boldsymbol{r}_2,z) = \langle\hat{a}^\dagger_i(\boldsymbol{r}_1)\hat{a}^\dagger_j(\boldsymbol{r}_2)\hat{a}_k(\boldsymbol{r}_1)\hat{a}_l(\boldsymbol{r}_2)\rangle\\
        &= I_0 \int d\boldsymbol{\rho}_1 d\boldsymbol{\rho}_2 d\boldsymbol{\rho}_3 d\boldsymbol{\rho}_4 F(\boldsymbol{r}_1,\boldsymbol{r}_2,\boldsymbol{\rho}_1,\boldsymbol{\rho}_2,\boldsymbol{\rho}_3,\boldsymbol{\rho}_4,z) \\
        &\times\Big[\delta(\boldsymbol{\rho}_1-\boldsymbol{\rho}_3)\delta(\boldsymbol{\rho}_2-\boldsymbol{\rho}_4) + \delta(\boldsymbol{\rho}_1-\boldsymbol{\rho}_4)\delta(\boldsymbol{\rho}_2-\boldsymbol{\rho}_3)\Big].
    \end{aligned}
    \label{eq4-ch6}
\end{equation}
Remarkably, the Dirac-delta functions in Eq. (\ref{eq4-ch6}) demonstrate the presence of nontrivial correlations. Given the complexity of $I_0$ and $F(\boldsymbol{r}_1,\boldsymbol{r}_2,\boldsymbol{\rho}_1,\boldsymbol{\rho}_2,\boldsymbol{\rho}_3,\boldsymbol{\rho}_4,z)$, their explicit expressions are given in the Appendix \ref{app:A:CorrelationMatrix}. These describe the coherence of a photon with itself, which existed prior to interacting with the grating, and the spatial coherence gained by multiphoton scattering. These terms unveil the possibility of modifying quantum coherence of multiphoton systems upon propagation \cite{You2023CittertZernike}. We use this approach to describe the correlation properties of the multiphoton wavepackets that constitute our light beam. This allows us to use an equivalent density matrix for the system $\hat{\rho}_{ijkl}(z)$ (see Appendix \ref{app:A:CorrelationMatrix}) at the detection plane to compute its corresponding joint photon-number distribution $p_{ijkl}(n_1,n_2,z)$ as
\begin{equation}
    \begin{aligned}
 p(n_1,n_2,z) = \text{Tr}\left[\hat{\rho}_{ijkl}(z) |n_1,n_2\rangle\langle n_1,n_2|\right].
    \end{aligned}
       \label{eq5-ch6}
\end{equation}
\begin{figure}[!t]
    \centering
\includegraphics[width=0.85\textwidth]{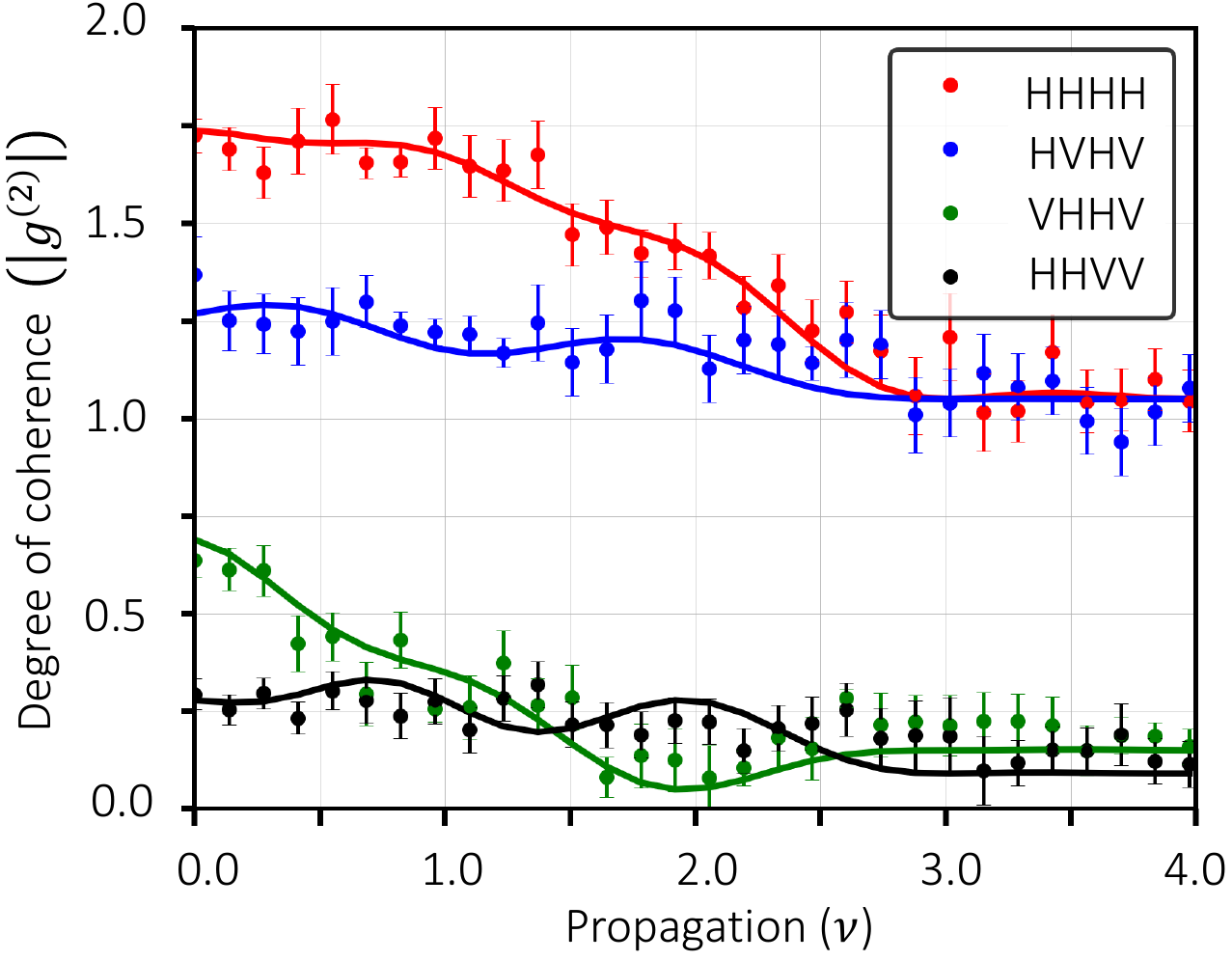}
    \mycaption{Measurement of Multiphoton Light with Sub-Shot-Noise Properties.}{ We isolate multiphoton subsystems with different polarization properties. These are characterized by the degree of second-order coherence $g^{(2)}_{ijkl}$. While the four multiphoton subsystems indicate the modification of quantum coherence with the $\nu$ parameter, it should be highlighted that the subsystems described by $g^{(2)}_{\text{VHHV}}$ and $g^{(2)}_{\text{HHVV}}$ show degrees of coherence below one. Notably,  quantum light sources with quantum statistical fluctuations below the shot-noise limit show degrees of coherence smaller than one. The continuous lines represent our theoretical predictions from Eq. (\ref{eq4-ch6}), whereas the dots indicate experimental data. These figures are taken from \cite{quantumcoherence2024}.}
    \label{Fig3-ch6}
\end{figure}
\noindent
As we shall see in the next section, these formulae allow for the prediction of very interesting correlation effects. Specifically, they predict that the statistical make-up of the light field is changing upon propagation in free space. The classical analogue to this behavior is explained by the van Cittert-Zernike theorem \cite{Cittert:34,ZERNIKE:38}, which predicts that the classical coherence properties of a light source can change upon free-space propagation. Therefore, we interpret our results in Eq. (\ref{eq4-ch6}) as those of a quantum van Cittert-Zernike theorem. This is because they predict the modification of quantum coherence upon free-space propagation, and that is directly analogous to the classical theorem's predictions. Specifically, Eq. (\ref{eq4-ch6}) predicts this free-space quantum modification through the nontrivial scattering effects induced by the Dirac-delta functions. Interestingly, these delta functions arise from the unique coherence properties of thermal light (see Appendix \ref{app:A:CorrelationMatrix}). Furthermore, Eq. (\ref{eq5-ch6}) allows us to study multiparticle quantum coherence, which is also changing upon free-space propagation. These quantum van Cittert-Zernike effects, therefore, are not only arising in polarization subsystems, but also in multiphoton subsystems. This showcases the fundamental and intrinsic quantum impacts of free-space propagation on our state.\\
\noindent
We explore the modification of the quantum coherence properties of propagating multiphoton systems using the experimental setup in Fig. \ref{Fig1-ch6}\text{b}. We use a combination of waveplates and a spatial light modulator (SLM) to rotate the polarization properties of our multiphoton system at any transverse position \cite{PRLMirhossein2014}. In addition, this arrangement enables us to characterize the polarization and photon-number distribution of multiphoton systems at different propagation planes. Specifically, we perform measurements at different propagation planes associated with the propagation parameter $\nu=L\Delta X/(\lambda z)$. Here, the transverse distance between detectors is described by $\Delta X$ and the wavelength of the beam by $\lambda$. As demonstrated in Fig. \ref{Fig2-ch6}, the many interference effects hosted by the propagating multiphoton system modify the photon-number distribution of the polarized components of the initial beam \cite{Loaiza2019, chenglongnature, npj2022}. These processes lead to multiphoton systems with different quantum fluctuations and quantum properties of coherence \cite{Loaiza2019, Hong2023}. Each multiphoton system is characterized through the degree of second-order self coherence
\begin{equation}
    g^{(2)}_{\nu}(0)=\frac{G^{(2)}_{\text{HHHH}} (\boldsymbol{r},\boldsymbol{r},z)}{G^{(1)}_{\text{HH}}(\boldsymbol{r},z)^2},
\end{equation}
where $G^{(1)}_{i,j}(\boldsymbol{r},z) = \braket{\hat{a}^\dagger_i(\boldsymbol{r})\hat{a}_j(\boldsymbol{r})} = \sqrt{I_0}L/(2 z^2 \lambda^2)$. Interestingly, the multiphoton system in Fig. \ref{Fig2-ch6}\text{a} is nearly thermal \cite{PhysRevCoherentIncoherent}. However, propagation leads to different kinds of multiphoton wavepackets. We show these from Fig. \ref{Fig2-ch6}\text{a} to \text{f}. The coherence properties of the multiphoton system in Fig. \ref{Fig2-ch6}\text{f} approach those observed in coherent light beams \cite{Gerry2004}. Remarkably, the conversion processes of the multiphoton system, described by Eq. (\ref{eq5-ch6}), take place in free space in the absence of light-matter interactions \cite{allen1987optical, venkataraman_phase_2013, Zasedatelev-2021, Snijders-2016, OLSEN2002373, chenglongnature,Kondakci2015}.
\begin{figure}[!hbpt]
    \centering \includegraphics[width=\textwidth] {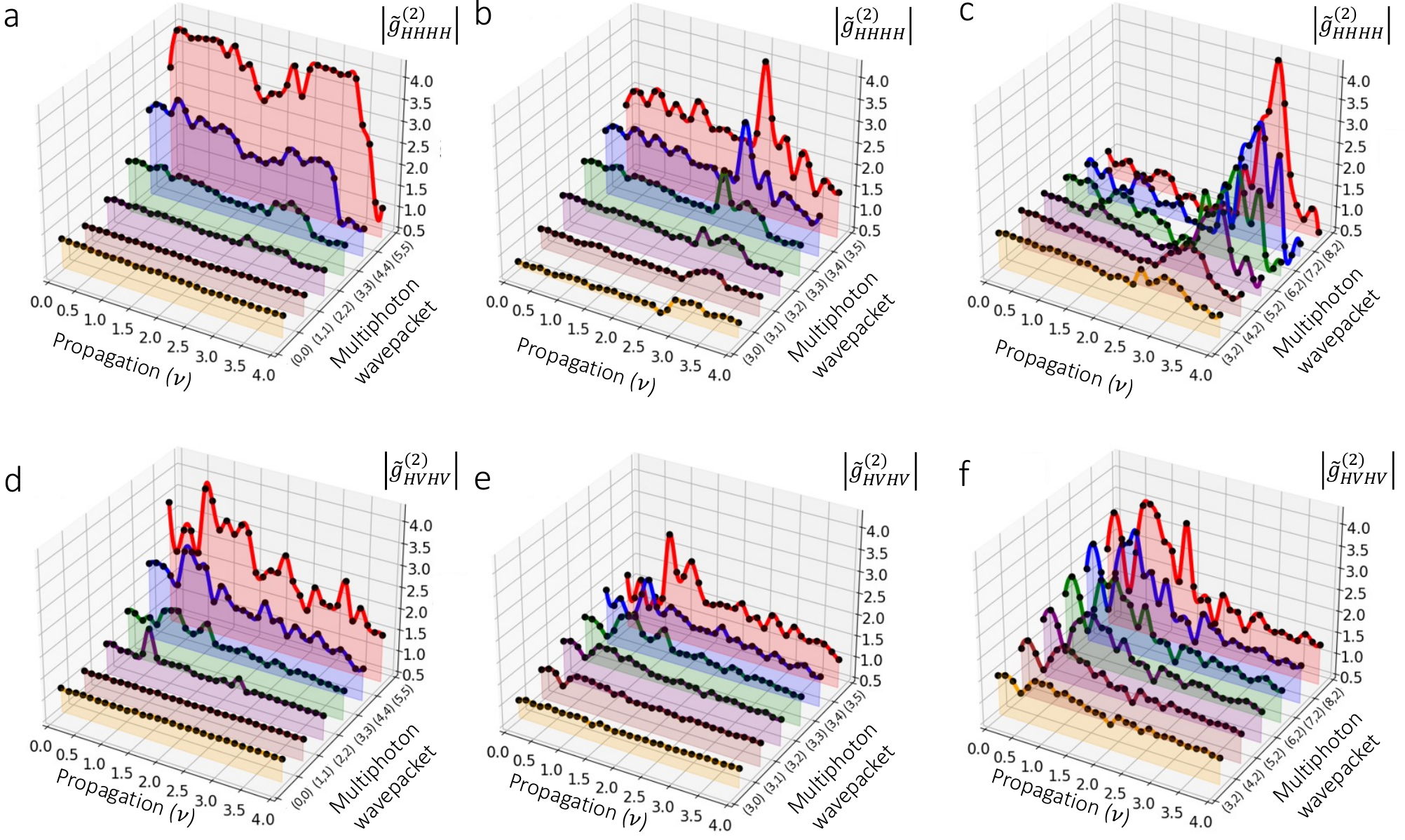}
    \mycaption{Quantum Coherence of Propagating Multiphoton Wavepackets.}{ The panels from (\text{a}) to (\text{c}) show the evolution of multiphoton wavepackets contained in the horizontally-polarized component of the initial thermal beam. We label the multiphoton wavepacket that leads to the detection of $n_1$ photons in arm 1 and $n_2$ photons in arm 2 with $(n_1,n_2)$. The results from (\text{a}) to (\text{c}) indicate that the multiphoton events that produce the degree of second-order coherence $g_{\text{HHHH}}^{(2)}$ in Fig. \ref{Fig3-ch6} follow distinct propagation dynamics. Although the contributions from the constituent wavepackets produce the trace described by $g_{\text{HHHH}}^{(2)}$, their individual propagation shows different coherence evolution. Specifically, we identify three representative dynamics. For example, multiphoton wavepackets with equal numbers for $n_1$ and $n_2$ exhibit the propagation dynamics in (\text{a}). In contrast, propagating wavepackets with different values of $n_1$ and $n_2$ show a different trend for the modification of quantum coherence, these are shown in (\text{b}) and (\text{c}). Moreover, the multiphoton wavepackets in the projected beam characterized by $g_{\text{HVHV}}^{(2)}$ exhibit the multiphoton dynamics reported from (\text{d}) to (\text{f}). The multiphoton dynamics in these panels also depend on the number of photons in each of the measured wavepackets. These figures are taken from \cite{quantumcoherence2024}. }
   \label{Fig4-ch6}
\end{figure} 
The polarization and photon-number properties of the propagating light beam at different transverse and longitudinal positions host many forms of interference effects \cite{ou2007multi, DellAnno2006}. We explore these dynamics by isolating the constituent multiphoton subsystems of the propagating beam. Each multiphoton subsystem, characterized by different polarization properties, exhibits different degrees of second-order coherence \cite{ChenglongAIP}. In the experiment, we perform projective measurements on polarization. These measurements unveil the possibility of extracting multiphoton subsystems with attenuated quantum fluctuations below the shot-noise limit \cite{ChenglongAIP, HashemiRafsanjani:17}. In this case, we use the four detectors depicted in the experimental setup in Fig. \ref{Fig1-ch6}\text{b} to perform full characterization of polarization \cite{Altepeter2005PhotonicST}. These measurements enable us to characterize correlations of multiphoton subsystems with different polarization properties, which are reported in Fig. \ref{Fig3-ch6}. We plot the degree of second-order mutual coherence
\begin{equation}
    g^{(2)}_{ijkl}(\boldsymbol{r}_1,\boldsymbol{r}_2,z)=\frac{G^{(2)}_{ijkl}(\boldsymbol{r}_1,\boldsymbol{r}_2,z)}{G^{(1)}_{i,j}(\boldsymbol{r}_1,z)G^{(1)}_{k,l}(\boldsymbol{r}_2,z)}.
\end{equation}
The propagation of the multiphoton subsystem described by $g^{(2)}_{\text{HHHH}}$ shows a modification of the quantum statistics from super-Poissonian to nearly Poissonian \cite{Hong2023, Gerry2004}. A similar situation prevails for the multiphoton subsystem described by $g^{(2)}_{\text{HVHV}}$. It is worth noticing that the multiphoton subsystems described by $g^{(2)}_{\text{VHHV}}$ and $g^{(2)}_{\text{HHVV}}$ unveil the possibility of extracting multiphoton subsystems with sub-shot-noise properties \cite{agarwal2012quantum}. This implies photon-number distributions narrower than the characteristic Poissonian distribution of coherent light \cite{PhysRevCoherentIncoherent,Mandel1979}. This peculiar feature might unlock novel paths towards the implementation of sensitive measurements with sub-shot-noise fluctuations \cite{Lawrie2019SqueezedLight}. 
We now turn our attention to describe the quantum coherence evolution of propagating multiphoton wavepackets. This is explored by projecting the polarized components of the initial thermal beam into its constituent multiphoton wavepackets \cite{PhysRevCoherentIncoherent}. In this case, we analyze wavepackets with $n_1+n_2$ number of photons.  The number of photons detected in arm 1 of our experiment is described by $n_1$, whereas $n_2$ indicates the number of photons detected in arm 2. Our findings unveil that despite the fact that the degree of second-order coherence $g^{(2)}_{\text{HHHH}}$ in Fig. \ref{Fig3-ch6} is produced by its constituent wavepackets, these show a completely different evolution of their properties of coherence. Our experimental measurements of these wavepackets can be found from Fig. \ref{Fig4-ch6}\text{a} to \text{c}. The results in Fig. \ref{Fig4-ch6}\text{a} indicate that multiphoton wavepackets, in which $n_1$ and $n_2$ are the same, show a particular evolution. In contrast, propagating wavepackets with asymmetric values of $n_1$ and $n_2$ show different trends in the modification of quantum coherence, these are shown in Fig. \ref{Fig4-ch6}\text{b} and \text{c}. The propagation of these wavepackets can be described using Eq. (\ref{eq5-ch6}). Specifically, we can calculate the multiphoton degree of second-order mutual coherence \cite{dawkins2024quantum}
\begin{equation}
    \Tilde{g}^{(2)}_{ijkl}(n_1,n_2,z)=\frac{p_{ijkl}(n_1,n_2,z)}{\sum_{n}p_{ijkl}(n,n_2,z)\sum_{m}p_{ijkl}(n_1,m,z)}.
\end{equation} 
Furthermore, the multiphoton wavepackets in the projected beam, characterized by $g^{(2)}_{\text{HVHV}}$, exhibit the multiphoton dynamics reported from Fig. \ref{Fig4-ch6}\text{d} to \text{f}. These results suggest that the multiphoton dynamics in Fig. \ref{Fig4-ch6} depend on the number of photons in each of the measured wavepackets.\\
\noindent
This quantum field theoretic approach to studying the quantum van Cittert-Zernike theorem provides us with the ability to describe the propagation dynamics of the multiphoton systems that constitute classical light beams. We used this formalism to extract propagating multiphoton subsystems, with quantum statistical properties, from unpolarized thermal light fields. While nonlinear light-matter interactions offer the possibility of engineering complex quantum systems \cite{venkataraman_phase_2013, Zasedatelev-2021, Snijders-2016}, our scheme exploits linear propagation of multiphoton systems \cite{HallajiNatPhys2017, Mirhosseini2016PRA}. This feature enabled us to exploit multiphoton scattering in free space to produce wavepackets with different quantum statistical properties \cite{Loaiza2019}. As such, our work combines the benefits of post-selective measurements with those of multiphoton scattering in propagating light beams, and it allows us to study the modification of the quantum statistical properties of multiphoton wavepackets in free space. Although, the incoherent combination of light beams with different polarization properties can lead to the modification of the degree of second-order coherence \cite{chenglongnature,Hong2024}, we performed direct measurements of polarized multiphoton systems with propagating quantum coherence properties (see Fig. \ref{Fig4-ch6}\text{a} to \text{c}). Interestingly, these processes are defined by the number of particles in the measured multiphoton system. Consequently, these findings have important implications for all-optical engineering of multiphoton quantum systems.\\
\noindent
We demonstrated the possibility of modifying the excitation mode of thermal multiphoton fields through free space propagation. This modification stems from the scattering of multiphoton wavepackets in the absence of light-matter interactions \cite{allen1987optical, venkataraman_phase_2013, Zasedatelev-2021, Snijders-2016, OLSEN2002373, chenglongnature, Kondakci2015,Wen}. The modification of the excitation mode of a photonic system and its associated quantum fluctuations result in the formation of different light fields with distinct quantum coherence properties \cite{PhysRevCoherentIncoherent, GlauberPR1963, PhysRevLettSudarshan}. The evolution of multiphoton quantum coherence is described through the nonclassical formulation of the van Cittert-Zernike theorem, unveiling conditions for the formation of multiphoton systems with attenuated quantum fluctuations below the sub-shot-noise limit \cite{ChenglongAIP,Batarseh:18, Lawrie2019SqueezedLight}. Notably, these quantum multiphoton systems emerge in the absence of optical nonlinearities, suggesting an all-optical approach for extracting multiphoton wavepackets with nonclassical statistics. We believe that the identification of this surprising multiphoton dynamics has important implications for multiphoton protocols quantum information \cite{Walmsley:23, aspuru-guzik_photonic_2012}.

%% file: chapter6.tex
\chapter{Multiphoton Quantum Imaging Using Natural Light}
\label{chap:chapter6}
\section{Motivation}

It is thought that schemes for quantum imaging are fragile against realistic environments in which the background noise is often stronger than the nonclassical signal of the imaging photons \cite{omar:2019,Genovese}. Unfortunately, it is unfeasible to produce brighter quantum light sources to alleviate this problem. Here, we overcome this paradigmatic limitation by developing a quantum imaging scheme that relies on the use of natural sources of light. This is achieved by performing conditional detection on the photon number of the thermal light field scattered by a remote object. Specifically, the conditional measurements in our scheme enable us to extract quantum features of the detected thermal photons to produce quantum images with improved signal-to-noise ratios. This technique shows a remarkable exponential enhancement in the contrast of quantum images.  Surprisingly, this measurement scheme enables the possibility of producing images from the vacuum fluctuations of the light field. This is experimentally demonstrated through the implementation of a single-pixel camera with photon-number-resolving capabilities. As such, we believe that our scheme opens a new paradigm in the field of quantum imaging. It also unveils the potential of combining natural light sources with nonclassical detection schemes for the development of robust quantum technologies.

\section{Background}
The use of nonclassical correlations of photons to produce optical images in a nonlocal fashion gave birth to the field of quantum imaging almost three decades ago \cite{PhysRevApticalimaging, PhysRevLettTwoPhotonCoincidence, omar:2019}. Interestingly, it was then discovered that exploiting the quantum properties of the light field enables improving the resolution of optical instruments beyond the diffraction limit \cite{TwoPhotonLithography, PhysRevLettScalableSpatial, PhysRevLettDowling, PhysRevXSuperresolution, npj2022}. It was also shown that schemes for quantum imaging allow for the formation of images with sub-shot-noise levels of precision \cite{Brida2010NaturePhotonics, Quantumsecuredimaging, Lemosundetectedphotons}. These features have been exploited to demonstrate the formation of few-photon images with high contrast \cite{APlMagana-Loaiza2013, Morris2015NatureComm, BarretoLemosmetrology}.  Furthermore, the compatibility of quantum imaging techniques with protocols for quantum cryptography have cast interest in the development of schemes for quantum-secured imaging \cite{Quantumsecuredimaging, omar:2019}. Despite the enormous potential of quantum imaging for microscopy, remote sensing, and astronomy, schemes for quantum imaging remain fragile against realistic conditions of loss and noise \cite{ref1Microscopy, PhysRevAGhostimaging, PhysRevLettPlenopticImaging, omar:2019, Genovese}. Unfortunately, these limitations render the realistic application of quantum imaging unfeasible \cite{ref1Sciencespatialcorrelations, omar:2019}.
Sharing similarities with other quantum technologies, existing techniques for quantum imaging rely on the use of nonclassical states of light \cite{OBrien2009NaturePhotonics, Lawrie2019SqueezedLight, midgal2013}. However, the brightness of available quantum light sources is generally low \cite{Loaiza2019, midgal2013, Hong2023}. For example, existing sources of nonclassical light allow for the preparation of few-photon states that exhibit fragile quantum correlations \cite{PhysRevLettThermalLight,PhysRevLett.TurbulenceFree, ref1Sciencespatialcorrelations}. This situation leads to common scenarios where environmental noise is typically larger than the signal of photons produced by processes of spontaneous parametric down-conversion or four-wave mixing \cite{DELLANNO200653, midgal2013}. Unfortunately, it is not feasible to produce brighter quantum light sources to overcome these limitations. Moreover, losses and noise cannot be avoided in realistic scenarios \cite{omar:2019}. Thus, any robust protocol for quantum imaging must rely on ubiquitous natural sources of light, such as thermal light. 
\section{Conceptual Overview}
Here we demonstrate the extraction of quantum images from classical noisy images produced by thermal sources of light \cite{You2023CittertZernike, Dakna}.  Specifically, our quantum imaging scheme isolates multiphoton subsystems of thermal light sources to dramatically improve the signal-to-noise ratio of imaging instruments. This robust protocol for quantum imaging is demonstrated through the implementation of a novel single-pixel camera with photon-number-resolving capabilities \cite{you2020identification}. Surprisingly, this quantum camera enables the extraction of information from the vacuum-fluctuation components of thermal light sources to produce quantum images with improved contrast. This technique shows a remarkable exponential improvement in the contrast of quantum images.
We also demonstrate the possibility of using correlated multiphoton subsystems to form high-contrast quantum images from images in which the background noise is comparable to the signal of thermal light sources. These surprising results can only be explained using quantum physics \cite{PhysRevCoherentIncoherent, PhysRevLettSudarshan}. Our work unveils the potential of combining natural light with nonclassical detection schemes for the development of robust quantum technologies. We believe that our findings open a new paradigm in the field of quantum imaging. 

\begin{figure}[!htb]
\centering	
\includegraphics[width=0.9\textwidth]{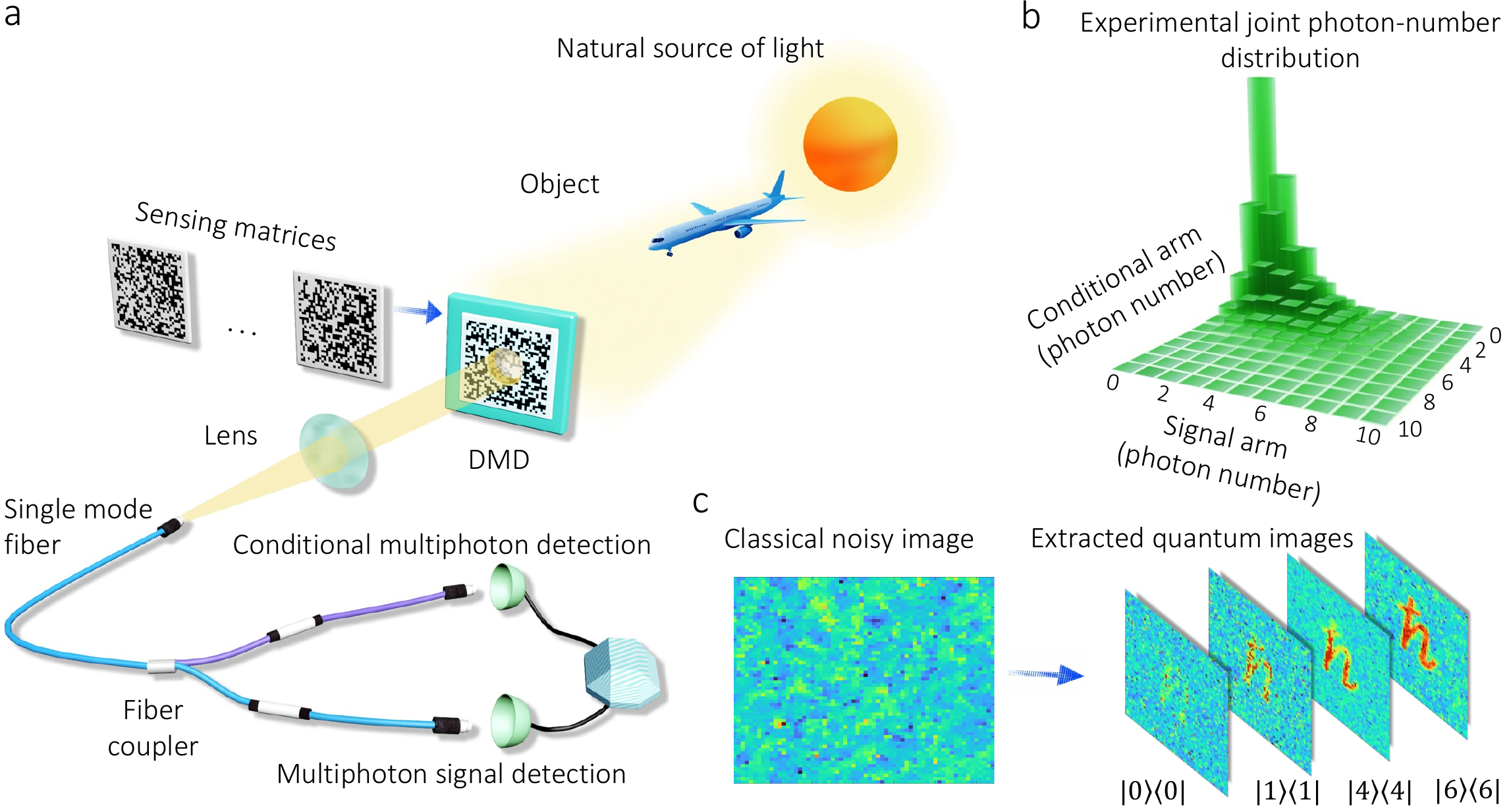}
	\mycaption{Multiphoton Quantum Imaging Using Natural Sources of Light.}{
The schematic in (\text{a}) depicts the implementation of a quantum camera with photon-number-resolving (PNR) capabilities. Here the thermal light field reflected off a target object is projected into a series of random binary matrices and then coupled into a single-mode fiber (SMF). The binary sensing matrices are displaced onto a digital micromirror device (DMD). Further, the thermal light field coupled into the SMF  is split by a 40:60 fiber coupler and measured by two PNR detectors. We report the experimental joint photon-number distribution of our thermal light source in (\text{b}). In this case, the degree of second-order coherence, $g^{(2)}(0)$, of the thermal light source is equal to 2. The series of PNR measurements for different binary sensing matrices enables us to use compressive sensing (CS) to demonstrate a single-pixel camera with PNR capabilities. As shown in (\text{c}), our ability to measure the multiphoton subsystems, represented by the elements of the joint photon-number distribution of the thermal source, enables us to demonstrate quantum imaging even in situations in which noise prevents the formation of the classical image of the object. Specifically, the environmental noise in (\text{c}) forbids the imaging of the character $\hbar$. However, the projection of the thermal light field into its vacuum  component reveals the presence of the object. Remarkably, the projection into larger multiphoton subsystems enables the extraction of quantum images of the object that was not visible in the classical image.} 
	\label{fig:figure1-ch5}
 \end{figure}
 \noindent 
In Fig. \ref{fig:figure1-ch5}\text{a} we illustrate the experimental implementation of our scheme for multiphoton quantum imaging. Here, the thermal light reflected off a target object is projected onto a digital micromirror device (DMD) where a series of random binary patterns are displayed. The thermal photons from the DMD are collected by a single-mode fiber (SMF) and then probabilistically split by a fiber coupler. The photons in each fiber are measured by two photon-number-resolving (PNR) detectors \cite{you2020identification, npj2022}. The random sensing matrices displayed on the DMD are used to implement a single-pixel camera \cite{PRLMirhossein2014, tutorial, Montaut2018, PhysRevAGhostimaging}. Further, our photon counting scheme enables us to project the coupled thermal light field into its constituent multiphoton subsystems. The joint photon-number distribution of the thermal source is reported in Fig. \ref{fig:figure1-ch5}\text{b}. The classical nature of the source is certified by the degree of second-order coherence $g^{(2)}$, which is equal to 2 in our experiment \cite{Gerry2004}. Each element in this joint probability distribution represents a multiphoton subsystem that we can isolate through the implementation of projective measurements \cite{PRLMirhossein2014, tutorial, Montaut2018, PhysRevAGhostimaging}. This measurement approach lies at the hearth of our protocol for multiphoton quantum imaging. 
\begin{figure*}[!htb]
\centering	
\includegraphics[width=0.87\textwidth]{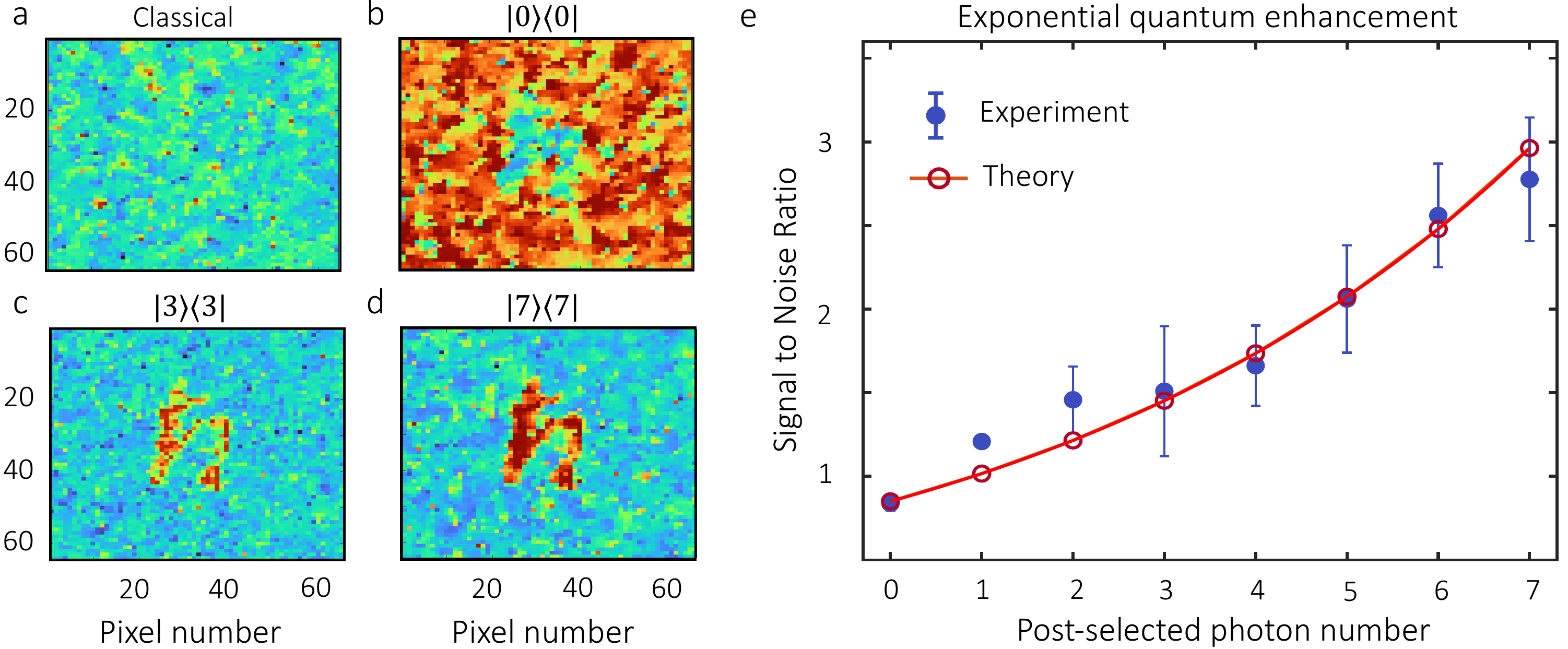}
\mycaption{Extraction of Quantum Images from a Classical CS Reconstruction.}{ The reconstructed image using our single-pixel camera for classical thermal light is shown in (\text{a}). In this case, environmental noise is higher than the signal and consequently the reconstructed image shows a low contrast that prevents the observation the object. Surprisingly, the projection of the light field into its vacuum component boosts the contrast of the image, this is reported in (\text{b}). Naturally, the formation of this image cannot be understood using classical optics. As demonstrated in (\text{c}), the projection of the thermal source of light into three-photon events enables the extraction of a quantum image with an improved signal-to-noise ratio (SNR). Remarkably, the projection of the detected thermal field into seven-particle subsystems leads to the formation of the high-contrast quantum image in (\text{d}). As reported in (\text{e}), and in agreement with Eq. (\ref{Eq:post}), the improvement in the SNR is exponential with the number of projected photons.  These results were obtained using 25$\%$ of the total number of measurements that can be used in our CS algorithm. Furthermore, the mean photon number $\bar{n}_t$ of the thermal light source is 0.8. 
} 
\label{fig:figure2-ch5}
\end{figure*}
As shown in Fig. \ref{fig:figure1-ch5}\text{c}, the projection of thermal light scattered by a target object into its constituent multiphoton subsystems enables the formation of high-contrast quantum images. This surprising effect enables extracting quantum images of a target object, even when environmental noise prevents the formation of its classical image through intensity measurements.
\section{Theory}
We now describe the quantum multiphoton processes that make this effect possible. For the sake of simplicity, we assume the uniform illumination of the object $\Vec{\boldsymbol{s}}_0$ by a thermal light field. As depicted in Fig. \ref{fig:figure1-ch5}\textbf{a}, 
the projection of the object into random sensing matrices, represented by the covector $\Vec{\boldsymbol{Q}}_t$, enables us to discretize the object into $X$ pixels. The label $t$ indexes the different sensing matrices. All such matrices can be represented by the $M\times X$ matrix  $\boldsymbol{Q} = \bigoplus_{t=1}^M \Vec{\boldsymbol{Q}}_t$, where $M$ is the number of sensing matrix configurations. Then, each filtering configuration results in a thermal state with a mean photon number given by $\bar{n}_t = \Vec{\boldsymbol{Q}}_t\cdot \Vec{\boldsymbol{s}}_0$. The multiphoton state after the fiber coupler can be written in terms of the Glauber-Sudarshan $P$ function as \cite{PhysRevLettSudarshan,PhysRevCoherentIncoherent}
\begin{equation}
\begin{aligned}
\label{Eq:state}
      \hat{\boldsymbol{\rho}}_{\boldsymbol{Q}} =&
      \bigoplus_{t=1}^M\int d^2\alpha \frac{1}{\pi\bar{n}_t}e^{-\frac{\left|\alpha\right|^2}{\bar{n}_t}}
      \left|\alpha \cos(\theta),i\alpha\sin(\theta)\rangle\langle\alpha \cos(\theta),i\alpha\sin(\theta)\right|_{a,b}.
\end{aligned}
\end{equation}
 \noindent 
The indices $a$ and $b$ denote the output modes of the fiber coupler. Furthermore, the parameter $\theta$ describes the splitting ratio between the two output ports. Next, we describe the signal-to-noise ratio (SNR) and how this quantity is modified by projecting the thermal field into its constituent multiphoton subsystems. To account for noise, we must consider photocounting with quantum efficiencies $\eta_{a/b}$ and noise counts $\nu_{a/b}$ \cite{Loaiza2019,PhysRevATruephoton,PhysRevAmultimodethermal}. Specifically, the joint photon-number distribution reported in Fig. \ref{fig:figure1-ch5}\textbf{b}, can be mathematically described as

   \begin{equation}
   \begin{aligned}
   \label{Eq:joint}
               \Vec{\boldsymbol{p}}_{\boldsymbol{Q}}(n, m)&=\bigoplus_{t=1}^M\left\langle: \frac{\left(\eta_a \hat{n}_a+\nu_a\right)^n}{n !} e^{-\left(\eta_a \hat{n}_a+\nu_a\right)} \otimes \frac{\left(\eta_b \hat{n}_b+\nu_b\right)^m}{m !} e^{-\left(\eta_b \hat{n}_b+\nu_b\right)}:\right\rangle\\
               &=\bigoplus_{t=1}^M\frac{e^{-\nu_a-\nu_b}}{\bar{n}_t n! m!}\sum_{i=0}^n\sum_{j=0}^m \binom{n}{i}\binom{m}{j}(i+j)!\frac{\eta_a^i\eta_b^j\nu_a^{n-i}\nu_b^{m-j}}{\left(\frac{1}{\bar{n}_t}+\eta_a\cos^2(\theta)+\eta_b\sin^2(\theta)\right)^{1+i+j}}\cos^{2i}(\theta)\sin^{2j}(\theta),
   \end{aligned}
    \end{equation} 
\noindent
where $\hat{n}_{a/b}$ is the photon number operator, and $:\cdot:$ represents the normal ordering prescription. We write the $t^{\text{th}}$ component of this vector as $p_{\boldsymbol{Q},t}(n,m)$. Additionally, when there is no signal and only noise is measured, we will have the probability distribution $p_{n,i}(k) = e^{-\nu_i}\frac{\nu_i^k}{k!}$ in each arm, where $i = a,b$. The joint probability distribution in this case is then given by $p_n(k,l) = p_{n,a}(k)p_{n,b}(l)$.
The two-mode multiphoton system described by Eq. (\ref{Eq:joint}) enables two schemes for projective measurements that lead to different scaling factors for the SNR of quantum images. First, we project one of the arms into a particular multiphoton subsystem. In other words, we ignore arm $b$ and implement a photon-number-projective measurement in arm $a$. For such post-selection on a multiphoton subsystem with $N$ photons, the SNR scales with 
\begin{equation}
      \label{Eq:post}
      \overrightarrow{\textbf{SNR}}_{\text{post}} = \frac{\sum_{m=0}^\infty \Vec{\boldsymbol{p}}_{\boldsymbol{Q}}(N,m)}{p_{n,a}(N)}=\frac{\Vec{\boldsymbol{p}}_{\boldsymbol{Q}}(N)}{p_{n,a}(N)}.
\end{equation} 
Remarkably, this expression follows an exponentially increasing trend with respect to $N$, meaning that post-selection can significantly reduce the noise of a quantum image.
\begin{figure*}[!tbp]
\centering	
\includegraphics[width=0.9\textwidth]{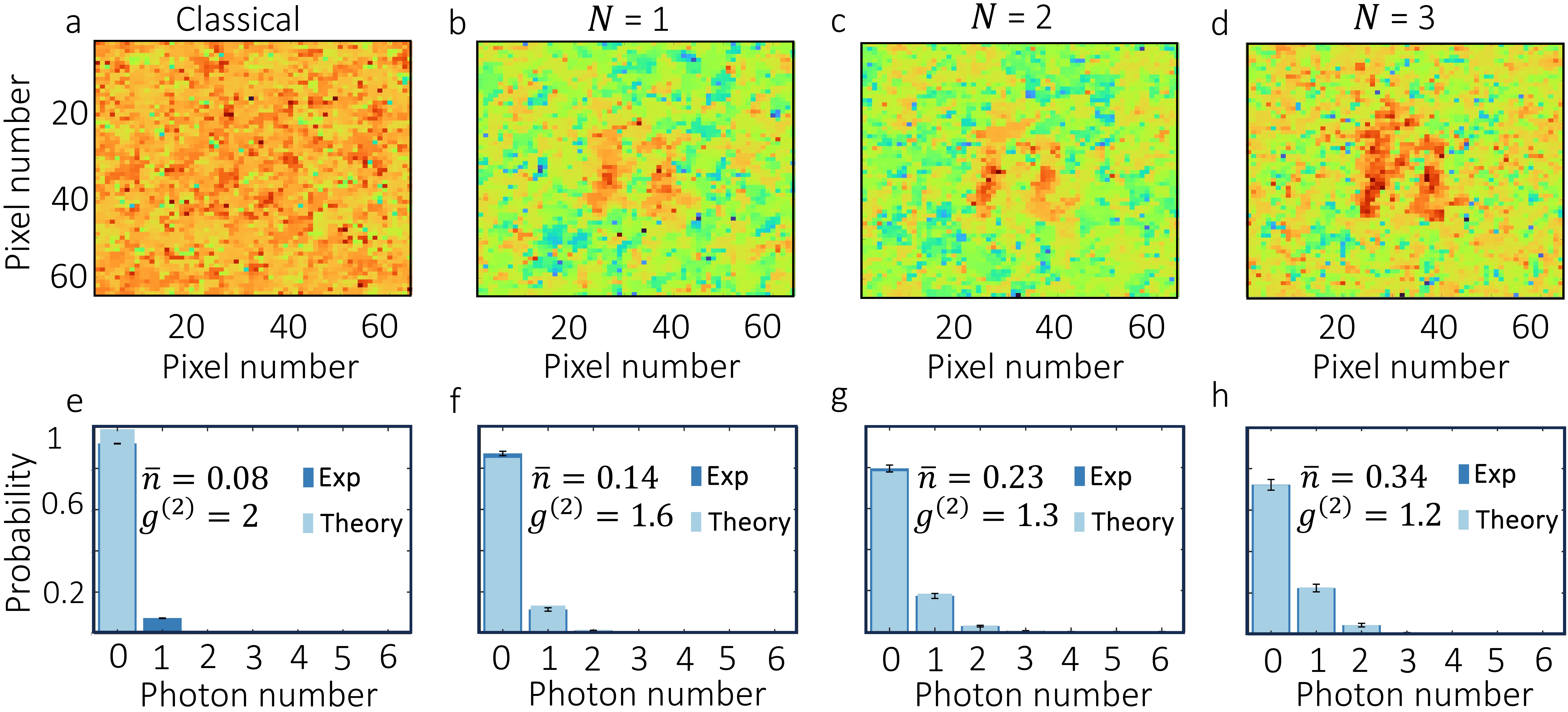}
\mycaption{Photon-Subtracted Multiphoton Quantum Imaging.}{ The noise accompanying a signal reflected off a target object produces the classical image reported in (\text{a}). Here, it is not possible to identify the object of interest with a classical single-pixel camera \cite{APlMagana-Loaiza2013}. The mean photon number $\bar{n}_t$ of our thermal light source is $0.08$. Interestingly, the subtraction of one photon improves the contrast of the image leading to the CS reconstruction in (\text{b}). Furthermore, our single-pixel camera with PNR capabilities enables multiphoton subtraction to produce the quantum images shown in  (\text{c}) and (\text{d}). In these cases, we subtracted two and three photons, respectively. These images were produced using only 12$\%$ of the total number of measurements that can be used in our CS algorithm. The advantage provided by our protocol for photon-subtracted quantum imaging can be understood through the photon-number distributions reported from (\text{e}) to (\text{h}). The unconditional detection of the weak thermal light signal produces the histogram in (\text{e}). This histogram unveils the overwhelming presence of vacuum and single-photon events used to reconstruct the image in (\text{a}). Furthermore, as shown in (\text{f}), the process of one-photon subtraction increases the mean photon number of the thermal signal while reducing its degree of second-order coherence $g^{(2)}$. The subtraction of two-photon events leads to a stronger signal characterized by the histogram in \text{g}. This conditional signal produces the enhanced image of the object in (\text{c}). Notably, the implementation of three-photon subtraction leads to the optical signal with nearly coherent statistics reported in (\text{h}). This boosted signal enables the reconstruction of the high-contrast image in (\text{d}). 
} 
\label{fig:figure3-ch5}
\end{figure*}
The second approach relies on the subtraction of $N$ photons from the thermal multiphoton system in Eq. (\ref{Eq:joint}) \cite{Dakna, Mostafavi2022, HashemiRafsanjani:17}. This procedure entails measuring photon events in arm $a$ conditioned on the detection of $N$ photons in arm $b$. Using Eq. (\ref{Eq:joint}), the intensity in arm $a$ is then given by $\langle\hat{\boldsymbol{n}}_a\rangle_N = \bigoplus_{t=0}^M\left(\sum_{k=0}^\infty k p_{\boldsymbol{Q},t}(k,N)\right)/\left(\sum_{k=0}^\infty p_{\boldsymbol{Q},t}(k,N)\right)$. Additionally, the photon-subtracted noise can be written as $\langle\hat{n}_a\rangle_{N,0} = \bigoplus_{t=0}^M\left(\sum_{k=0}^\infty k p_n(k,N)\right)/\left(\sum_{k=0}^\infty p_n(k,N)\right)$. This scheme leads to the following expression for the SNR:
\begin{equation}
 \label{Eq:sub}
    \overrightarrow{\textbf{SNR}}_{\text{sub}} = \frac{\langle\hat{\boldsymbol{n}}_a\rangle_N}{\langle\hat{n}_a\rangle_{N,0}}.
\end{equation}
The quantum enhancement for the SNR in this case is linearly increasing with respect to $N$. Therefore, photon-subtraction is also an effective means for noise-reduction. 
The series of spatial projective measurements described by the vector $\Vec{\boldsymbol{Q}}_t$ enables implementing a single-pixel camera with photon-number resolving capabilities via compressive sensing (CS) \cite{PRLMirhossein2014, tutorial, Montaut2018, PhysRevAGhostimaging}. This technique permits the reconstruction of  multiphoton quantum images described by $\Vec{\boldsymbol{s}}'$ via the minimization of the following quantity with respect to $\Vec{\boldsymbol{s}}'$: 
\begin{equation}
\label{eq:cs}
     \sum_{i=0}^X \lVert\nabla s_i'\rVert_{l_1} + \frac{\mu}{2}\lVert \boldsymbol{Q}\Vec{\boldsymbol{s}}' - \langle\hat{\boldsymbol{n}}\rangle\rVert_{l_2}.
\end{equation}
\noindent
As described above, $\langle\hat{\boldsymbol{n}}\rangle$ could be either $\Vec{\boldsymbol{p}}_{\boldsymbol{Q}}(N)$ or $\langle\hat{\boldsymbol{n}}_a\rangle_N$. Moreover, the $1$- and $2$-norm are denoted by $\lVert\cdot\rVert_{l_1}$ and $\lVert\cdot\rVert_{l_2}$, respectively. The discrete gradient operator is described by $\nabla$, and the penalty factor by $\mu$  \cite{tutorial, PRLMirhossein2014,Montaut2018, PhysRevAGhostimaging}.
\section{Experimental Results and Discussions}
We now discuss the experimental process of quantum-image extraction from classical images. This was implemented using one PNR detector. In Fig. \ref{fig:figure2-ch5}\text{a}, we show the CS reconstruction of a classical image for a situation in which environmental noise is comparable to the signal. In this case, the level of noise forbids the observation of the object. The mean photon number $\bar{n}_{t}$ of the thermal light source is 0.8.  Surprisingly, the projection of the thermal signal into its vacuum component reveals the presence of the object. As such, the quantum image in Fig. \ref{fig:figure2-ch5}\text{b} is formed by the vacuum-fluctuation component of the electromagnetic field and cannot be explained using classical physics \cite{Genovese,PhysRevLettSudarshan,PhysRevCoherentIncoherent, You2023CittertZernike}. This nonclasical reconstruction, obtained from the measurement of vacuum events, demonstrates that the process of projecting the thermal light signal into one of its constituent quantum subsystems, such as the vacuum, modifies the SNR as established by Eq. (\ref{Eq:post}). As suggested by the reconstruction in Fig. \ref{fig:figure2-ch5}\text{c}, the post-selection on higher multiphoton events leads to quantum images with an improved contrast. Interestingly, the projection of the thermal light signal into seven-photon subsystems leads to a dramatic improvement of the contrast of the image. This effect becomes evident in the quantum image shown in Fig. \ref{fig:figure2-ch5}\text{d}. Remarkably, the exponential growth of the SNR with the number of projected multiphoton subsystems is summarized in Fig. \ref{fig:figure2-ch5}\text{e}. These results demonstrate that our single-pixel camera with PNR capabilities enables the extraction of quantum multiphoton images with unprecedented degrees of contrast \cite{Brida2010NaturePhotonics,npj2022, Morris2015NatureComm, Genovese, omar:2019}. 
\begin{figure}[!htb]
\centering
\includegraphics[width=0.8\textwidth]{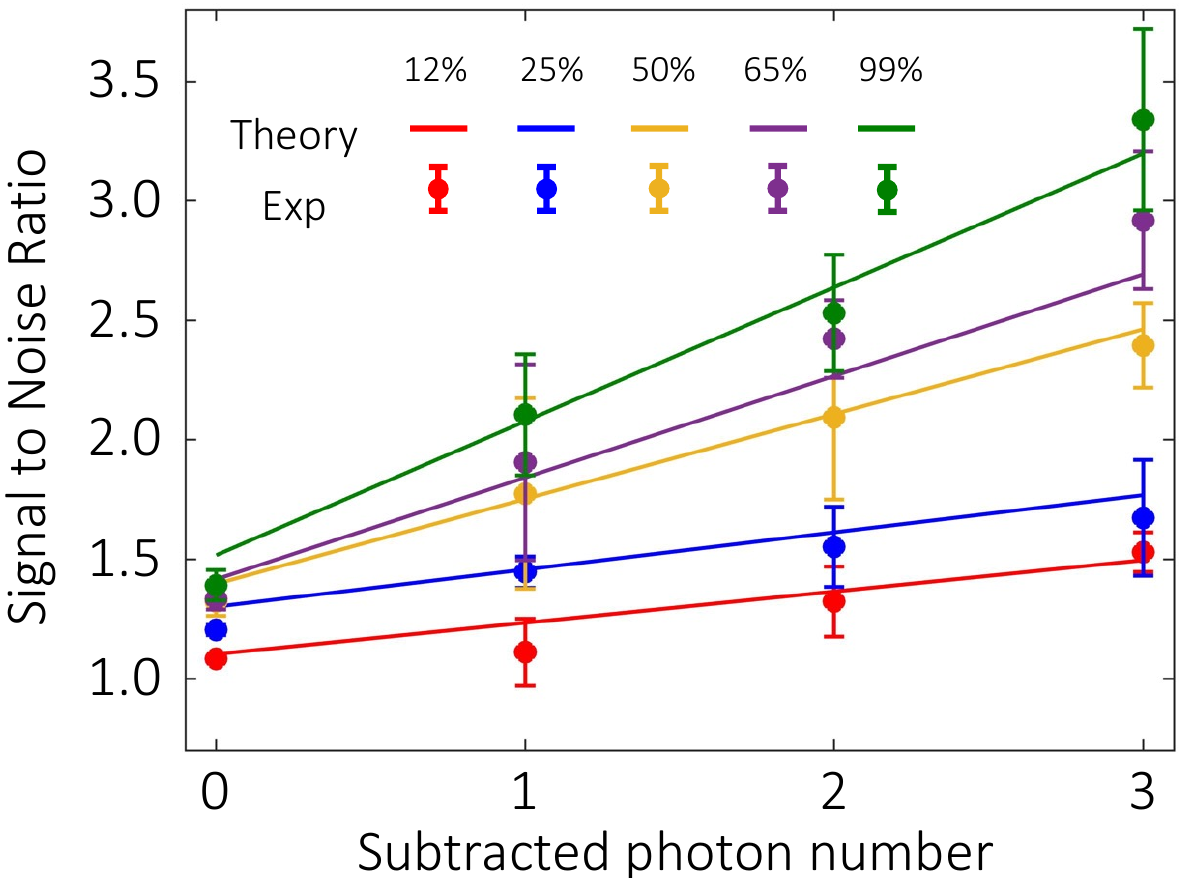}
\mycaption{Performance of Photon-Subtracted Multiphoton Quantum Imaging.}{ The SNR of the photon-subtracted quantum images shows a linear dependence on the number of subtracted photons. This behavior is in good agreement with Eq. (\ref{Eq:sub}). Interestingly, the collection of larger sets of data leads to faster improvements of the SNR for our multiphoton quantum imaging scheme. The percentages represent the number of CS measurements with respect to the total number of pixels in the image. The error bars represent the standard deviation of the SNR of image reconstructions using five different datasets, where each dataset contains millions of PNR measurements. } 
\label{fig:figure4-ch5}
\end{figure}
While the projection of thermal light into its constituent multiphoton subsystems enables the extraction of quantum images with high contrast, it is also possible to correlate photon events to improve the SNR of a quantum imaging protocol. This feature also enables us to perform quantum imaging at low light levels. We now experimentally demonstrate this possibility by implementing a scheme for photon subtraction on our single-pixel camera with PNR capabilities. In this case, the mean photon number $\bar{n}_t$ is equal to 0.08,
one order of magnitude lower than the brightness of the source used for the experiment discussed in Fig. \ref{fig:figure2-ch5}. As illustrated in Fig. \ref{fig:figure1-ch5}\text{a}, this quantum imaging scheme utilizes two PNR detectors \cite{Loaiza2019,You2021Scalablemultiphoton}. First, we use the noisy thermal signal to reconstruct the classical image shown in Fig. \ref{fig:figure3-ch5}\text{a}. Here, the large levels of noise forbid the observation of the target object. Remarkably, the subtraction of one photon from the thermal noisy signal reveals the presence of the object in Fig. \ref{fig:figure3-ch5}\text{b}. As predicted by Eq. (\ref{Eq:sub}), the process of multiphoton subtraction leads to enhanced quantum images. Specifically, two-photon subtraction leads to the improved image in Fig. \ref{fig:figure3-ch5}\text{c}. Furthermore, the CS reconstruction of the three-photon subtracted quantum image reported in Fig. \ref{fig:figure3-ch5}\text{d} shows a significant  improvement of the contrast with respect to the classical image in Fig. \ref{fig:figure3-ch5}\text{a}. The physics behind our scheme for quantum imaging can be understood through the increasing mean photon number that characterizes the histograms shown from Fig. \ref{fig:figure3-ch5}\text{e} to \text{h}. Moreover, the thermal fluctuations of the detected field are reduced by subtracting photons \cite{Dakna, omar:2019, HashemiRafsanjani:17}. This effect is indicated by the decreasing degree of second-order coherence $g^{(2)}$ corresponding to the photon-number distributions in Fig. \ref{fig:figure3-ch5}.
\noindent
The improvement in the SNR of the experimental photon-subtracted quantum images is quantified in Fig. \ref{fig:figure4-ch5}. In agreement with Eq. (\ref{Eq:sub}), the contrast of the filtered images, as a function of the number of subtracted photons, follows a linear dependence. Although, the benefits of our photon-subtracted scheme for multiphoton quantum imaging are evident for small and incomplete sets of data, the rate at which the SNR increases can be further amplified by collecting larger sets of data. It is worth noting that the exponential and linear mechanisms, reported in Fig. \ref{fig:figure2-ch5} and Fig. \ref{fig:figure4-ch5}, for improving the SNR of weak and noisy imaging signals have the potential to enable the realistic application of robust quantum cameras with PNR capabilities \cite{npj2022,Cheng2023}. As such, these findings could lead to novel quantum techniques for multiphoton microscopy and remote sensing \cite{Morris2015NatureComm, omar:2019, Genovese}.
Quantum imaging schemes have been demonstrated to be fragile against realistic environments in which the background is comparable to the nonclassical signal of the imaging photons 
\cite{ref1Microscopy, PhysRevAGhostimaging, PhysRevLettPlenopticImaging, omar:2019, Genovese, ref1Sciencespatialcorrelations, omar:2019}. This issue prevents the realistic application of quantum imaging techniques for microscopy, remote sensing, and astronomy \cite{Morris2015NatureComm, omar:2019, Genovese}. In this chapter, we overcome this paradigmatic limitation by developing a multiphoton quantum imaging scheme that relies on the use of natural sources of light.  This is demonstrated through the implementation of a single-pixel camera with photon number resolving capabilities that enables the projection of classical thermal light fields into its constituent multiphoton subsystems. This kind of quantum measurement enables us to extract high-contrast quantum images from noisy classical images of target objects. Our technique shows a remarkable exponential enhancement in the contrast of quantum images.
Surprisingly, we demonstrated the formation of quantum images produced by the vacuum-fluctuation components of thermal light sources. We also demonstrate the possibility of using correlated multiphoton subsystems to form high-contrast quantum images from images in which the background noise is comparable to the signal of thermal light sources. Thus, we believe that our scheme opens a new paradigm in the field of quantum imaging \cite{omar:2019, Genovese,OBrien2009NaturePhotonics, Lawrie2019SqueezedLight}. Furthermore, it unveils the potential of combining natural light sources with nonclassical detection schemes for the development of robust quantum technologies \cite{omar:2019, Genovese,OBrien2009NaturePhotonics, Lawrie2019SqueezedLight, DELLANNO200653}.
\subsection{Experimental Setup}
Our proof-of-principle quantum imaging setup utilizes pseudo-thermal light, which has the same properties of coherence as natural sources of light \cite{Arecchi65PRL}. This source is generated by passing the coherent light from a continuous-wave laser at 633 nm through a rotating ground glass \cite{Arecchi65PRL,you2020identification}. The thermal light is then collected into a single-mode fiber and collimated with a lens ($f=5$ cm) to illuminate the target object. Here, the target object ``$\hbar$'' is generated using a digital micro-mirror device (DLP6500 DLP$\textsuperscript{\textregistered}$ DMD). Then, the reflected ``$\hbar$'' is projected onto a second DMD with a 4-f system comprised of two lenses, each with a focal length of 10 cm. This second DMD facilitates compressive sensing by displaying a series of random binary matrices \cite{PRLMirhossein2014, APlMagana-Loaiza2013}. Next, the reflected light from this DMD is imaged using a another 4-f system comprised of two lenses, with focal lengths of $25$ and $10$ cm. Then, we couple the reflected light into a 1$\times$2 50:50 fiber beam splitter (Thorlabs TW630R5F1) using a Rochester lens ($f=4.5$ mm). The split beams are detected by two fiber-coupled avalanche photodiodes (APDs, Excelitas SPCM-AQRH-13-FC), where photon-number-resolving detection is implemented \cite{npj2022, you2020identification}. Finally, these detection events are recorded by a time tagger (PicoQuant MultiHarp 150) and analyzed. This experimental setup allows us to accurately measure the joint photon-number distribution at both outputs of the fiber beam-splitter. This enables us to perform photon subtraction and post-selection for image reconstruction.
\subsection{Data Analysis and Image Reconstruction}
In the compressive sensing process, we sequentially display some percentage of 4096 unique random matrices on the second DMD, with the measurement time for each matrix fixed at one second. For example, the reconstructions for Fig. \ref{fig:figure2-ch5} utilized 1025 projective measurements, which corresponds to 25\% of the 4096 measurements that our CS algorithm can use. The images reported on Fig. \ref{fig:figure3-ch5} were obtained with 12\% of the measurements, which means 512 binary matrices. 
To reconstruct the image, we apply the TVAL3 algorithm \cite{li2010efficient}. Due to the long coherence time of our pseudo-thermal light source, we bin each one-second measurement into 1 $\mu$s intervals \cite{you2020identification, HashemiRafsanjani:17}. Then we count the number of detections in each bin, and this defines a photon-number resolving event. For each event, we then denote the photon-counts in the two arms as $n_1$ and $n_2$. To achieve $N$-photon subtraction, we isolate the events where $n_2 = N$. In other words, we filter $n_1$ by only considering the events where $n_2 = N$ in the second arm. We then perform the reconstruction process with this conditional dataset to obtain an enhanced image quality. Conversely, for post-selection on $n$ and $m$ photons, we compute the probability that $n_1=n$ and $n_2=m$. Then, we perform image reconstruction using this probability to obtain an improvement in image quality.

%% file: chapter7.tex
\chapter{Conclusion}
This dissertation investigated the complex dynamics of multiphoton quantum systems for applications in quantum sensing. Chapters 2 and 3 explored light-matter interaction in plasmonic platforms, revealing how these platforms can manipulate quantum statistics and control fluctuations \cite{You2020,Mostafavi2022}. This paves the way for enhanced precision in photonic sensors. Notably, the optical near fields allow for the exploitation of quantum correlations and spatial coherence manipulation. These advancements in understanding the link between spatial coherence and quantum statistics have significant implications for quantum plasmonic sensing, information processing, and many-body systems.
Building upon these insights, future work will focus on expanding experimental applications in quantum plasmonics, particularly by integrating structured light and projective measurement techniques. \\
Chapter 4 presented innovative approaches to information encoding in fiber optic communication using structured light and machine learning, highlighting the synergy between quantum principles and modern techniques. Our encryption protocol leverages high-dimensional spatial modes for secure data transmission \cite{Lollie2022}. Additionally, as part of our current investigation, we are exploring the exciting potential of quantum random walks utilizing structured light. 
Furthermore, Chapter 5 demonstrated the manipulation of thermal multiphoton fields' excitation mode through free-space propagation. This modification arises solely from the scattering of multiphoton wavepackets, without any light-matter interaction. This phenomenon can be described by the nonclassical van Cittert-Zernike theorem, revealing conditions to generate multiphoton systems with remarkably low quantum fluctuations, even surpassing the shot-noise limit. Notably, these advancements were achieved without the need for optical nonlinearities, paving the way for an all-optical method to extract multiphoton wavepackets exhibiting nonclassical statistics. Our findings hold significant implications for the future development of quantum metrology techniques \cite{quantumcoherence2024}. 
Finally, Chapter 6 tackled a challenge in quantum imaging: overcoming strong background noise in real-world environments.  This chapter showcases the potential of merging classical and quantum techniques by successfully utilizing natural light sources. Our innovative approach paves the way for the development of more robust quantum technologies. Our contribution demonstrates a single-pixel camera capable of resolving individual photons. This camera essentially breaks down classical light into its fundamental quantum components, enabling the capture of high-contrast quantum images.